\documentclass
[superscriptaddress,secnumarabic,amssymb,amsmath,nobibnotes,aps,prd,nofootinbib,nopacsnumber,onecolumn,10pt]{revtex4}%
\usepackage{graphicx}
\usepackage{epsf}
\usepackage{bm}
\usepackage{placeins}
\usepackage{amsmath}
\usepackage{amsfonts}
\usepackage{amssymb}%
\usepackage{csquotes}
\providecommand{\U}[1]{\protect\rule{.1in}{.1in}}

\newcommand{\be}{\begin{equation}}
\newcommand{\ee}{\end{equation}}

\newcommand{\mincir}{\raise
-3.truept\hbox{\rlap{\hbox{$\sim$}}\raise4.truept\hbox{$<$}\ }}
\newcommand{\magcir}{\raise
-3.truept\hbox{\rlap{\hbox{$\sim$}}\raise4.truept\hbox{$>$}\ }}

\RequirePackage{latexsym}
\usepackage{hyperref}
\hypersetup{
    colorlinks=true,
    linkcolor=red,
    citecolor=blue,
    filecolor=magenta,      
    urlcolor=cyan,
    pdftitle={Generalized Scalar Field Cosmologies: a Perturbative Analysis},
    pdfpagemode=FullScreen}  
\newcommand{\bea}{\begin{eqnarray}}
\newcommand{\eea}{\end{eqnarray}}

\newcommand{\bef}{\begin{figure}}  \newcommand{\eef}{\end{figure}}
\newcommand{\bec}{\begin{center}}  \newcommand{\eec}{\end{center}}

\usepackage{url} 
\usepackage{pdflscape} 
\usepackage{amsmath}
\usepackage{amssymb} 
\usepackage{epsfig} 
\usepackage{amsfonts}
\usepackage{graphicx}
\usepackage{subfigure}
\usepackage{graphicx}
\usepackage{times,fancyhdr}
\usepackage{enumitem}
\setlist[enumerate,2]{label=\roman*)}

\def\case#1/#2{\textstyle\frac{#1}{#2}}

\newcommand{\ben}{\begin{eqnarray}}
\newcommand{\een}{\end{eqnarray}}
\usepackage{mdframed}
\usepackage{grffile}
\usepackage{amssymb}
\usepackage[T1]{fontenc}

\usepackage{amssymb,amsmath,amsfonts,amsthm,graphicx, graphics, setspace}
\usepackage{bigints}
\usepackage{anysize}
\usepackage{float}
\usepackage{fancyhdr}
\usepackage{subfigure}
\usepackage{pdflscape}
\usepackage{epsfig}
\usepackage{epsf}
\usepackage{epstopdf}
\usepackage{hyperref} 
\theoremstyle{remark}
\newtheorem{example}{Example}[section]
\newtheorem{theorem}{Theorem}[section]
\newtheorem{proposition}{Proposition}

\providecommand{\U}[1]{\protect\rule{.1in}{.1in}}

\usepackage{tikz,xcolor,hyperref}

\definecolor{lime}{HTML}{A6CE39}
\DeclareRobustCommand{\orcidicon}{%
	\begin{tikzpicture}
	\draw[lime, fill=lime] (0,0) 
	circle [radius=0.16] 
	node[white] {{\fontfamily{qag}\selectfont \tiny ID}};
	\draw[white, fill=white] (-0.0625,0.095) 
	circle [radius=0.007];
	\end{tikzpicture}
	\hspace{-2mm}
}

\foreach \x in {A, ..., Z}{%
	\expandafter\xdef\csname orcid\x\endcsname{\noexpand\href{https://orcid.org/\csname orcidauthor\x\endcsname}{\noexpand\orcidicon}}
}

\begin{document}

\title[Averaging Dynamics of Scalar Field-Matter Interacting Models in Anisotropic Universes: The Locally Rotationally Symmetric Bianchi I Spacetime]{Averaging Dynamics of Scalar Field-Matter Interacting Models in Anisotropic Universes: The Locally Rotationally Symmetric  Bianchi I Spacetime}
\author{Alfredo D. Millano\orcidC{}}
\address{Departamento  de  Matem\'aticas,  Universidad  Cat\'olica  del  Norte, Avda. Angamos  0610,  Casilla  1280  Antofagasta,  Chile}
\email{alfredo.millano@ce.ucn.cl}
\author{Genly Leon\orcidA{}}
\address{Departamento  de  Matem\'aticas,  Universidad  Cat\'olica  del  Norte, Avda. Angamos  0610,  Casilla  1280  Antofagasta,  Chile}
\address{Department of Mathematics, Faculty of Applied Sciences, Durban University of Technology, Durban 4000, South Africa}
\email{genly.leon@ucn.cl}
\author{Andronikos Paliathanasis\orcidB{}}
\address{Department of Mathematics, Faculty of Applied Sciences, Durban University of Technology, Durban 4000, South Africa}
\affiliation{National Institute for Theoretical and Computational Sciences (NITheCS), South Africa}
\email{anpaliat@phys.uoa.gr}

\begin{abstract}
We consider an anisotropic cosmological model based on the locally rotational Bianchi I spacetime, incorporating a scalar field and a non-zero cosmological interaction term. The framework of averaging theory is employed to study the associated non-linear differential equations. Through a qualitative analysis of the gravitational field equations, we obtain valuable insights into the structure of the solution space for the anisotropic scalar field model with a generalized harmonic potential. The interaction between the scalar field and matter is described by a general expression that depends on the Hubble parameter, the time derivative of the scalar field, and the energy densities of cold dark matter and dark energy. This formulation involves real parameters that modulate the interaction, as well as a coupling constant with the dimensions of the Hubble parameter. We show that the Hubble parameter serves as a time-dependent perturbation parameter, controlling the discrepancy between the full system and its time-averaged counterpart. As this parameter decreases, both systems converge to the same asymptotic behaviour. This enables the suppression of oscillatory effects, significantly simplifying the dynamical analysis. Finally, we identify conditions on the interaction parameters that ensure the regularity of the system's evolution by preventing the emergence of singularities.
\end{abstract}


\maketitle
\section{Introduction}
\label{intro}
Cosmological models with interaction between dark matter and dark
energy \cite{Amendola:1999er,Clemson:2011an,Wang:2006qw,Cai:2004dk,SantanaJunior:2024cug,Paliathanasis:2024abl,Munyeshyaka:2022fck,Benetti:2019lxu,Luongo:2013eza} are of particular interest, as they offer a simple mechanism to
alleviate current cosmological tensions \cite{CosmoVerse:2025txj,Abdalla:2022yfr,An:2018vzw,Gomez-Valent:2020mqn,Amendola:2003wa,DiValentino:2019ffd,Kumar:2019wfs,Carloni:2025jlk,Luongo:2013eza,Yang:2019jwn,Lucca:2021eqy,Guo:2021rrz,Dahmani:2023bsb,Gao:2021xnk,Mishra:2023ueo,vanderWesthuizen:2023hcl,Gariazzo:2021qtg,Bhattacharya:2024nua,Shah:2024rme,Bernui:2023byc} as also other cosmological problems. The analysis of recent cosmological
observations \cite{Costa:2019uvk,Joseph:2022khn,Brax:2023tvn,Chakraborty:2024xas,Giare:2024smz,Montani:2024pou,Li:2024qso,Sabogal:2024yha,Lima:2024wmy,Yang:2025vnm,Shah:2025ayl,Silva:2025hxw,You:2025uon,Yang:2025boq,vanderWesthuizen:2025iam,Sahoo:2025cvz,Paliathanasis:2025dcr,Paliathanasis:2025xxm,Li:2025ula} does not rule out interacting models; on the contrary, some
studies suggest that these observations may even favor interacting models over
the $\Lambda$-Cosmology. 

Only a few cosmological interacting models arise from
theoretical considerations, such as those based on Weyl Integrable Spacetime \cite{Salim:1996ei,Romero:2012hs,Chatzidakis:2022mpf,Konstantinov:1994jg},
quintom scalar fields \cite{Yang:2024kdo,Cai:2006dm,Li:2025cxn,Yang:2025mws,Basilakos:2025olm,Cai:2009zp,Tot:2022dpr,Leon:2018lnd,Leon:2012vt,Lazkoz:2007mx}, Chiral models \cite{Chervon:2013btx,Brown:2017osf,Paliathanasis:2023moe,Paliathanasis:2022mnb,Giacomini:2021kuf,Paliathanasis:2020wjl,Christodoulidis:2025wew,Sa:2023coi,DAgostino:2021vvv,Anguelova:2023dui,Iacconi:2023slv}, and the Chameleon mechanism \cite{Khoury:2003aq,Khoury:2003rn,Paliathanasis:2024sle,Altarawneh:2025mig,Zaregonbadi:2022lpw,Singh:2022afk,Biswas:2022udk}. However,
over the past decades, a wide variety of phenomenological models have also
been proposed in the literature. The nature of the interaction is essential
for the determination of the cosmological fluid.

Inflation is the acceleration era of the early universe which has been
introduced solve long-standing problems about the structure of the universe
\cite{Guth:1980zm,Linde:2007fr,Barrow:1990td,Achucarro:2022qrl} that would otherwise require special initial conditions
\cite{Planck:2015sxf}. According to the cosmic "no-hair" \cite{Barrow:1987ia}
the rapid expansion described by the de Sitter solution causes the universe to
effectively lose memory of its initial conditions.

This cosmic acceleration epoch is attributed to a scalar field known as
inflaton \cite{Guth:1980zm}. Within the framework of General Relativity, the inflaton contributes
dynamically to the energy-momentum tensor, providing a natural mechanism to
drive the early accelerated expansion. This scalar field could also serve as a
major candidate for the dark sector of the universe, particularly as a source
of dark energy \cite{Ratra:1987rm,Tsujikawa:2013fta,Gialamas:2025pwv}. Hence, scalar fields can be used to connect the two
acceleration eras.

Although nowadays, in large scales, the universe is considered isotropic and
homogeneous described by the spatially flat
Friedmann--Lema\^{\i}tre--Robertson--Walker (FLRW) geometry. Anisotropies may played an important role in the early stages of the cosmic history \cite{Ryan:1975jw}.  The Bianchi classification includes the nine possible four-dimensional homogeneous anisotropic spaces with a three-dimensional space-like isometries. Bianchi I is a geometry with three isometries which form the $T_{3}$ Lie algebra. In the limit of
isotropization Bianchi I model is reduced to the spatially flat FLRW
spacetime. 

In the vacuum, the field equations of General Relativity
for the Bianchi I geometry lead to the anisotropic Kasner universe \cite{Kasner:1921zz}. It is a
singular solution which describe the behaviour of the anisotropies near to the
singularity. However, when the cosmological constant is introduced the Bianchi
I universe lead to an asymptotically isotropic FLRW universe. The Kasner universe has been played an important role as a paradigm for the study of the observational consequences of anisotropic expansion during the cosmic history, including quantum particle creation, baryosynthesis, and others \cite{Zeldovich:1971mw,Barrow:1981bv,Barrow:1982ei,Hawking:1966vx,Gupt:2012vi,Rennert:2013qsa}.  Due to the
importance of the Bianchi cosmologies on the description of the early stages
of the universe, there is a plethora of studies in the literature on Bianchi
models \cite{Paliathanasis:2022mnb,Giacomini:2021kuf,Liu:2017edh,Paliathanasis:2017htk,Skugoreva:2017vde,Paliathanasis:2022mrp,Ritchie:2022bvk,Sharma:2023abm,deCesare:2019suk,Neves:2022qyb,Frolov:2001wz,Toporensky:2018xpo,Paliathanasis:2024qkh,Murtaza:2025gme,Adhav:2012zz,FarasatShamir:2016ipk}.  

In this study we study the cosmological dynamics and the evolution of the
anisotropies on the homogeneous Bianchi I geometry, where we assume the
presence of scalar field for the description of the dark energy, and a
pressureless fluid for the description of dark matter. Furthermore, we
consider nonzero interaction in the dark sector.\ We employ the method of
averaging for the analysis of the non-linear cosmological field equations. The
averaging method is an analytic technique for the analysis of non-linear differential equations \cite{grimshaw2017nonlinear,guckenheimer1983nonlinear,sanders2010averaging,strogatz2024nonlinear} and in recent years, it has been systematically applied to scalar field cosmologies characterized by generalized harmonic potentials and oscillatory matter dynamics, yielding powerful asymptotic results.

A multi-part series by Leon et al. \cite{Leon:2021lct,Leon:2021rcx, Leon:2021hxc} established that systems of differential equations defined on LRS Bianchi I, III, and both flat and closed FLRW models have identical late-time attractors when compared to their time-averaged counterparts, confirming that the averaged dynamical system captures the asymptotic fate of the full model. Additionally, the averaging method was extended to the LRS Bianchi V and their isotropic limits, demonstrating error control via the decaying Hubble parameter and preserving asymptotic equivalence as the fourth instalment of the averaging series \cite{Millano:2023vny}. These previous models did not consider interaction between the matter components. However, in an earlier work \cite{Leon:2020pvt} the authors considered the FLRW models and the LRS Bianchi I model in the presence of scalar-matter interactions and applied the averaging method to that context. In particular the interaction was described by a simple function $Q(H,\rho_m,\rho_{\phi},\dot{\phi}):=\frac{\lambda}{2}\rho_m\dot{\phi}.$ Our interest lies in the asymptotic behaviour of the LRS Bianchi I spacetime, subject to a generalized harmonic potential and incorporating interactions among matter components.

The structure of the paper is as follows. In Section \ref{averaging-sect}, we present the basic properties and
definitions for the averaging method. The cosmological model of our
consideration is presented in Section \ref{Sect:3}. Specifically, we consider
a homogeneous and locally rotational anisotropic universe described by the
Bianchi I geometry, with a scalar field which plays the role of the inflaton,
nonminimally coupled to the dust fluid source. We consider a phenomenological
interacting term $Q$ which defines the interaction between the scalar field
with energy density $\rho_{\phi}$ and the matter source with energy density
$\rho_{m}$. We assume that $Q(H,\rho_m,\rho_{\phi},\dot{\phi}) := \Gamma\left( \frac{H}{H_0} \right)^{1-\delta} \rho_m^{\alpha} \rho_\phi^{1-\alpha-\beta} (\rho_m + \rho_\phi)^{\beta} \dot{\phi}^{\delta}$, where $H$ is the Hubble function given by the scale factor in the Misner variables, and $\alpha$, $\beta$, and $\delta$ are dimensionless parameters, and $\Gamma$ the coupling parameter. This generic interaction term $Q(H,\rho_m,\rho_\phi,\dot{\phi})$ recovers several phenomenological models previously studied in the literature. Assuming all densities scale as $[H]^2$ and that $Q/H^3$ is dimensionless, a dimensional analysis shows that $[\Gamma] = [H] \cdot [\dot{\phi}]^{-\delta}$, and if the scalar field $\phi$ is dimensionless (so that $[\dot{\phi}]= [H]$), then $[\Gamma] = [H]^{1-\delta}$.

In Section \ref{Sect:4}, we consider the harmonic potential for the scalar
field and we perform a detailed analysis of the averaged cosmological dynamics
for nine different sets of the free parameters $\alpha$, $\beta$, $\delta$.
This analysis provides us with important information regarding the effects of
the non-linear components in the interaction on the cosmological dynamics.
Furthermore, we determine criteria for the free parameters $\alpha$, $\beta$,
$\delta$ so as to avoid singularities in the phase space for the cosmological
field equations. In Section \ref{sect-5}, we study the value of the deceleration parameter evaluated at the equilibrium points obtained in the previous section. Additionally we calculate an expression for the Hubble function $H$ for each equilibrium point. In Section \ref{int:numerics}, we evaluate the deceleration parameter in some numerical solutions of the original and averaged systems to study the discuss transitions between a decelerated stage into an accelerated one. We generate numerical
solutions of the cosmological field equations and we compare the results with
the analysis using the averaging method. Finally, in Section \ref{conclu-4} we
summarize the results of this study and draw our conclusions.
\section{Averaging method}
\label{averaging-sect}
In this section, we motivate the use of the averaging method by highlighting the limitations of alternative approaches, such as regular perturbation techniques, which often fail to capture the long-term behaviour of systems with rapidly oscillating dynamics. We then provide a brief overview of the averaging method, emphasizing its theoretical foundations and practical advantages. Finally, we discuss its relevance and applicability to cosmological models, particularly those involving scalar fields and anisotropic geometries.

We begin by analysing second-order differential equations that model a mass--spring system subject to weak damping, characterized by a small positive parameter \( \varepsilon \), with \( 0 < \varepsilon \ll 1 \). The first approach we consider is the \textbf{regular asymptotic expansion method}~\cite{strogatz2024nonlinear}, a classical technique often employed to study the effect of small perturbations on the behaviour of dynamical systems.

\begin{example}
    Determine a function \( y(t) \) that satisfies the equation of a weakly damped oscillator, subject to the given initial conditions
    \begin{equation}
\label{1eq1}
    y''+\varepsilon y'+y=0, \quad \text{for} \quad t \geq 0, \quad      y(0)=0,\quad  y'(0)=1,  \quad \text{where}\; \varepsilon \rightarrow 0^+.
\end{equation}

\end{example} 
The regular expansion method consists in proposing a solution to the problem in the form of a regular power series expansion:
\begin{equation}
\label{eq3}
    y(t)\sim y_0(t)+\varepsilon y_1(t)+\cdots
\end{equation}
Substituting equation~\eqref{eq3} into equation~\eqref{1eq1} yields
 \begin{equation}
       (y_0+\varepsilon y_1+\cdots)''+\varepsilon (y_0+\varepsilon y_1+\cdots)'+(y_0+\varepsilon y_1+\cdots)=0+0 \varepsilon+\cdots.
   \end{equation}
Note that the terms involving \( \varepsilon^2 \) (and higher powers) are negligible. Grouping terms according to powers of \( \varepsilon \), we obtain

\begin{equation}
    (y_0''+y_0)\varepsilon^0+(y_1''+y_0'+y_1)\varepsilon+\mathcal{O}(\varepsilon^2)=0\varepsilon^0+0\varepsilon+\mathcal{O}(\varepsilon^2).
\end{equation}
 Evaluating equation \eqref{eq3} at the initial conditions, we have
  \begin{equation*}
      0+0 \varepsilon =y(0)=y_0(0)+\varepsilon y_1(0),
  \end{equation*}
    \begin{equation*}
             1+0 \varepsilon =y'(0)= y_0'(0)+\varepsilon y_1'(0).
  \end{equation*}
     We now solve each of the following differential equations corresponding to each power of \( \varepsilon \):
\begin{enumerate}
    \item Order \( 1 = \varepsilon^0 \) problem: \begin{equation*}
        y_0''(t)+y_0(t)=0, \quad y_0(0)=0, \quad y_0'(0)=1.
    \end{equation*} Note that this problem is obtained from the differential equation~\eqref{1eq1} by assuming \( \varepsilon = 0 \); for this reason, it is also referred to as the \textit{unperturbed problem}. The solution to the initial value problem is \( y_0(t) = \sin(t) \).
\item Order \( \varepsilon \) problem:
     \begin{equation*}
        y_1''(t)+y_1(t)=-y_0'(t), \quad y_1(0)=0, \quad y_1'(0)=0.
    \end{equation*}
To solve this problem, the previously derived function \( y_0(t) \) is needed. The solution with initial conditions is  
\( y_1(t) = -\frac{1}{2} t \sin(t). \)  
\end{enumerate}

\begin{figure}[H]
       \centering
       \includegraphics[scale=0.7]{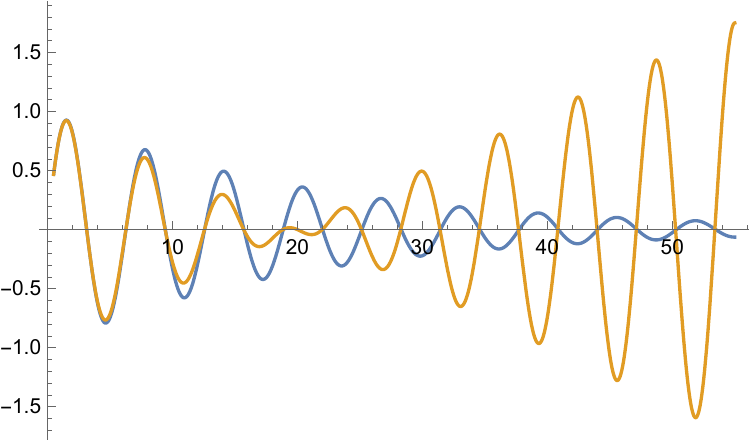}
       \caption{Exact solution (blue) vs. regular asymptotic approximation (orange), for \( \varepsilon = 0.1 \).}
       \label{fig1}
   \end{figure}

Essentially, substituting \( y_0 \) and \( y_1 \) into equation~\eqref{1eq1} yields the regular solution  
\[
y_{\text{regular}}(t) = \sin(t) - \frac{1}{2} \varepsilon t \sin(t).
\]  
Setting \( \varepsilon = 0.1 \), in Figure~\ref{fig1} we show the exact solution 
\[
y_{\text{exact}}(t) = 1.00125\, e^{-0.05 t} \sin(0.998749 t)
\]  
(in blue) and the regular solution (in orange) for $t>0$.  
While \( y_{\text{exact}} \) tends to zero as \( t \to \infty \), the regular solution \( y_{\text{regular}} \) diverges due to the secular term \( -\frac{1}{2} \varepsilon t \sin(t) \). This demonstrates that the regular expansion method is not suitable for this type of analysis.

In the following example, we examine the \textbf{averaging method} \cite{grimshaw2017nonlinear,guckenheimer1983nonlinear,sanders2010averaging,strogatz2024nonlinear} applied to the periodic differential equation \( \dot{\mathbf{x}} = \mathbf{f}(t, \mathbf{x}, \varepsilon) \). This approach begins by solving the unperturbed case, \( \dot{\mathbf{x}} = \mathbf{f}(t, \mathbf{x}, \varepsilon = 0) \). From this initial solution, variational equations are derived in standard form, allowing the averaging process to be applied.

This technique is designed to approximate solutions of initial value problems in ordinary differential equations with perturbations. Its importance lies in the simplification of periodic dynamical systems and in the understanding of their long-term behavior, as described in Chapter~11 of~\cite{verhulst2006methods}. This makes it an essential tool for the analysis of complex systems.

\begin{example}
Determine a function $\phi(t)$ that satisfies the equation of a weakly damped oscillator, subject to the given initial conditions:
 \begin{equation}
 \label{harm-osc}
 \ddot \phi + \omega^2 \phi = -2\varepsilon  \dot{\phi}, \quad \text{for} \quad  t\geq 0,  \quad \phi(0)=\phi_0,\quad  \dot\phi(0)=\phi_1 \quad \text{where}\; \varepsilon \rightarrow 0^+.
 \end{equation}
The unperturbed problem,  
$
\ddot{\phi} + \omega^2 \phi = 0,
$
admits the solution  
$
\phi(t) = \phi_0 \cos(\omega t) + \frac{\phi_1}{\omega} \sin(\omega t),$
which can also be rewritten as
\begin{align}
    \phi(t)=& r_0 \sin (\omega t-\Phi_0),\\
    \dot\phi(t)=& r_0 \omega \cos (\omega t-\Phi_0).
\end{align}
 Where $r_0$ and $\Phi_0$ are constants that are related to the initial conditions through 
 \begin{small}
 \begin{equation}
     r_0= \frac{\sqrt{\omega ^2 \phi_0^2+\phi_1^2}}{\omega },\quad 
     \phi_0=\tan
   ^{-1}\left(\frac{\phi_1}{\sqrt{\phi_1^2+\phi_0^2 \omega ^2}},-\frac{\phi_0
   \omega }{\sqrt{\phi_1^2+\phi_0^2 \omega ^2}}\right)+2 \pi  c_1\text{ with }c_1\in
   \mathbb{Z}.
 \end{equation} 
 \end{small}
 Using the method of variation of constants, $r_0$ and $\Phi_0$ are promoted to functions depending on $t$. This approach is also known as the amplitude-phase transformation, as defined in \cite[Chapter 11]{verhulst2006methods}:
 \begin{equation}
 \label{amplitud-fase}
 \dot{\phi}(t)= r(t) \omega \cos (\omega t-\Phi(t)), \;  \phi(t)  = r(t) \sin (\omega t-\Phi(t)),
 \end{equation}
where the amplitude function $r(t)$ and the phase function $\Phi(t)$ are given by:
\begin{equation}
 \label{eqAA25}
 r=\frac{\sqrt{\dot{\phi}^2(t)+\omega ^2 \phi^2(t)}}{\omega }, \;  \Phi =\omega t-\tan ^{-1}\left(\frac{\omega \phi
   (t)}{\dot \phi(t)}\right). 
 \end{equation}
 Using this transformation, equation \eqref{harm-osc} can be rewritten as the following system of two differential equations:
 \begin{equation}
 \label{eq4}
 \dot r= -2 r \varepsilon  \cos ^2(\omega t-\Phi), \;  \dot\Phi = - \varepsilon \sin (2 (\omega t-\Phi )).
 \end{equation}
For values close to zero ($\varepsilon \to 0^+$), the system \eqref{eq4} behaves approximately as:  
\begin{equation}  
\label{slowly-var-with-time}  
\dot{r} \approx 0, \quad \dot{\Phi} \approx 0.  
\end{equation}  
Hence, $r$ and $\Phi$ are functions that \textbf{vary slowly with time}, also referred to as \textbf{quasi-constants}. The averaging method focuses on the non-zero average of the system's terms, treating $r$ and $\Phi$ as constant and neglecting their time dependence. 

Assume that $\mathbf{f}(\cdot, t)$ is an $L$-periodic function in the variable $t$. The \textbf{averaged function} $\bar{\mathbf{f}}$ is defined as:
\begin{equation}
\label{timeavrg}
\bar{\mathbf{f}}(\cdot) := \frac{1}{L} \int_{0}^L \mathbf{f}(\cdot, t) \, dt, \quad L = \frac{2\pi}{\omega}, \quad \omega \text{ is the angular frequency}.
\end{equation}
It can be shown that the time derivative commutes with the averaging operator, i.e., $\dot{\bar{\mathbf{f}}} = \overline{\dot{\mathbf{f}}}$. Therefore, by averaging the right-hand side of system \eqref{eq4}, we obtain the following system:
 \begin{align}
 \label{eq5}
 & \dot {\bar{r}} =\frac{1}{L}\int_0^L(-2r\varepsilon \cos^2(\omega t-\Phi))dt=-\frac{2r\varepsilon}{L}\int_0^L\cos^2(\omega t-\Phi)dt= - \varepsilon  \bar{r},\\
 \label{eq6}
 &\dot{\bar{\Phi}}=\frac{1}{L}\int_0^L (-\varepsilon \sin(2(\omega t-\Phi)))dt=-\frac{\varepsilon}{L}\int_0^L sin(2(\omega t-\Phi))dt= 0. 
 \end{align}
Here we used the facts that $\overline{\cos^2(\omega t - \Phi)} = \frac{1}{2}$ and $\overline{\sin(2(\omega t - \Phi))} = 0$.  
Solving the system \eqref{eq5}–\eqref{eq6} with initial conditions $\bar{r}(0) = r_0$ and $\bar{\Phi}(0) = \Phi_0$, we obtain:
\begin{align*}
& \bar{r}(t) = r_0 e^{- \varepsilon t}, \quad \bar{\Phi}(t) = \Phi_0.
\end{align*}

Therefore, the averaged solution of the original problem is obtained by substituting these expressions into  
$\bar{\phi}(t) = \bar{r}(t) \sin(\omega t - \bar{\Phi}(t))$. That is,
\[
\bar{\phi}(t) = r_0 e^{-\varepsilon t} \sin(\omega t - \Phi_0),
\]
which provides a good approximation to the exact solution given by:
   \begin{small}
   \begin{equation*}
      \phi(t)=  -r_0  e^{-t \varepsilon }  \sin (\Phi_0) \cos
   \left(t \sqrt{\omega ^2-\varepsilon ^2}\right) \nonumber  -\frac{r_0 e^{-t
   \varepsilon } \sin \left(t \sqrt{\omega ^2-\varepsilon ^2}\right) (\varepsilon 
   \sin (\Phi_0)-\omega  \cos (\Phi_0))}{\sqrt{\omega
   ^2-\varepsilon ^2}},
   \end{equation*}
   \end{small}
   because the difference 
   \begin{small}
\begin{align*}
& \phi(t)-\bar{\phi}(t)  =  r_0  e^{-t \varepsilon } \left(-\frac{\varepsilon  \sin (\Phi_0) \sin
   (t \omega )}{\omega }+\varepsilon ^2 \left(\frac{\cos (\Phi_0) \sin (t \omega
   )}{2 \omega ^2}-\frac{t \cos (\Phi_0-t \omega )}{2
   \omega }\right)+O\left(\varepsilon ^3\right)\right)\\
   & = -\frac{\varepsilon 
   r_0 \sin (\Phi_0) \sin (t \omega )}{\omega }+ \varepsilon ^2 \left(\frac{r_0 \cos (\Phi_0) \sin
   (t \omega )}{2 \omega ^2}-\frac{r_0 t \cos (t \omega
   +\Phi_0)}{2 \omega }\right)+O\left(\varepsilon ^3\right) 
\end{align*}
\end{small}
goes to zero as $\varepsilon \rightarrow 0^+.$ This confirms that the averaged solution $\bar{\phi}(t)$ provides an increasingly accurate approximation to $\phi(t)$ as the perturbation parameter $\varepsilon$ becomes small.
In Figure~\ref{fig2}, the exact solution of the differential equation~\eqref{harm-osc} is shown in red, while the averaged solution is shown in blue. Both are plotted for specific values of the parameters \( \omega \), \( r_0 \), and \( \Phi_0 \), with \( \varepsilon = 0.1 \).
    \begin{figure}[H]
        \centering
        \includegraphics[scale=0.5]{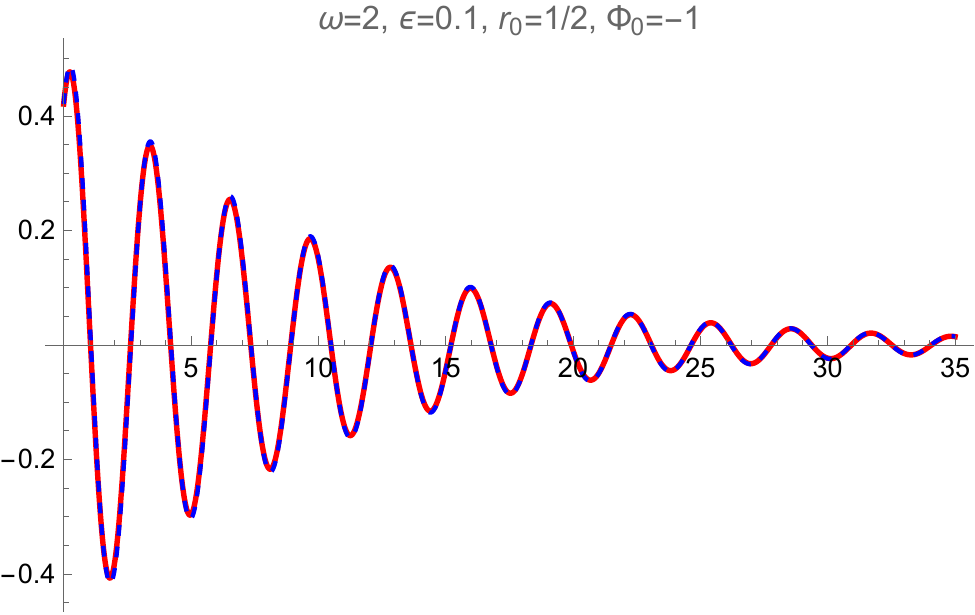}
        \caption{Exact solution (red) and averaged solution (blue dashed line) for equation~\eqref{harm-osc}, showing decaying oscillations as \( t \rightarrow \infty \).}
        \label{fig2}
    \end{figure}
\end{example}

Both solutions exhibit similar behavior: the oscillations decay as \( t \to \infty \) due to the exponential factor \( e^{-\varepsilon t} \), which causes the damping. The effectiveness of each method depends on the specific problem to which it is applied. However, for this type of problem, the \textbf{averaging method} stands out because it avoids generating secular terms that could alter the long-term behavior of the solution for certain values of \( t \).
\subsection{General class of systems with a time-dependent perturbation parameter}

A possible generalization of equation \eqref{harm-osc} is the Klein–Gordon (KG) system:
\begin{align}
\label{KGharmonic}
   & \ddot \phi + \omega^2 \phi = -3 H \dot \phi, \\
   &\dot{H}= -\frac{1}{2}\dot\phi^2. \label{Friedmann}
\end{align}
Under certain conditions, the similarity between equations \eqref{harm-osc} and \eqref{KGharmonic} suggests that the system can be treated as a perturbation of a weakly damped oscillator, making the \textbf{averaging method} a useful tool.

In this case, the perturbation parameter is represented by the function $H(t)$, which decreases strictly as $t \to \infty$ according to equation \eqref{Friedmann}. This function can be used to measure and control the error between the original solution and its averaged approximation. As $H$ decays, the error is also expected to diminish. This approach enables the asymptotic behavior of the KG system to be studied through a simplified version, making it easier to extract meaningful information.

Reference \cite{Fajman:2021cli} investigates the late-time behavior of solutions within a general class of systems written in standard form as
\begin{equation}
\label{standard51}
\left(\begin{array}{c}
       \dot{H} \\
        \dot{\mathbf{x}}
  \end{array}\right)= H \left(\begin{array}{c}
       0 \\
       \mathbf{f}^1 (\mathbf{x}, t)
  \end{array}\right) + H^2\left(\begin{array}{c}
       f^{[2]} (\mathbf{x}, t) \\
       \mathbf{0}
  \end{array}\right),
\end{equation}
where $H > 0$ is strictly decreasing in $t$ and satisfies $\lim_{t \to \infty} H(t) = 0$.

Let $\|\cdot\|$ denote the discrete $\ell^1$-norm, defined for $\mathbf{u} \in \mathbb{R}^n$ by $\|\mathbf{u}\| := \sum_i^n |u_i|$. Also, let $L_{\mathbf{x}, t}^\infty$ be the $L^\infty$ space in both variables $t$ and $\mathbf{x}$, with norm defined as $\|\mathbf{f}\|_{L_{\mathbf{x}, t}^\infty} := \sup_{\mathbf{x}, t} |\mathbf{f}(\mathbf{x}, t)|$.

The following result characterizes the local asymptotic behavior of the system \eqref{standard51}.

\begin{theorem}[Theorem 3.1 of \cite{Fajman:2021cli}]
\label{localintime}Suppose  $H(t)>0$ is strictly decreasing in $t$ and $\lim_{t\rightarrow \infty} H(t)=0.$ Fix any $\epsilon>0$ with $\epsilon<H(0)$ and define $t_*>0$ such that $\epsilon=H(t_*).$ Suppose that $\|\mathbf{f}^1\|_{L_{\mathbf{x}, t}^\infty},\quad \|f^{[2]}\|_{L_{\mathbf{x}, t}^\infty}<\infty$ and that $\mathbf{f}^1(\mathbf{x}, t)$ is Lipschitz continuous and $f^{[2]}$ is continuous with respect to $x$ for all $t\geq t_*.$ Also, assume that $\mathbf{f}^1$ and $f^{[2]}$ are $T$-periodic for some $T>0.$ Then for all $t>t_*$ with $t=t_*+\mathcal{O}\Big(H(t_*)^{-\delta}\Big)$ for any given $\delta \in (0,1)$ we have $$\mathbf{x}(t)-\mathbf{z}(t)=\mathcal{O}\Big(H(t_*)^{\min\{1,2-2\delta\}}\Big),$$  where $\mathbf{x}$ is the solution of system \eqref{standard51} with initial condition $\mathbf{x}(0)=\mathbf{x}_0$ and $\mathbf{z}(t)$ is the solution of the time-averaged system 
\begin{equation*}
    \dot{\mathbf{z}}=H(t_*)\bar{\mathbf{f}}^1(\mathbf{z}),\quad \text{for} \quad t>t_*,
\end{equation*} 
with initial condition $\mathbf{z}(t_*)=\mathbf{x}(t_*)$ where the time-averaged vector $\bar{\mathbf{f}}^1$ is defined as 
\begin{equation*}
   \bar{\mathbf{f}}^1(\mathbf{z})=\frac{1}{T}\int_{t_*}^{t_*+T}\mathbf{f}^1(\mathbf{z},s)ds.
\end{equation*}
\end{theorem}
We emphasize that the key properties of Theorem \ref{localintime} are closely tied to the choice of the truncation time $t^*$ for the first-order system: the longer one waits before truncating, the more accurate the estimate becomes, and the longer the estimate remains valid. This result highlights that the Hubble function $H$ acts as a time-dependent perturbation parameter, enabling the use of first-order approximations and error estimates, as well as periodic averaging, to study the asymptotic behaviour of the system. To approximate the solution $\mathbf{x}(t)$ of system \eqref{standard51}, we consider a truncated first-order system with initial data prescribed at $t > t_{*}$, namely $\mathbf{y}(t^*) = \mathbf{x}(t^*)$, providing a simplified yet effective description of the system’s dynamics.

Let us consider the problem
\begin{equation}
\left(\begin{array}{c}
       \dot{\mathcal{H}} \\
        \dot{\mathbf{y}}
  \end{array}\right)= 
\mathcal{H}(t) \left(\begin{array}{c}
      0 \\
      \mathbf{f}^{1}(\mathbf{y}, t)
  \end{array}\right), \label{tag4.1}
\end{equation}
with initial conditions $\mathcal{H}(t_{*}) = H(t_{*})$ and $\mathbf{y}(t_{*}) = \mathbf{x}(t_{*})$, so that both systems share the same initial data. Since $\dot{\mathcal{H}} = 0$, it follows that $\mathcal{H}(t) \equiv H(t_{*})$ is constant.

We now present the following proposition, which provides error estimates for the first-order approximation.
\begin{proposition}[Proposition 4.1 of \cite{Fajman:2021cli}]\label{PropFajmann}
Fix any \(t_{*} > 0\). Suppose that \(\mathbf{x}\) and \(\mathbf{y}\) are solutions of \eqref{standard51} and \eqref{tag4.1}, respectively, with initial conditions \(\mathbf{x}(0) = \mathbf{x}_0\) and \(\mathbf{y}(t_{*}) = \mathbf{x}(t_{*})\). Suppose further that \(H = H(t) > 0\) is strictly decreasing in \(t\) and  \(\lim_{t\to\infty} H(t) = 0\). In addition, suppose that 
\[
\|\mathbf{f}^1\|_{L_{\mathbf{x},t}^\infty},\;\|f^{[2]}\|_{L_{\mathbf{x},t}^\infty} < \infty,
\]
and that \(\mathbf{f}^1(\mathbf{x},t)\) is Lipschitz continuous in \(\mathbf{x}\) for all \(t \ge t_{*}\). Then, for all \(t > t_{*}\) with
$
t = t_{*} + O\bigl(H(t_{*})^{-\gamma}\bigr)
$
for some \(\gamma \in (0,1)\), we have
\begin{equation}
\mathbf{x}(t) - \mathbf{y}(t) = O\bigl(H(t_{*})^{2 - 2\gamma}\bigr).
\end{equation}
\end{proposition}

Following the references \cite{Leon:2021lct,Leon:2021rcx, Leon:2021hxc, Millano:2023vny}, we analyze systems which, although not initially in the standard form \eqref{standard51}, can be expressed through a Taylor expansion around \(H=0\):
\begin{equation}
\label{nonstandtard}
 \left(\begin{array}{c}
       \dot{H} \\
        \dot{\mathbf{x}}
  \end{array}\right)= \left(\begin{array}{c}
    0 \\
       \mathbf{f}^0 (\mathbf{x}, t; \omega)
  \end{array}\right)+ H \left(\begin{array}{c}
       0 \\
       \mathbf{f}^1 (\mathbf{x}, t)
  \end{array}\right) 
   + H^2\left(\begin{array}{c}
       f^{[2]} (\mathbf{x}, t)  \\
       \mathbf{0}
  \end{array}\right)+ \mathcal{O}(H^3),
  \end{equation}
where an additional term appears, depending on the function \(\mathbf{f}^0 (\mathbf{x}, t; \omega)\). However, this function depends on a free frequency parameter \(\omega\), which can be tuned so that \(\mathbf{f}^0(\mathbf{x}, t; \omega) = \mathbf{0}\). Therefore, the systems can be recast in the standard form \eqref{standard51}.
\section{Interacting Anisotropic Scalar Field Cosmology}
\label{Sect:3}
For a scalar field theory with exponential potential Hubble-normalized quantities are used. Moreover,  the evolution equation for $H$, given by the Raychaudhuri equation, decouples. Then, one can work in reduced phase space. The equilibrium points typically give the asymptotic of the remaining reduced system, and often a dynamical system analysis can determine it \cite{Coley:1999uh, Coley:2003mj, Wainwrightellis1997}. This remarkable reduction is because the exponential potential has symmetry such that its derivative is also an exponential function. The harmonic potential $V(\phi)= \mu^2 \phi^2$ does not satisfy the above symmetry; the Raychaudhuri equation fails to decouple \cite{Alho:2014fha}. Therefore, Hubble-normalized equations are often complicated to analyze using the standard dynamical systems approach. That is due to oscillations entering the system via the Klein-Gordon equation \cite{Fajman:2020yjb, Leon:2021lct, Leon:2021rcx}. Oscillations in scalar-field cosmologies with generalized harmonic potentials of type $V(\phi)= \mu^2 \phi^2 + \text{cosine corrections}$ are extended here using averaging techniques similar to those used in \cite{Fajman:2020yjb, Leon:2021lct, Leon:2021rcx} for a family of generalized harmonic potentials when $H$ monotonically decreases.   Continuing the analysis of models like the introduced in \cite{Leon:2020pvt}, we consider interacting models based on the interactive matter-scalar field schemes where the conservation equations have the structure \cite{Leon:2020pvt}
\begin{equation}
\dot {\phi} \left [\ddot {\phi} +3 H \dot {\phi} + V'(\phi) \right] =  Q, \quad \dot{\rho}_ {m} + 3H \left(\rho_ {m} + p_ {m} \right)= - Q,
 \label{interacting-scheme}
\end{equation}%
where a dot means derivative with respect to cosmic time $t$, comma derivative with respect to $\phi$, $ \rho_m $ is the energy density of matter, $\phi $ is the scalar field, $V(\phi ) $ its the generalized harmonic potential defined by \begin{align}
V(\phi)= \frac{\phi ^2}{2}+f\left[1- \cos \left(\frac{\phi }{f}\right)\right], \quad f> 0, \label{pot1}
\end{align}
$ Q $ is the interaction term, and $H=\dot{a}/a$ stands for the  \emph{ Hubble parameter} (which is a general measure of the isotropic rate of spatial expansion), where $a$ denotes the scale factor of the universe. 
An exciting research program is to investigate the dynamics and asymptotic behaviour of the solutions of the equations of the gravitational field for various interacting functions $ Q = Q \left (\phi, \dot{\phi}, \rho_ {m}\right) $. In particular, we generalize \cite{Leon:2020pvt} by defining the general interaction term
\begin{equation}
\label{int}
    Q=\Gamma \left(\frac{H}{H_0}\right)^{1-\delta}\rho_m^{\alpha}\rho_\phi^{1-\alpha -\beta}(\rho_m +\rho_\phi)^{\beta}{\dot{\phi}}^{\delta},
\end{equation}
where $\rho_m$ is the cold dark matter density and $\rho_\phi:= \frac{1}{2}{\dot{\phi}}^2 + V(\phi)$, with $V(\phi)$ defined in \eqref{pot1} represents the dark energy density. Aditionally, $\alpha$, $\beta$, $\delta$  are real numbers, and $\Gamma$ is a coupling parameter such that $\Gamma {\dot{\phi}}^{\delta}$ has dimensions of $H$. The equation of state for the matter components is $p_m=(\gamma-1)\rho_m$, where the barotropic index is $\gamma$ defined in  $0 \leq \gamma \leq 2$. On the other hand, note that the inclusion of the interaction function \eqref{int} generalizes de results published in \cite{Leon:2021rcx} for the LRS Bianchi I model. In that research, the authors considered the usual Klein-Gordon and matter conservation equations defined as
\begin{equation}
\ddot {\phi} +3 H \dot {\phi} + V'(\phi) = 0, \quad \dot{\rho}_ {m} + 3H \left(\rho_ {m} + p_ {m} \right)= 0.
\end{equation}
where the function $Q$ presented in \eqref{interacting-scheme} is set to zero. The function $Q$ represents the energy exchange between the matter sources, introducing more complex dynamics into the model. We consider nine different interactions according to the following choice of the real $\alpha$, $\beta$ and $\delta$,
\begin{align}
 &\textbf{Interaction 1:}\label{int-1} (\alpha,\beta,\delta)=(1,0,1): Q \left(\dot{\phi},\rho_{\phi }, \rho _{m}\right)=\Gamma\rho_m \dot\phi.\\
    &\textbf{Interaction 2:}\label{int-2} (\alpha, \beta, \delta)=(1,-1,0): Q \left(\dot{\phi},\rho_{\phi }, \rho _{m}\right)= \frac{\Gamma}{H_0} H\frac{  \rho_{m} \rho_\phi}{\rho_{m}+  \rho_\phi}.\\
    &\textbf{Interaction 3:}\label{int-3} (\alpha,\beta, \delta)=(1,0,0): Q\left(\dot{\phi},\rho_{\phi }, \rho _{m}\right)=\frac{\Gamma}{H_0} H \rho_m.\\
    &\textbf{Interaction 4:}\label{int-4} (\alpha, \beta, \delta)=(0,0,0):   Q\left(\dot{\phi},\rho_{\phi }, \rho _{m}\right)=\frac{\Gamma}{H_0} H \rho_\phi.\\
    &\textbf{Interaction 5:}\label{int-5} (\alpha, \beta, \delta)=(0,1,0):  Q\left(\dot{\phi},\rho_{\phi }, \rho _{m}\right)=\frac{\Gamma}{H_0} H \left(\rho_{m}+\rho_\phi\right).\\
    &\textbf{Interaction 6:}\label{int-6} (\alpha, \beta, \delta)=(0,0,2):  Q\left(\dot{\phi},\rho_{\phi }, \rho _{m}\right)=\frac{\Gamma  H_0 \rho_{\phi }  \dot{\phi}^2}{H}.
    \end{align}
    \begin{align}
    &\textbf{Interaction 7:}\label{int-7} (\alpha, \beta, \delta)=(1,0,2):  Q\left(\dot{\phi},\rho_{\phi }, \rho _{m}\right)=\frac{\Gamma  H_0 \rho_{m } \dot{\phi}^2}{H}.\\
    &\textbf{Interaction 8:}\label{int-8} (\alpha, \beta, \delta)=(1,-1,1):  Q\left(\dot{\phi},\rho_{\phi }, \rho _{m}\right)=\frac{\Gamma  \rho_{m } \rho_{\phi }  \dot{\phi}}{\rho_{m }+\rho_{\phi } }.\\
    &\textbf{Interaction 9:}\label{int-9} (\alpha, \beta, \delta)=(1,-1,2):  Q\left(\dot{\phi},\rho_{\phi }, \rho _{m}\right)=\frac{\Gamma  H_0 \rho_{m } \rho_{\phi }  \dot{\phi}^2}{H
   (\rho_{m }+\rho_{\phi } )}.
\end{align}
For each of these interactions, we shall derive the field equations and define Hubble normalized variables to study de late-time dynamics of each model using the averaging techniques first presented in \cite{Leon:2021lct, Leon:2021rcx, Leon:2021hxc, Millano:2023vny}. 
According to \cite{Nilsson:1995ah}, the general line element for the LRS Bianchi III, LRS Bianchi I Kantowski-Sachs models is
\begin{align}
\label{metric}
   &  ds^2= - dt^2 + \left[{e_1}^1(t)\right]^{-2} dr^2 + \left[{e_2}^2(t)\right]^{-2}  \left[ d \vartheta^2 + k^{-1} \sin^2 (\sqrt{k} \vartheta)d \zeta^2\right],
\end{align}
where ${e_1}^1$, ${e_2}^2$ y ${e_3}^3 = \sqrt{k} {e_2}^2/\sin(\sqrt{k} \vartheta)$ are functions depending only on $t$, that represent the frame vectors \cite{Coley:2008qd}:
\[\mathbf{e}_0 = \partial_t, \quad \mathbf{e}_1 = {e_1}^1 \partial_r, \quad \mathbf{e}_2 = {e_2}^2 \partial\vartheta, \quad \mathbf{e}_3 = {e_3}^3 \partial\zeta. \]
The scalar $k$ in the definition of the line element determines which model is being considered. As it was discussed in \cite{Fadragas:2013ina}, the following limits are satisfied 
\begin{align}
  & \lim_{k\rightarrow -1}  k^{-1} \sin^2 (\sqrt{k} \vartheta)=\sinh^2 (\vartheta), \\
  & \lim_{k\rightarrow 0}  k^{-1} \sin^2 (\sqrt{k} \vartheta)= \vartheta^2,\\
   & \lim_{k\rightarrow 1}  k^{-1} \sin^2 (\sqrt{k} \vartheta)=\sin^2 (\vartheta),
\end{align}
therefore, we obtain the following configurations:
i) LRS Bianchi I for $k = 0$, ii) LRS Bianchi III for $k = -1$, iii) Kantowski-Sachs for $k = +1$. As stated before, in this work we will consider the first case, LRS Bianchi I, and will study the Bianchi III model in a future research.

The field equations for the LRS Bianchi I model with interaction are
\begin{subequations}
	\begin{align}
	&\ddot\phi+3 H \dot \phi +\phi + \sin\left( \frac{\phi}{f}\right)= Q(\phi, {\dot\phi}, \rho_m)/\dot{\phi} ,\\
	&\dot{\rho}_m+3\gamma H\rho_m=-  Q(\phi, {\dot\phi}, \rho_m),\\
	&\dot a = a H, \\
	& \dot{H}=-\frac{1}{2}\left(\gamma \rho_m+{\dot \phi}^2\right)-\frac{\sigma_0^2}{a^6},\\
	& \label{friedmann-bianchi-i-Interaction}3H^2=\rho_m+\frac{1}{2}\dot\phi^2+\frac{\phi ^2}{2}+f\left[1- \cos \left(\frac{\phi }{f}\right)\right]+\frac{\sigma_0^2}{a^6}.
	\end{align}
\end{subequations}
By setting $Q=0$ in these equations, one recovers the non-interacting model examined in \cite{Leon:2021rcx}.
The Hubble normalized variables are
\begin{align} \label{Bianchi_I_vars}
\Omega= \frac{\sqrt{\dot{\phi}^2+\omega^2  \phi^2}}{\sqrt{6} H}, \quad \Sigma=\frac{\sigma_0}{\sqrt{3} a^3 H}, \quad  \varphi= \omega t-\arctan\left(\frac{\omega\phi}{\dot{\phi}}\right), \quad
 \Omega_m= \frac{\rho_m}{3 H^2},  
\end{align}
we obtain the system
\begin{small}
\begin{align}
\dot{\Omega} & =\frac{3}{2} H \Omega  \left(2 \Sigma ^2+2 \Omega ^2 \cos ^2(\varphi -t \omega ) + \gamma  \Omega_m\right) \nonumber \\
   & +\frac{\cos (\varphi -t \omega )}{2 \sqrt{6} H} 
   \Bigg\{\Gamma H_0^{\delta -1} 6^{\frac{1}{2} (2 \alpha +\delta -1)} \Omega_m^{\alpha } \Omega ^{\delta -1} H^{2 \alpha }  \cos ^{\delta -1}(\varphi -t \omega ) \nonumber \\
   & \times 
   \left[\frac{6 H^2 \left(\omega ^2 \cos ^2(\varphi -t \omega )+\sin ^2(\varphi -t \omega )\right) \Omega ^2}{\omega ^2}+4 f \sin ^2\left(\frac{\sqrt{\frac{3}{2}} H \Omega  \sin (\varphi -t \omega
   )}{f \omega }\right)\right]^{-\alpha -\beta +1} \nonumber \\
   & \times \Bigg[6 \Bigg(\Omega ^2 \cos ^2(\varphi -t \omega )+\frac{\Omega ^2 \sin ^2(\varphi -t \omega )}{\omega ^2} +\Omega_m \Bigg) H^2
   +4 f \sin ^2\left(\frac{\sqrt{\frac{3}{2}} H \Omega  \sin (\varphi -t \omega )}{f \omega }\right)\Bigg]^{\beta }  \nonumber \\
   & -6 \sqrt{6} H^2 \Omega  \cos (\varphi -t \omega )-2 \sqrt{6} H \omega  \Omega  \sin (\varphi -t \omega )+\frac{2 \sqrt{6} H \Omega  \sin (\varphi -t \omega)}{\omega } \nonumber \\
   & +2 \sin \left(\frac{\sqrt{6} H \Omega  \sin (\varphi -t \omega )}{f \omega }\right)\Bigg\},\label{gen-redu-1}
\end{align}
\begin{align}
\dot{\Sigma} & =  \frac{3}{2} H \Sigma  \left(2 \Sigma ^2+2 \Omega ^2 \cos ^2(\varphi -t \omega ) -2  + \gamma \Omega_m\right), \label{gen-redu-2}
\\
\dot{\varphi}  &=  -\frac{\sin (\varphi -t \omega )}{2 \sqrt{6} H \Omega} \Bigg\{\Gamma H_0^{\delta -1} 6^{\frac{1}{2} (2 \alpha +\delta -1)} \Omega_m^{\alpha } \Omega ^{\delta -1} H^{2 \alpha }  \cos ^{\delta -1}(\varphi -t \omega ) \nonumber \\
   & \times \left(-2 f \cos \left(\frac{\sqrt{6} H \Omega  \sin (\varphi -t \omega )}{f \omega }\right)+2 f+\frac{6 H^2 \Omega ^2 \sin ^2(\varphi -t \omega )}{\omega ^2}+6 H^2 \Omega ^2 \cos ^2(\varphi -t \omega )\right)^{-\alpha -\beta +1} \nonumber \\
   & \times  \left[-2 f \cos \left(\frac{\sqrt{6} H \Omega  \sin (\varphi -t \omega )}{f \omega }\right)+2 f+\frac{6 H^2 \Omega ^2 \sin ^2(\varphi -t \omega
   )}{\omega ^2}+6 H^2 \Omega ^2 \cos ^2(\varphi -t \omega )+6 H^2 \Omega_m\right]^{\beta } \nonumber \\
   & -6 \sqrt{6} H^2 \Omega  \cos (\varphi -t \omega )-2 \sqrt{6} H \omega  \Omega  \sin (\varphi -t \omega ) 
   + \frac{2 \sqrt{6} H \Omega  \sin (\varphi -t \omega )}{\omega }\nonumber \\
   & +2 \sin \left(\frac{\sqrt{6} H \Omega  \sin (\varphi -t \omega )}{f \omega }\right) \Bigg\}, \label{gen-redu-3}
   \\
\dot{H} & =   -\frac{3}{2} H^2 \left(\gamma  \Omega_{m}+2 \Sigma ^2+\Omega ^2 \cos (2 (\varphi -t \omega ))+\Omega ^2\right), \label{gen-redu-4} 
   \end{align}
\end{small}
\noindent where the Friedmann equation \eqref{friedmann-bianchi-i-Interaction} was used to reduce the dimension of the system deriving the following expression for $\Omega_m$
\begin{small}
\begin{align}
\Omega_m&= \frac{1}{3} \left(\frac{f \cos \left(\frac{\sqrt{6} H \Omega  \sin (\varphi -t \omega )}{f \omega
   }\right)}{H^2}-\frac{f}{H^2}-3 \Sigma ^2-\frac{3 \Omega ^2 \left(\omega ^2 \cos ^2(\varphi -t \omega )+\sin ^2(\varphi -t
   \omega )\right)}{\omega ^2}+3\right). \label{8N163}
\end{align}
\end{small}
In the following Section we study the dynamics of the latter system by using the Averaging approach.
\section{Averaging Dynamics of Interactions}
\label{Sect:4}
For the study of the anisotropic cosmological field equations we follow the approach established before in \cite{Leon:2021lct, Leon:2021rcx, Leon:2021hxc, Millano:2023vny}. This means that for each of the interactions \eqref{int-1}-\eqref{int-9}, we obtain a different dynamical system when the corresponding values of $\alpha, \beta, \delta$ are substituted into \eqref{gen-redu-1}-\eqref{gen-redu-4}. Each of these systems is studied by taking a Taylor series around $H=0$ and keeping only up to second order terms that provide significant contribution to the dynamics. 
In order to operate within a a physically acceptable regime, we set $f=(\omega^2-1)^{-1}$, with $\omega^2 > 1$. The next step is to apply the averaging method discussed in Section \ref{averaging-sect} to obtain averaged versions of the original dynamical systems. We will average the right-hand side term of every equation, neglecting the time-dependence of the dynamical variables to suppress the oscillations of the model. 
The use of the averaging method ensures that the simplified equations preserve the fundamental dynamics of the original system, eliminating short-term oscillations that could complicate the identification of equilibrium points and the analysis of trajectories toward them.
This technique is crucial for assessing the stability of equilibrium points and their significance in the cosmological context, as it provides a clear and precise framework for their study. In the following lines, we investigate the dynamics for each interaction.
\subsection{Interaction 1: $Q(\dot{\phi},\rho_\phi,\rho_m)=\Gamma \rho_m \dot{\phi}$}
\label{int:1}
\begin{figure}[H]
    \centering
    \includegraphics[scale=0.4]{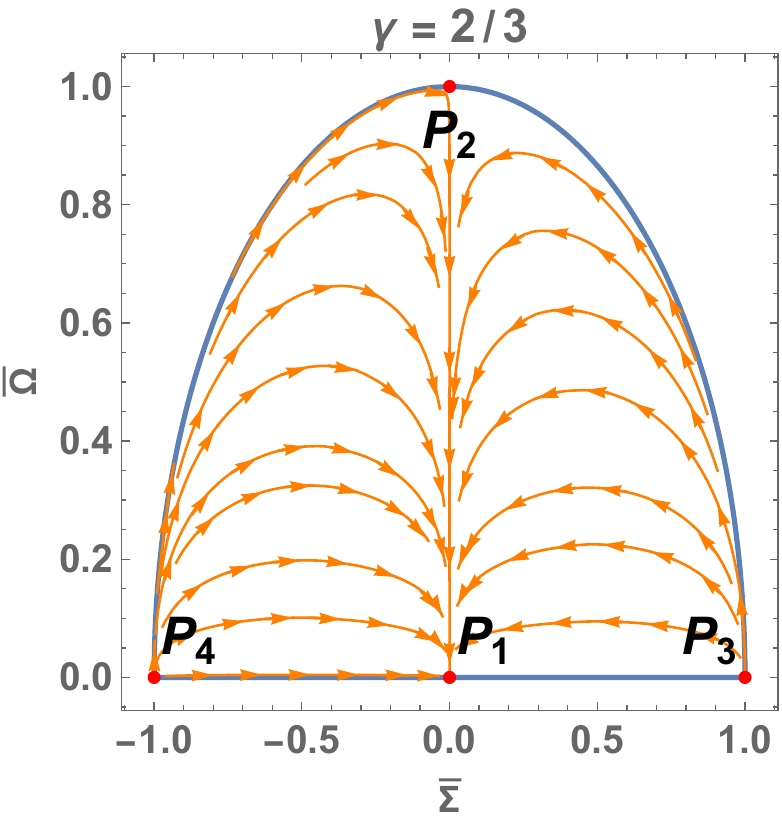}
    \includegraphics[scale=0.4]{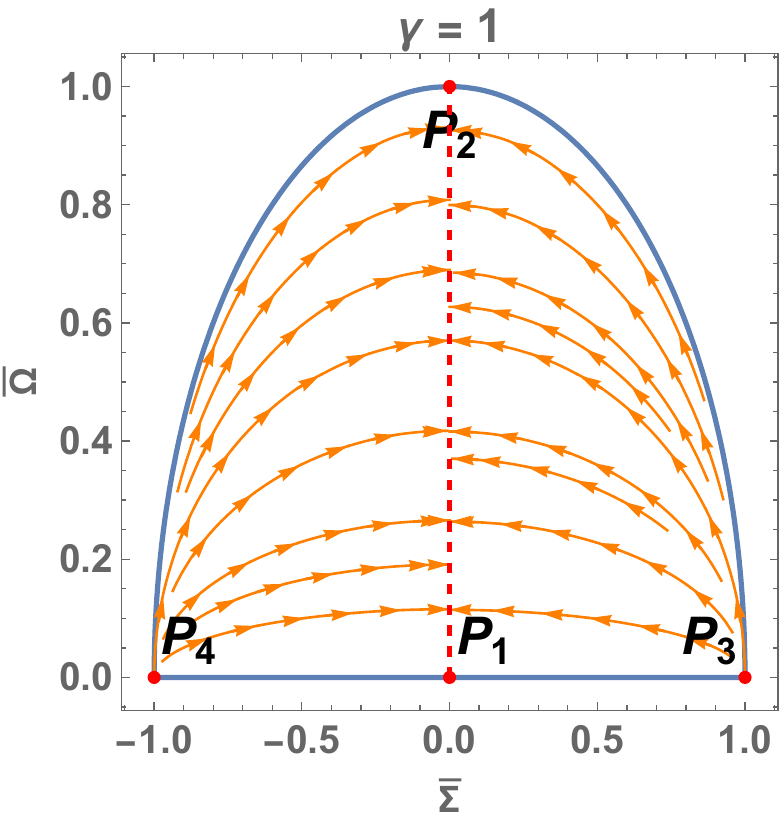}
    \includegraphics[scale=0.4]{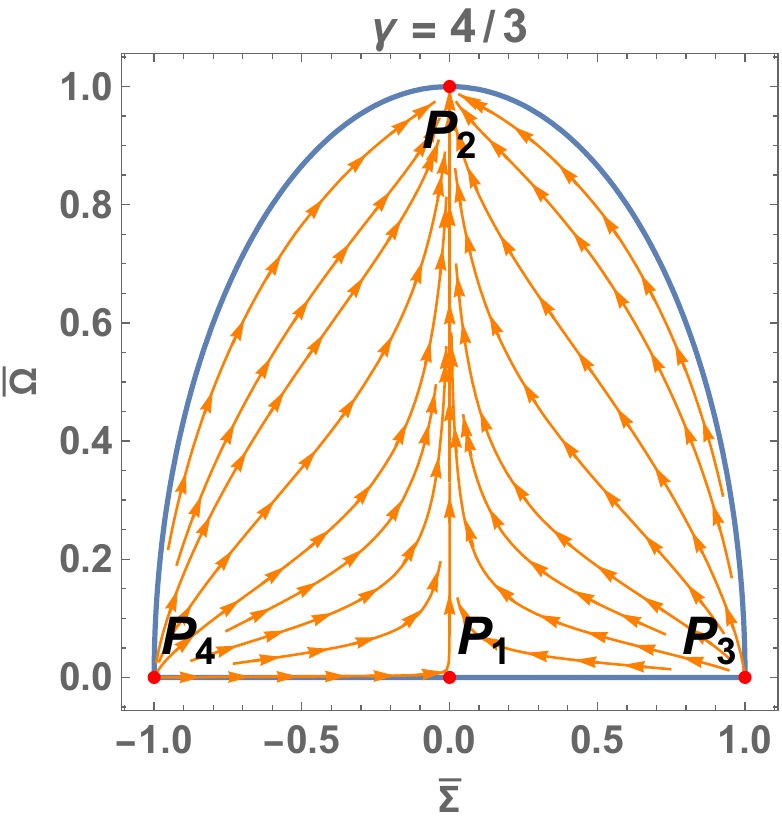}
    \includegraphics[scale=0.4]{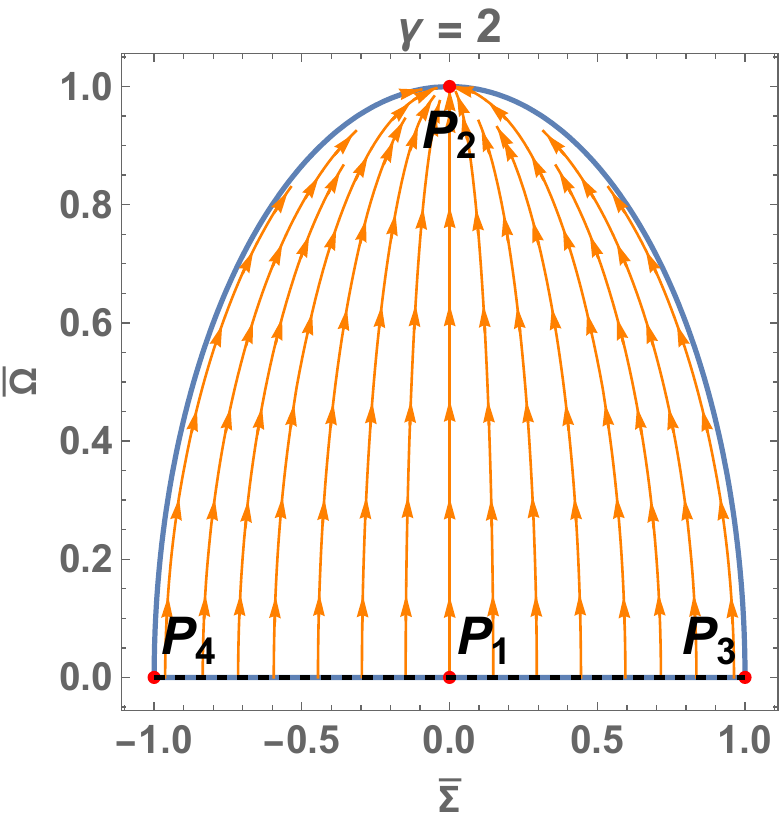}
    \caption{Phase-plane plots for system \eqref{int-1-2D-a}-\eqref{int-1-2D-b} with $\gamma = 2/3, 1, 4/3 , 2$.}
    \label{Fig1-c} 
   \end{figure}

Setting $\alpha=1,$ $\beta=-1,$ $\delta=0$ in system \eqref{gen-redu-1}-\eqref{gen-redu-4}, and expanding in a Taylor series centred at $H = 0$ and keeping terms up to $H^2$, the truncated system is
\begin{align}
\label{comp-int-1-a}
    &\dot{H}=-\frac{3}{2} \left(2 \Sigma ^2+\Omega ^2-\gamma  \left(\Sigma ^2+\Omega ^2-1\right)+\Omega ^2 \cos (2 (\varphi -t
   \omega ))\right) H^2+O\left(H^3\right),\\
    &\dot{\Omega}=\frac{1}{2} \left(3 \Omega  \left(2 \Sigma ^2+\Omega ^2-\gamma 
   \left(\Sigma ^2+\Omega ^2-1\right)+\left(\Omega ^2-1\right) \cos (2 (\varphi -t \omega ))-1\right)\right) H\\ \nonumber
   &-\frac{1}{2} \left(\sqrt{6} \Gamma 
   \left(\Sigma ^2+\Omega ^2-1\right) \cos (\varphi -t \omega )\right) H \\ \nonumber
   &-\frac{\left(\omega ^2-1\right)^3 \Omega ^3 \cos
   (\varphi -t \omega ) \sin ^3(\varphi -t \omega ) H^2}{\omega ^3}+O\left(H^3\right),\\
    &\dot{\Sigma}=\frac{3}{2} \Sigma  \left(2 \Sigma
   ^2+\Omega ^2-\gamma  \left(\Sigma ^2+\Omega ^2-1\right)+\Omega ^2 \cos (2 (\varphi -t \omega ))-2\right)
   H+O\left(H^3\right),\\
\label{comp-int-1-d}&\dot{\varphi}=\frac{\sqrt{\frac{3}{2}} \left(\Gamma  \left(\Sigma ^2+\Omega ^2-1\right)+\sqrt{6} \Omega 
   \cos (\varphi -t \omega )\right) \sin (\varphi -t \omega ) H}{\Omega }\\ \nonumber &+\frac{\left(\omega ^2-1\right)^3 \Omega ^2 \sin
   ^4(\varphi -t \omega ) H^2}{\omega ^3}+O\left(H^3\right).
\end{align}
Taking the average of each equation, the averaged system is
\begin{align}
\label{BI-prom-1}
    &\dot{H}=\frac{3}{2} H^2 \left(\gamma  \left(\bar{\Sigma} ^2+\bar{\Omega} ^2-1\right)-2 \bar{\Sigma} ^2-\bar{\Omega} ^2\right),\\
    \label{BI-prom-2}
    &\dot{\bar{\Omega}}=-\frac{3}{2}
   \bar{\Omega} H \left(\gamma  \left(\bar{\Sigma} ^2+\bar{\Omega} ^2-1\right)-2 \bar{\Sigma} ^2-\bar{\Omega} ^2+1\right),\\
    \label{BI-prom-3}&
    \dot{\bar{\Sigma}}=\frac{3}{2} \bar{\Sigma} H
   \left(-\gamma  \left(\bar{\Sigma} ^2+\bar{\Omega} ^2-1\right)+2 \bar{\Sigma} ^2+\bar{\Omega} ^2-2\right),\\
    \label{BI-prom-4}&\dot{\bar{\varphi}}=0.
\end{align}
From this system, dividing by $H$ and defining a new time derivative $f'= \frac{1}{H}\dot{f}$, the guiding system is
\begin{align}
  \label{int-1-2D-a}  &\bar{\Sigma}'=\frac{3}{2} \bar{\Sigma} 
   \left(-\gamma  \left(\bar{\Sigma} ^2+\bar{\Omega} ^2-1\right)+2 \bar{\Sigma} ^2+\bar{\Omega} ^2-2\right),\\
  \label{int-1-2D-b} &\bar{\Omega}'=-\frac{3}{2}
   \bar{\Omega}  \left(\gamma  \left(\bar{\Sigma} ^2+\bar{\Omega} ^2-1\right)-2 \bar{\Sigma} ^2-\bar{\Omega} ^2+1\right),
\end{align}
 which is defined in the compact phase-space
\begin{equation}
    \left\{(\bar{\Sigma},\bar{\Omega})\in \mathbb{R}^2| \;  \bar{\Omega}^2 + \bar{\Sigma}^2 \leq 1, \bar{\Omega}\geq 0, -1\leq \bar{\Sigma} \leq 1 \right\}.
\end{equation}
The equilibrium points of the guiding system \eqref{int-1-2D-a}-\eqref{int-1-2D-b} for the coordinates $(\bar{\Sigma},\bar{\Omega})$ together with their stability conditions are the following
\begin{enumerate}
    \item $P_1=(0, 0)$ with eigenvalues $\lbrace \frac{3 (\gamma -2)}{2},\frac{3 (\gamma -1)}{2}\rbrace$. It always exists and it verifies $\Omega_m=1$. It describes a matter-dominated flat FLRW solution. The point is 
    \begin{enumerate}
        \item sink for $0\leq \gamma <1$, 
        \item saddle for $1<\gamma \leq 2$,
        \item non-hyperbolic for $\gamma=1,2$.
    \end{enumerate}
    \item $P_2=(0,1)$ with eigenvalues $\lbrace -\frac{3}{2},-3 (\gamma -1) \rbrace$. It always exists and represents a scalar field dominated flat FLRW solution. The point is 
    \begin{enumerate}
        \item saddle for $0\leq \gamma <1$, 
        \item sink for $1<\gamma \leq 2$, 
        \item non-hyperbolic for $\gamma=1$.
    \end{enumerate}
    \item $P_{3,4}=(\pm 1,0)$ with eigenvalues $\lbrace \frac{3}{2},-3 (\gamma -2) \rbrace$. They always exist and both represent an anisotropic Bianchi I vacuum solution. They are
    \begin{enumerate}
        \item sources for $0\leq \gamma <2$, 
        \item non-hyperbolic for $\gamma=2$. 
    \end{enumerate}
\end{enumerate}
For this first interaction, the results are very similar to those obtained for the non-interacting model investigated in \cite{Leon:2021rcx} because in the averaging process, the parameter $\Gamma$ cancels out. In TABLE \ref{Tabla2DBianchiI} we present a summary of the analysis of the equilibrium points for system \eqref{int-1-2D-a}-\eqref{int-1-2D-b}. Aditionally, in FIG. \ref{Fig1-c} we present the dynamics of the model depicted in the phase-plane for different values of the barotropic index $\gamma:$  $\gamma=1$ (dust);  $\gamma=4/3$ (radiation); and $\gamma=2$ (stiff matter).
\begin{table}[H]
\centering
\setlength{\tabcolsep}{1mm}
\begin{tabular}{|c|c|c|c|}
\hline
Label. & $(\bar{\Sigma},\bar{\Omega})$   & Stability & Solution \\ \hline 
$P_1$ &  $(0, 0)$ & sink for $0\leq \gamma <1$ &\\ 
&& saddle for $1<\gamma \leq 2$ &\\ 
&&no h. for $\gamma=1,2$ & matter-dominated flat FLRW \\ \hline 
$P_2$ & $(0,1)$   & saddle for $0\leq \gamma <1$ &\\  
&& sink for $1<\gamma \leq 2$ &\\ 
&& no h. for $\gamma=1$ & scalar field dominated flat FLRW\\ \hline
$P_{3,4}$ & $(\pm 1,0)$  & sources for $0\leq \gamma <2$ &\\ 
&& no h. for $\gamma=2$ & anisotropic Bianchi I vacuum \\ \hline
\end{tabular}
\caption{Summary of the stability analysis for \eqref{int-1-2D-a}-\eqref{int-1-2D-b}.}
\label{Tabla2DBianchiI}
\end{table}
\subsection{Interaction 2: $ Q(\phi, {\dot\phi}, \rho_m)=\frac{\Gamma}{H_0} H\frac{  \rho_{m} \rho_\phi}{\rho_{m}+  \rho_\phi}$}
\label{int:2}
\begin{figure}[H]
    \centering
    \includegraphics[scale=0.4]{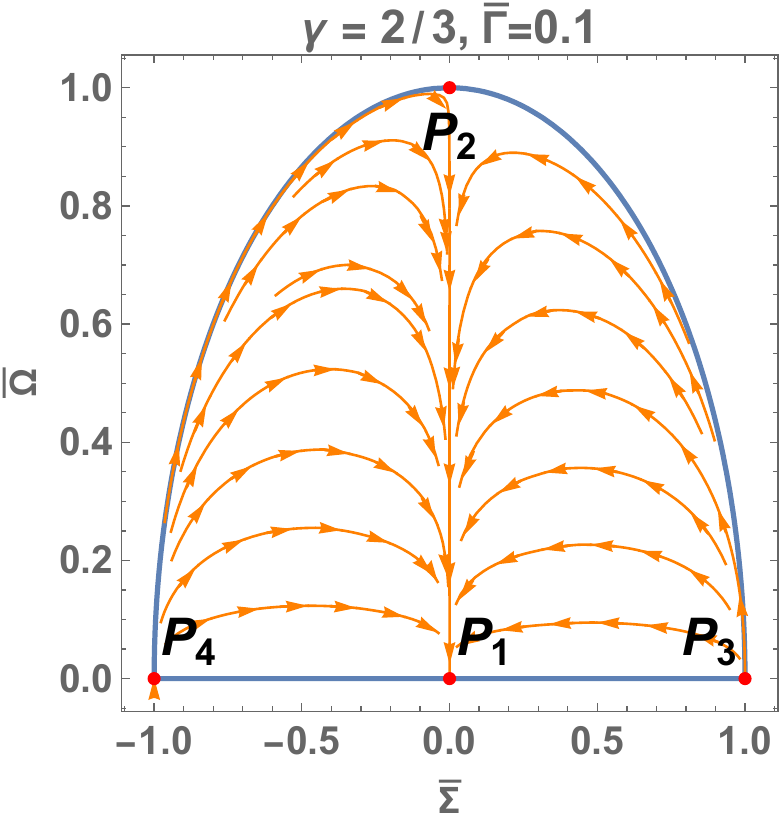}
    \includegraphics[scale=0.4]{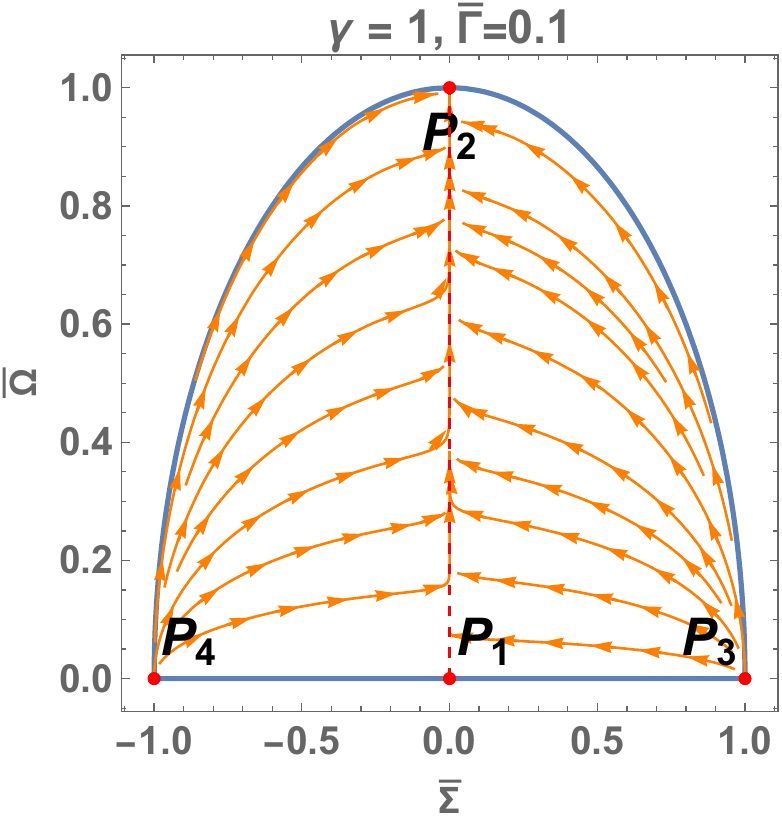}
    \includegraphics[scale=0.4]{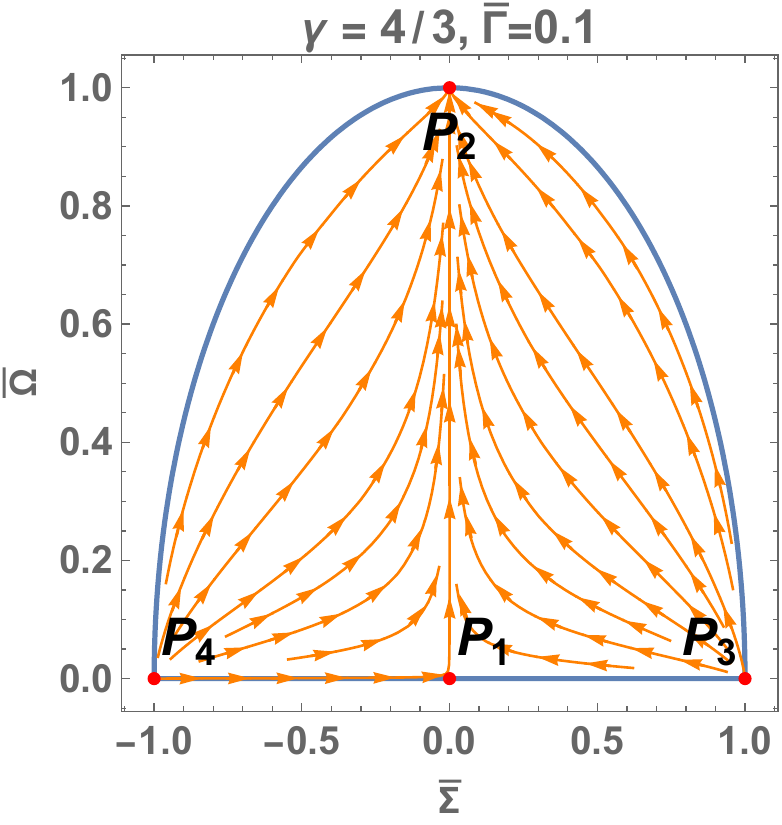}
    \includegraphics[scale=0.4]{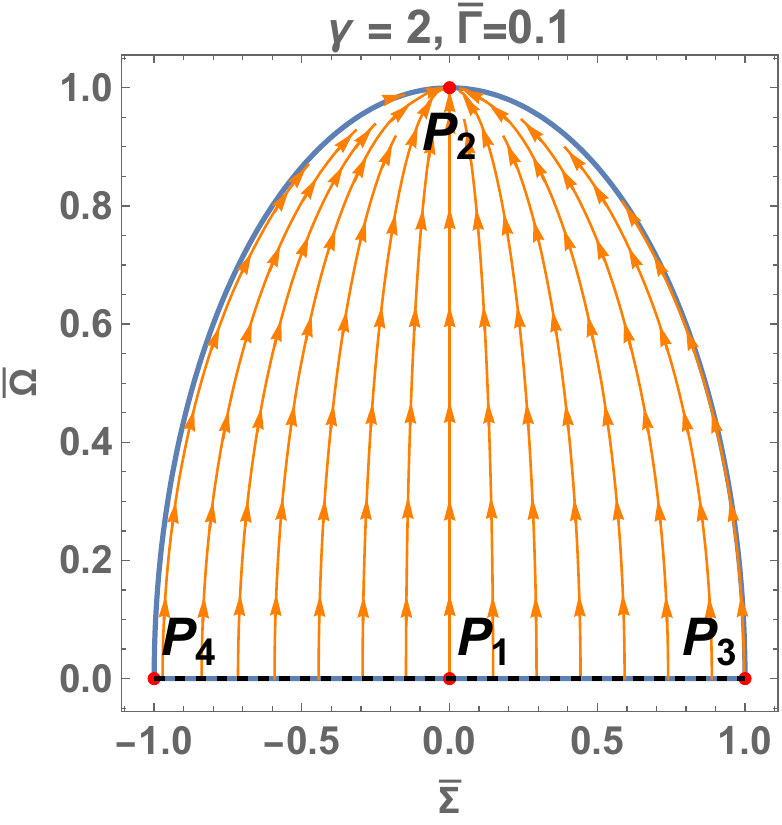}
    \caption{Phase-plane diagrams for system \eqref{int-2-2D-a}-\eqref{int-2-2D-b} setting $\bar{\Gamma}=0.1$ for different values of $\gamma$.}
    \label{Fig2}
\end{figure}

\begin{figure}[H]
    \centering
    \includegraphics[scale=0.4]{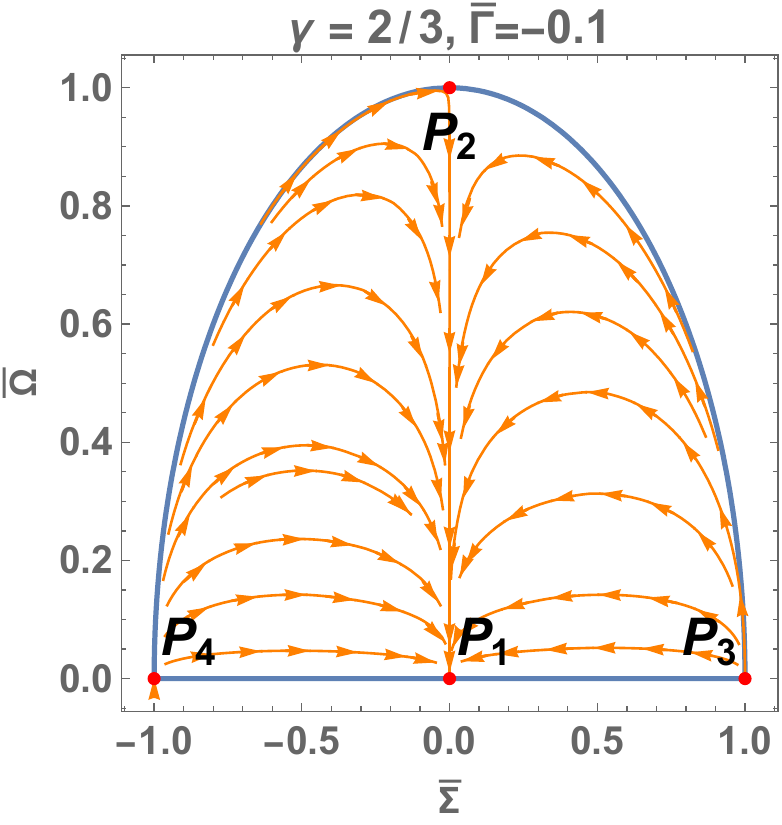}
    \includegraphics[scale=0.4]{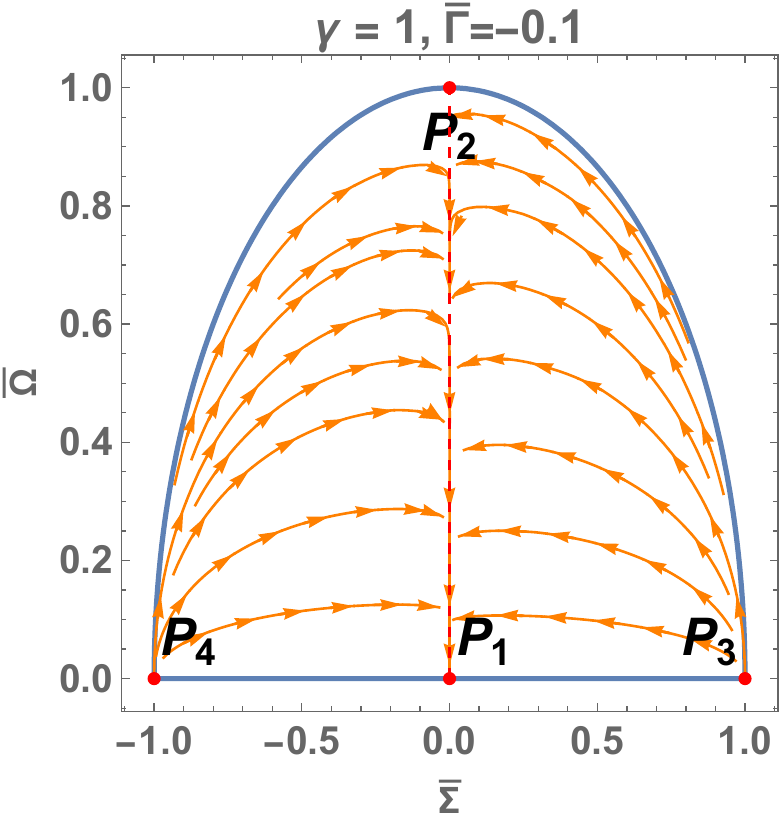}
    \includegraphics[scale=0.4]{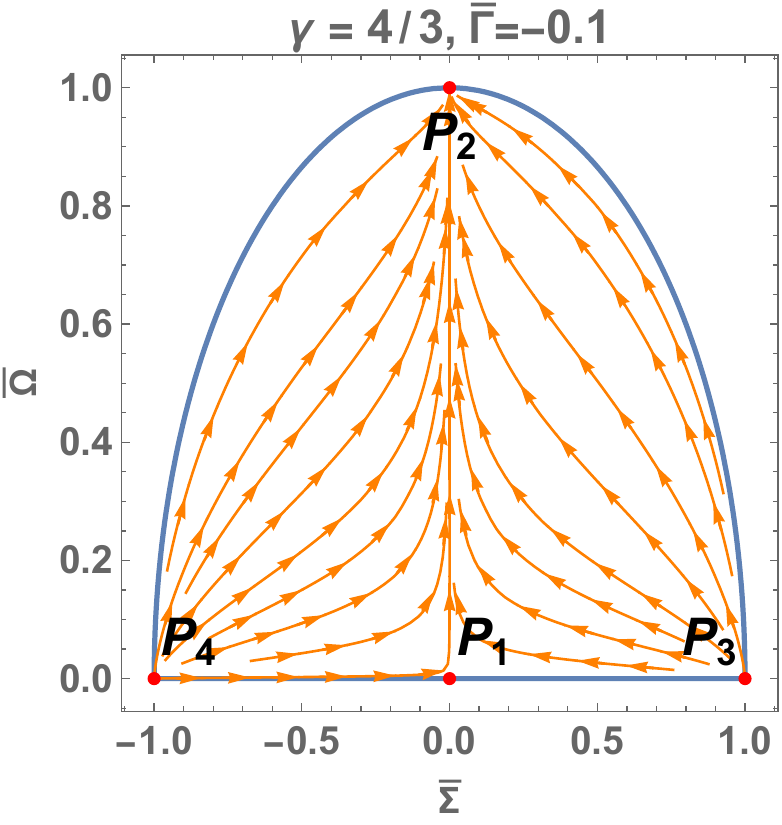}
    \includegraphics[scale=0.4]{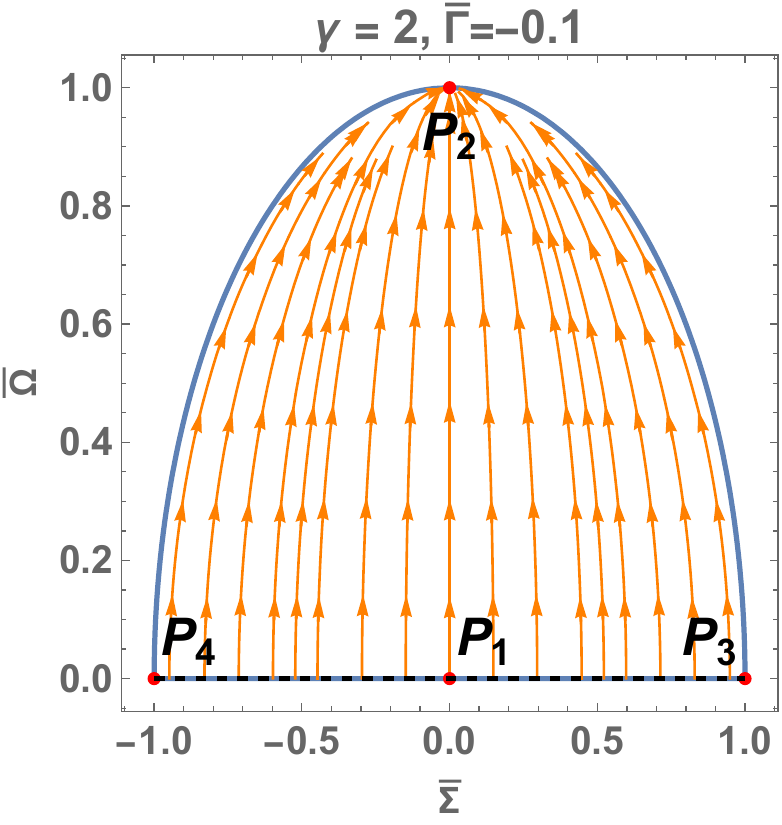}
    \caption{Phase-plane diagrams for system \eqref{int-2-2D-a}-\eqref{int-2-2D-b} setting $\bar{\Gamma}=-0.1$ for different values of $\gamma$.}
    \label{Fig3}
\end{figure}

\begin{figure}[H]
    \centering
    \includegraphics[scale=0.5]{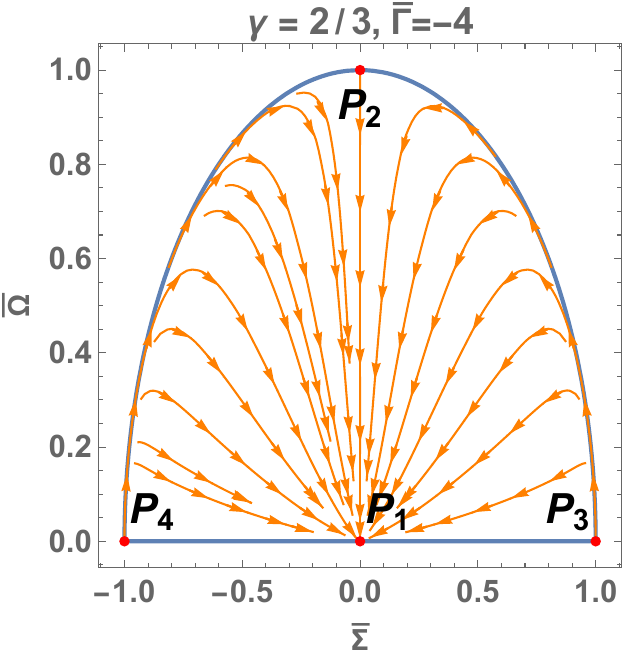}
    \includegraphics[scale=0.5]{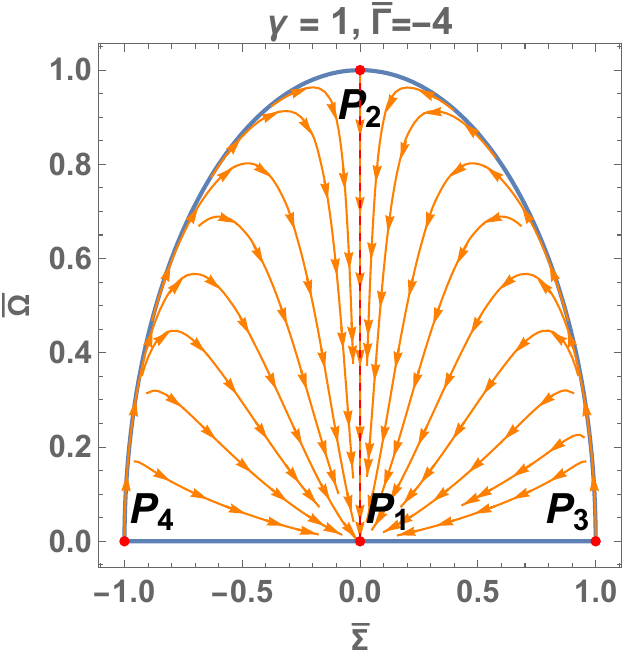}
    \includegraphics[scale=0.5]{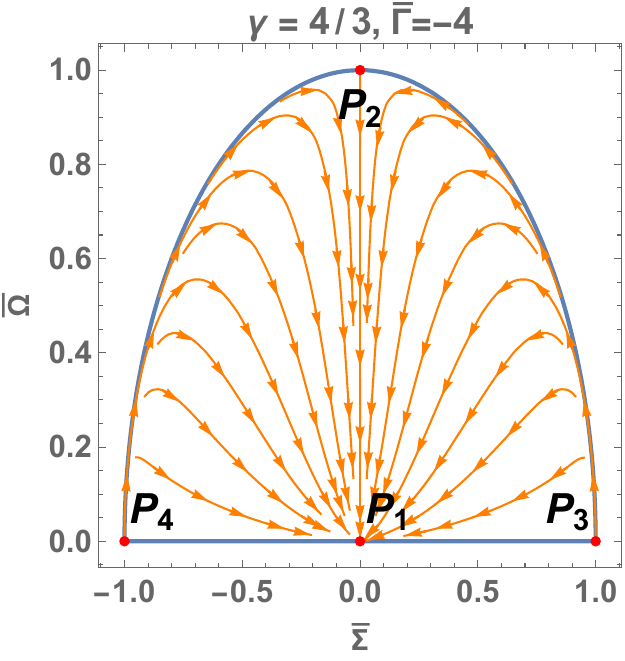}
    \includegraphics[scale=0.5]{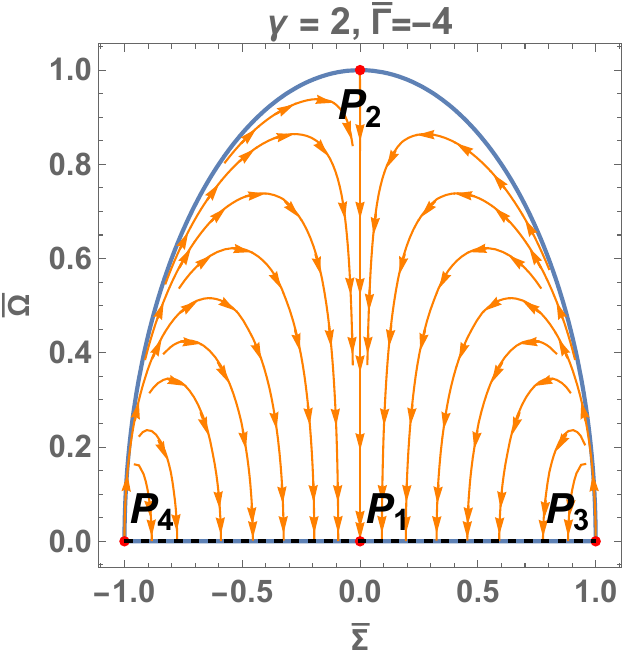}
    \caption{Phase-plane diagrams for system \eqref{int-2-2D-a}-\eqref{int-2-2D-b} setting $\bar{\Gamma}=-4$ for different values of $\gamma$.}
    \label{Fig3-b}
\end{figure}

As for the previous interaction, setting $\alpha=1,$ $\beta=-1,$ $\delta=0$ in system \eqref{gen-redu-1}-\eqref{gen-redu-4}, and expanding in a Taylor series centred at \( H = 0 \) and retaining terms up to \( H^2 \), and introducing the rescaling \( \Gamma = H_0 \bar{\Gamma} \), the late-time truncated system is:
\begin{small}
\begin{align}
&\dot{H}=-\frac{3}{2} \left(2 \Sigma ^2+\Omega
   ^2-\gamma  \left(\Sigma ^2+\Omega
   ^2-1\right)+\Omega ^2 \cos (2 (\varphi -t \omega
   ))\right) H^2+O\left(H^3\right),\\
    &\dot{\Omega}=\frac{1}{2}
   \Omega  \left(3 \left(2 \Sigma ^2+\Omega
   ^2-\gamma  \left(\Sigma ^2+\Omega
   ^2-1\right)+\left(\Omega ^2-1\right) \cos (2
   (\varphi -t \omega
   ))-1\right)+\frac{\left(\Sigma ^2+\Omega
   ^2-1\right) \bar{\Gamma}}{\Sigma
   ^2-1}\right) H\nonumber\\
   &-\frac{\left(\omega
   ^2-1\right)^3 \Omega ^3 \cos (\varphi -t \omega
   ) \sin ^3(\varphi -t \omega ) H^2}{\omega
   ^3}+O\left(H^3\right),     \label{comp-int-2-a}\\
    &\dot{\Sigma}=\frac{3}{2} \Sigma 
   \left(2 \Sigma ^2+\Omega ^2-\gamma 
   \left(\Sigma ^2+\Omega ^2-1\right)+\Omega ^2
   \cos (2 (\varphi -t \omega ))-2\right)
   H+O\left(H^3\right),\\
   &\dot{\varphi}=\left(\frac{3}{2} \sin
   (2 (\varphi -t \omega ))-\frac{\left(\Sigma
   ^2+\Omega ^2-1\right) \bar{\Gamma} \tan
   (\varphi -t \omega )}{2 \left(\Sigma
   ^2-1\right)}\right) H \nonumber\\
   &+\frac{\left(\omega
   ^2-1\right)^3 \Omega ^2 \sin ^4(\varphi -t
   \omega ) H^2}{\omega
   ^3}+O\left(H^3\right).  \label{comp-int-2-d}
\end{align}
\end{small}
The averaged equations are
\begin{align}
\label{1-prom-int2}
    &\dot{H}=\frac{3}{2} H^2 \left(\gamma 
   \left(\bar{\Sigma} ^2+\bar{\Omega} ^2-1\right)-2
   \bar{\Sigma} ^2-\bar{\Omega} ^2\right),\\
  \label{2-prom-int2}
    &\dot{\bar{\Omega}}=\frac{1}{2} H
   \bar{\Omega}  \left(\frac{\bar{\Gamma}
   \left(\bar{\Sigma} ^2+\bar{\Omega} ^2-1\right)}{\bar{\Sigma}
   ^2-1}-3 \gamma  \left(\bar{\Sigma} ^2+\bar{\Omega}
   ^2-1\right)+6 \bar{\Sigma} ^2+3 \bar{\Omega}
   ^2-3\right),\\
   \label{3-prom-int2} &\dot{\bar{\Sigma}}=\frac{3}{2} H \bar{\Sigma} 
   \left(-\gamma  \left(\bar{\Sigma} ^2+\bar{\Omega}
   ^2-1\right)+2 \bar{\Sigma} ^2+\bar{\Omega}
   ^2-2\right),\\
   \label{4-prom-int2} &\dot{\bar{\varphi}}=0.
\end{align}
from which we obtain the following guiding system 
\begin{align}
      \label{int-2-2D-a}&\bar{\Sigma}'=\frac{3}{2} \bar{\Sigma} 
   \left(-\gamma  \left(\bar{\Sigma} ^2+\bar{\Omega}
   ^2-1\right)+2 \bar{\Sigma} ^2+\bar{\Omega}
   ^2-2\right),\\
      \label{int-2-2D-b}&\bar{\Omega}'=\frac{1}{2}
   \bar{\Omega}  \left(\frac{\bar{\Gamma}
   \left(\bar{\Sigma} ^2+\bar{\Omega} ^2-1\right)}{\bar{\Sigma}
   ^2-1}-3 \gamma  \left(\bar{\Sigma} ^2+\bar{\Omega}
   ^2-1\right)+6 \bar{\Sigma} ^2+3 \bar{\Omega}
   ^2-3\right)
\end{align}
where once again, the new derivative is $f'= \frac{1}{H}\dot{f}$. Additionally, the system is defined in the compact phase-space 
\begin{equation}
    \left\{(\bar{\Sigma},\bar{\Omega})\in \mathbb{R}^2| \;  \bar{\Omega}^2 + \bar{\Sigma}^2 \leq 1, \bar{\Omega}\geq 0, -1\leq \bar{\Sigma} \leq 1 \right\}.
\end{equation}
Note that system \eqref{int-2-2D-a}-\eqref{int-2-2D-b} has a singularity in the denominator for the values $\bar{\Sigma} = \pm 1$, therefore, it is necessary to properly analyse the equilibrium points involving that coordinate \cite{Paliathanasis:2024jxo,Leon:2025sfd,Papagiannopoulos:2025zku}. The equilibrium points for system \eqref{int-2-2D-a}-\eqref{int-2-2D-b} in the coordinates $(\bar{\Sigma}, \bar{\Omega})$ are the following
\begin{enumerate}
    \item $P_1=(0,0),$ with eigenvalues 
    $\left\{\frac{3 (\gamma -2)}{2},\frac{1}{2} \left(\bar{\Gamma}+3 \gamma -3\right)\right\}.$ It always exists and we verify that $\Omega_m=1.$ It describes a matter-dominated flat FLRW solution. The point is
    \begin{enumerate}
        \item sink for $0\leq \gamma <2,$ $\bar{\Gamma}<-3 (\gamma -1),$
        \item saddle for $0\leq \gamma <2,$ $\bar{\Gamma}>-3 (\gamma -1),$
        \item non-hyperbolic for $\bar{\Gamma}=-3 (\gamma -1)$ or $ \gamma =2.$
    \end{enumerate}
    \item $P_2=(0,1),$ with eigenvalues $\left\{-\frac{3}{2},-\bar{\Gamma}-3 \gamma +3\right\}.$ It always exists It always exists and represents a scalar field dominated flat FLRW solution. The point is
    \begin{enumerate}
        \item sink for $0\leq \gamma \leq 2,$ $ \bar{\Gamma}>-3 (\gamma -1),$
         \item saddle for $0\leq \gamma \leq 2,$ $ \bar{\Gamma}<-3 (\gamma -1),$
         \item non-hyperbolic for
         $\bar{\Gamma}=-3 (\gamma -1).$
    \end{enumerate}
    \item $P_{3,4}=(\pm 1,0)$ with eigenvalues $\{0,0\}.$ These points are anisotropic vacuum Bianchi I solutions. These points make the numerators of the system vanish, but they also make the denominator of the equation for $\Omega$ vanish. To study their stability, the limit of the eigenvalues is considered when $\Sigma\rightarrow \pm 1$. For that prurpose, we evaluate the Jacobian matrix in $\Omega=0$, the eigenvalues are $$\left\{-\frac{3}{2} \left(3 \gamma  \Sigma
   ^2-\gamma -6 \Sigma ^2+2\right),\frac{1}{2}
   \left(\bar{\Gamma}-3 \gamma  \Sigma ^2+3
   \gamma +6 \Sigma ^2-3\right)\right\}.$$ Taking the limit, we verify that the eigenvalues are $$\left\{-\frac{3}{2} (2 \gamma -4),\frac{1}{2}
   \left(\bar{\Gamma}+3\right)\right\}.$$
 The stability of the points is as follows
   \begin{enumerate}
       \item source for $0\leq \gamma <2, \bar{\Gamma}>-3$
       \item saddle for $0\leq \gamma <2, \bar{\Gamma}<-3$,
       \item non-hyperbolic for $\gamma =2 \bar{\Gamma}+3\neq
   0$ or $\bar{\Gamma}=-3,
   0\leq \gamma \leq 2.$
   \end{enumerate}
\end{enumerate}
For the second interaction~\eqref{int-2}, the influence of the parameter \( \bar{\Gamma} \) on the stability of the equilibrium points is highlighted. In the full system~\eqref{comp-int-2-a}--\eqref{comp-int-2-d}, the presence of the tangent function alters the behaviour of the numerical solutions. FIG.~\ref{Fig2} shows the phase plane of the guiding system~\eqref{int-2-2D-a}--\eqref{int-2-2D-b} for different values of \( \gamma \) with \( \bar{\Gamma} = 0.1 \), while FIG.~\ref{Fig3} presents the phase plane of the same system for various values of \( \gamma \) with \( \bar{\Gamma} = -0.1 \). Additionally, FIG. \ref{Fig3-b} depicts a the phase plane setting $\bar{\Gamma}=-4.$
\subsection{Interaction 3: $Q\left(\dot{\phi},\rho_{\phi }, \rho _{m}\right)=\frac{\Gamma}{H_0} H \rho_m$}
\label{int:3}
Setting $\alpha=1,$ $\beta=0,$ $\delta=0$ in system \eqref{gen-redu-1}-\eqref{gen-redu-4}, and expanding in a Taylor series centred around  $H=0$, keeping up to second order terms and using the rescaling $\Gamma =H_0 \bar{\Gamma}$, the truncated system is
\begin{small}
\begin{align}
    &\dot{H}=-\frac{3}{2} \left(2 \Sigma ^2+\Omega
   ^2-\gamma  \left(\Sigma ^2+\Omega
   ^2-1\right)+\Omega ^2 \cos (2 (\varphi -t \omega
   ))\right)
   H^2+O\left(H^3\right), \label{comp-int-3-a}\\
    &\dot{\Omega}=\left(\frac{3}{2}
   \Omega  \left(2 \Sigma ^2+\Omega ^2-\gamma 
   \left(\Sigma ^2+\Omega
   ^2-1\right)+\left(\Omega ^2-1\right) \cos (2
   (\varphi -t \omega
   ))-1\right)-\frac{\left(\Sigma ^2+\Omega
   ^2-1\right) \bar{\Gamma}}{2 \Omega }\right)
   H\nonumber\\
   &-\frac{\left(\omega ^2-1\right)^3 \Omega ^3
   \cos (\varphi -t \omega ) \sin ^3(\varphi -t \omega
   ) H^2}{\omega
   ^3}+O\left(H^3\right),
   \end{align}
  \begin{align}
    &\dot{\Sigma}=\frac{3}{2} \Sigma 
   \left(2 \Sigma ^2+\Omega ^2-\gamma 
   \left(\Sigma ^2+\Omega ^2-1\right)+\Omega ^2
   \cos (2 (\varphi -t \omega ))-2\right)
   H+O\left(H^3\right),\\
 &\dot{\varphi}=\frac{\left(6 \Omega ^2
   \cos ^2(\varphi -t \omega )+\left(\Sigma
   ^2+\Omega ^2-1\right) \bar{\Gamma}\right)
   \tan (\varphi -t \omega ) H}{2 \Omega
   ^2}+\frac{\left(\omega ^2-1\right)^3 \Omega
   ^2 \sin ^4(\varphi -t \omega ) H^2}{\omega
   ^3}+O\left(H^3\right).    \label{comp-int-3-d}
\end{align}
\end{small}
The averaged equations are
\begin{align}
\label{1-prom-int3}
    &\dot{H}=\frac{3}{2} H^2 \left(\gamma 
   \left(\bar{\Sigma} ^2+\bar{\Omega} ^2-1\right)-2 \bar{\Sigma}
   ^2-\bar{\Omega} ^2\right),\\
   \label{2-prom-int3} &\dot{\bar{\Omega}}=H\left[-\frac{\bar{\Gamma} \left(\bar{\Sigma} ^2+\bar{\Omega} ^2-1\right)}{\bar{\Omega} }-3 \bar{\Omega}  \left(\gamma  \left(\bar{\Sigma}
   ^2+\bar{\Omega} ^2-1\right)-2 \bar{\Sigma} ^2-\bar{\Omega} ^2+1\right)\right],\\
    \label{3-prom-int3}&\dot{\bar{\Sigma}}=\frac{3}{2} H \bar{\Sigma}  \left(-\gamma 
   \left(\bar{\Sigma} ^2+\bar{\Omega} ^2-1\right)+2 \bar{\Sigma}
   ^2+\bar{\Omega} ^2-2\right),\\
   \label{4-prom-int3} &\dot{\bar{\varphi}}=0.
\end{align}
The guiding system is
\begin{align}
\label{int-3-2D-a}&\bar{\Sigma}'=\frac{3}{2}  \bar{\Sigma}  \left(-\gamma \left(\bar{\Sigma} ^2+\bar{\Omega} ^2-1\right)+2 \Sigma
   ^2+\bar{\Omega} ^2-2\right),\\
\label{int-3-2D-b}&\bar{\Omega}'=-\frac{\bar{\Gamma} \left(\bar{\Sigma} ^2+\bar{\Omega} ^2-1\right)}{\bar{\Omega} }-3 \bar{\Omega}  \left(\gamma  \left(\bar{\Sigma}
   ^2+\bar{\Omega} ^2-1\right)-2 \bar{\Sigma} ^2-\bar{\Omega} ^2+1\right).    
\end{align}
with a new derivative defined as $f'= \frac{1}{H}\dot{f}$. With the same compact phase-space as before  
\begin{equation}
    \left\{(\bar{\Sigma},\bar{\Omega})\in \mathbb{R}^2| \;  \bar{\Omega}^2 + \bar{\Sigma}^2 \leq 1, \bar{\Omega}\geq 0, -1\leq \bar{\Sigma} \leq 1 \right\}.
\end{equation}
The previous interaction had a singularity when $\bar{\Sigma}=\pm 1$, now we observe the presence of another singularity in system \eqref{int-3-2D-a}-\eqref{int-3-2D-b} in the line $\bar{\Omega}=0$ \cite{Paliathanasis:2024jxo,Leon:2025sfd,Papagiannopoulos:2025zku}. The equilibrium points for system \eqref{int-3-2D-a}-\eqref{int-3-2D-b} in the coordinates $(\bar{\Sigma},\bar{\Omega})$ are
\begin{enumerate}
    \item $P_2=(0,1),$ with eigenvalues $\left\{-3,-2 \left(\bar{\Gamma}+3 \gamma -3\right)\right\}.$ It always exists, it represents a scalar field dominated solution and is 
    \begin{enumerate}
        \item sink for $0\leq \gamma \leq 2$, $ \bar{\Gamma}>-3 (\gamma -1),$
        \item saddle for $0\leq \gamma \leq 2$, $\bar{\Gamma}<-3 (\gamma -1),$
        \item non-hyperbolic for $\bar{\Gamma}=-3 (\gamma -1).$
    \end{enumerate}
 \item $P_{3,4}=(\pm 1,0),$ with eigenvalues $\{0,0\}.$ They belong to the singular line $\bar{\Omega}=0$ and are non-hyperbolic.
    \item $P_5=(0,\sqrt{\frac{\bar{\Gamma}}{3 (1-\gamma )}})$ with eigenvalues $\left\{\bar{\Gamma}+3 \gamma -6,2
   \left(\bar{\Gamma}+3 \gamma
   -3\right)\right\}$. This point exist for $0\leq \gamma <1, 0\leq \bar{\Gamma}\leq
   3-3 \gamma$ or $1<\gamma \leq 2, 3-3 \gamma \leq \bar{\Gamma}\leq 0$ and is
   \begin{enumerate}
       \item sink for 
       $0\leq \gamma <1, 0\leq \bar{\Gamma}<3-3
   \gamma,$
   \item saddle for $1<\gamma <2, 3-3 \gamma <\bar{\Gamma}\leq 0$ or $\gamma =2, -3<\bar{\Gamma}<0$
       \item non-hyperbolic for $0\leq \gamma <1, 0\leq \bar{\Gamma}=3-3
   \gamma.$
   \end{enumerate}
\end{enumerate}
The analysis of Interaction 3~\eqref{int-3} shows that \( \bar{\Gamma} \) influences the stability of the equilibrium points and the existence of point \( P_5 \). In the full system~\eqref{comp-int-3-a}--\eqref{comp-int-3-d}, the tangent function alters the numerical behavior. Figure~\ref{Fig5} illustrates the phase plane of the guiding system~\eqref{int-3-2D-a}--\eqref{int-3-2D-b} for \( \bar{\Gamma} = 0.1 \) and \( \bar{\Gamma} = -0.1 \). For \( \bar{\Gamma} = -0.1 \), \( \bar{\Omega} = 0 \) is an attractor, but its singular nature makes it physically unacceptable \cite{Paliathanasis:2024jxo,Leon:2025sfd,Papagiannopoulos:2025zku}.
\begin{figure}[H]
    \centering
    \includegraphics[scale=0.4]{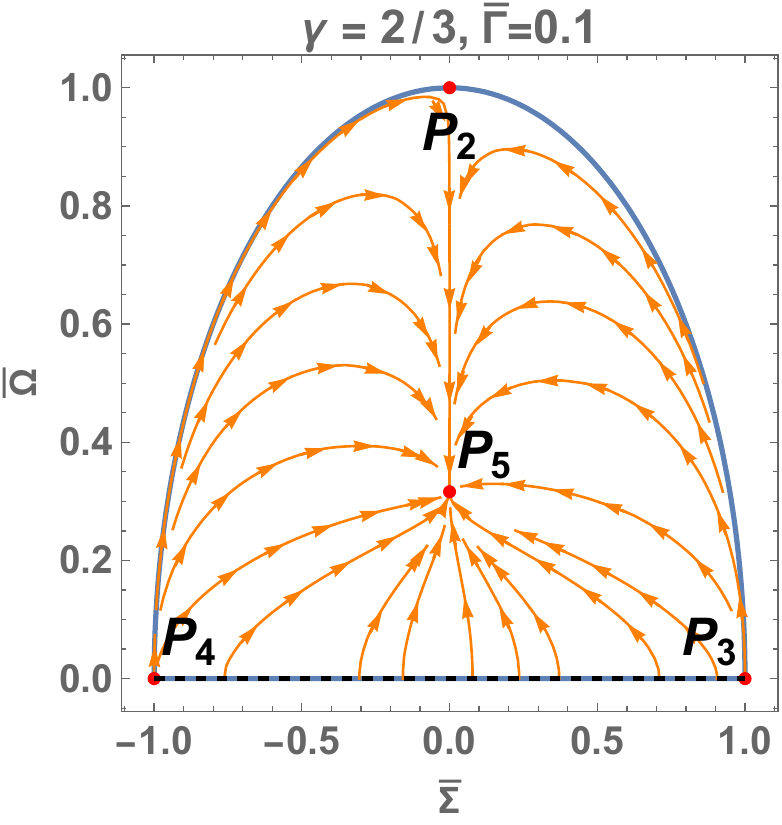}
    \includegraphics[scale=0.4]{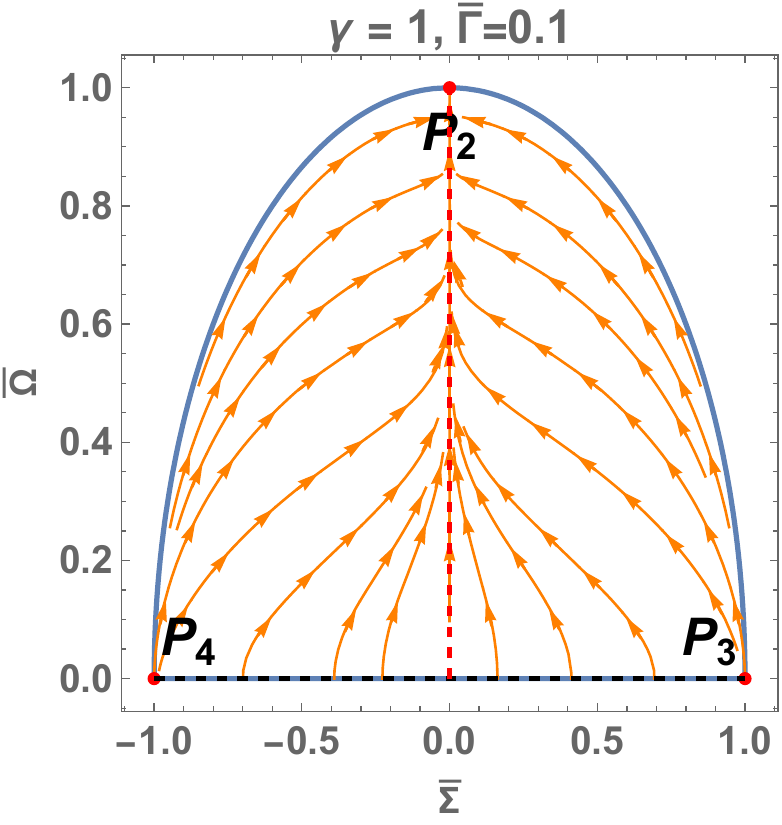}
    \includegraphics[scale=0.4]{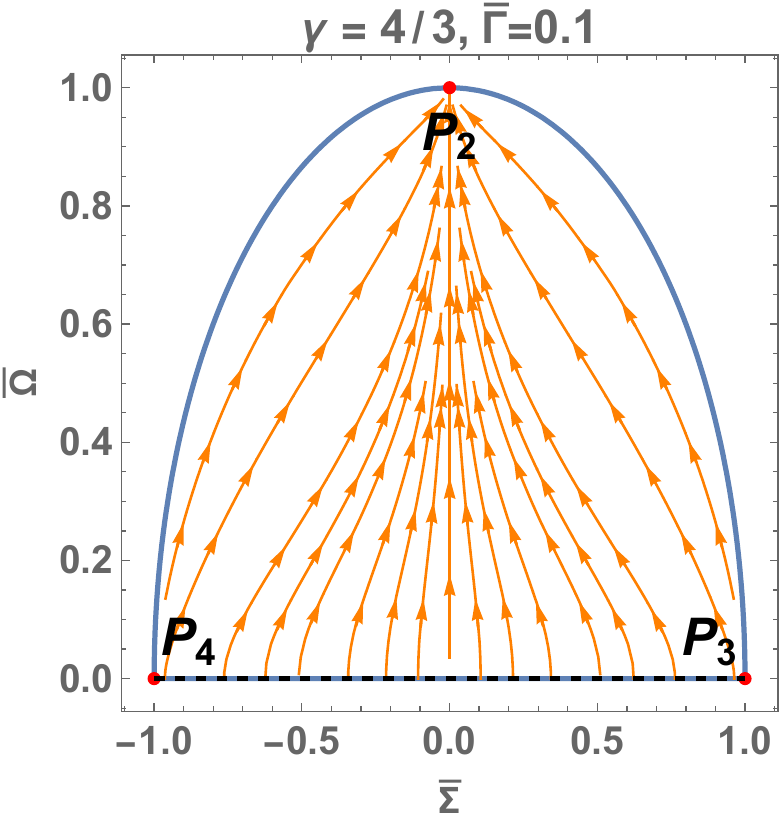}
    \includegraphics[scale=0.4]{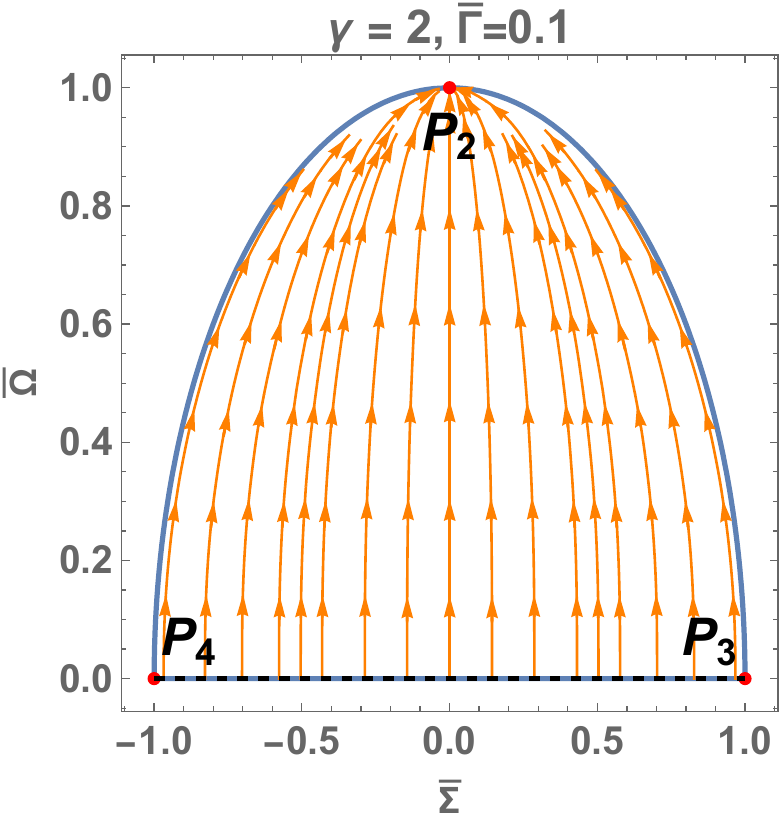}
    \caption{Phase-plane diagrams for system \eqref{int-3-2D-a}-\eqref{int-3-2D-b} setting $\bar{\Gamma}=0.1$ for different values of $\gamma$.}
    \label{Fig4}
\end{figure}
\begin{figure}[H]
    \centering
    \includegraphics[scale=0.4]{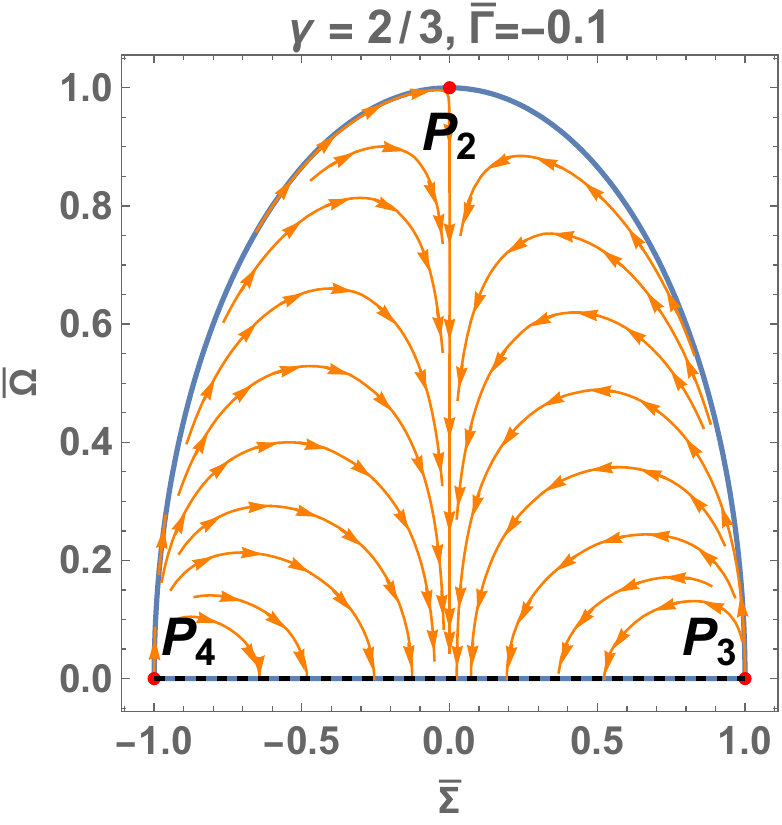}
    \includegraphics[scale=0.4]{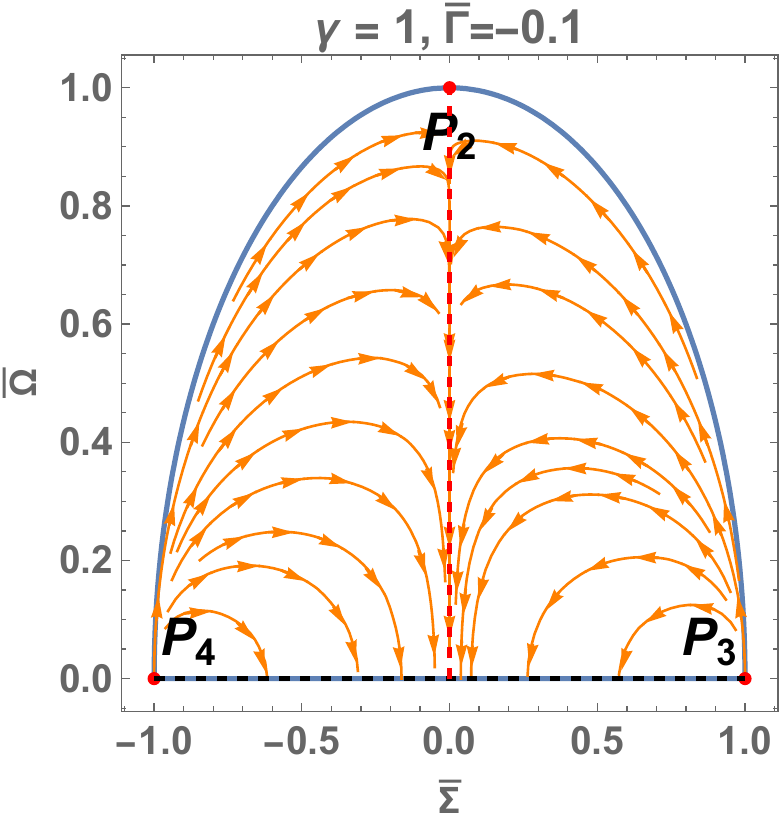}
    \includegraphics[scale=0.4]{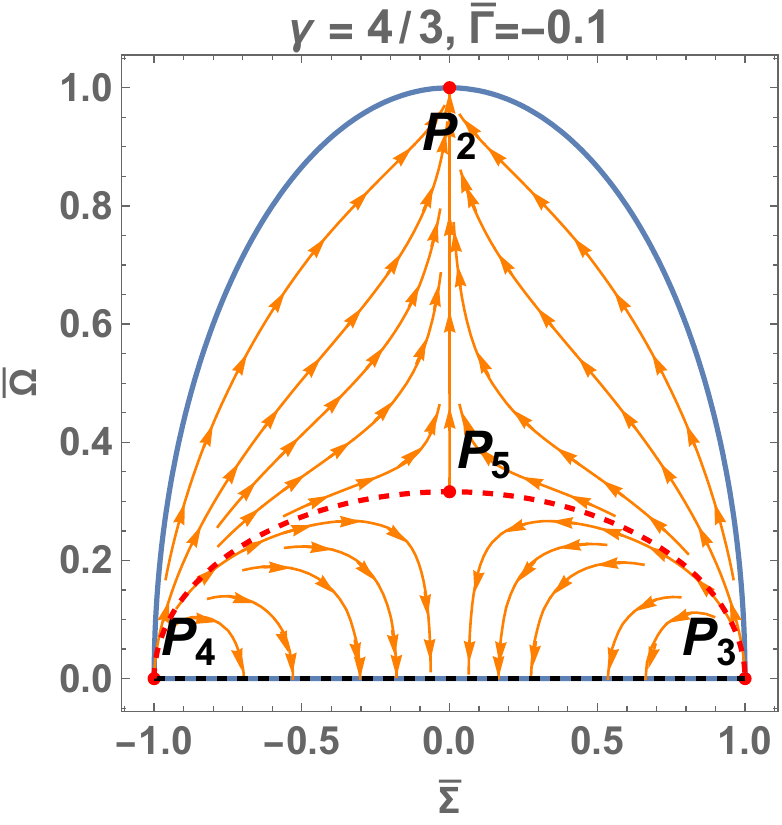}
    \includegraphics[scale=0.4]{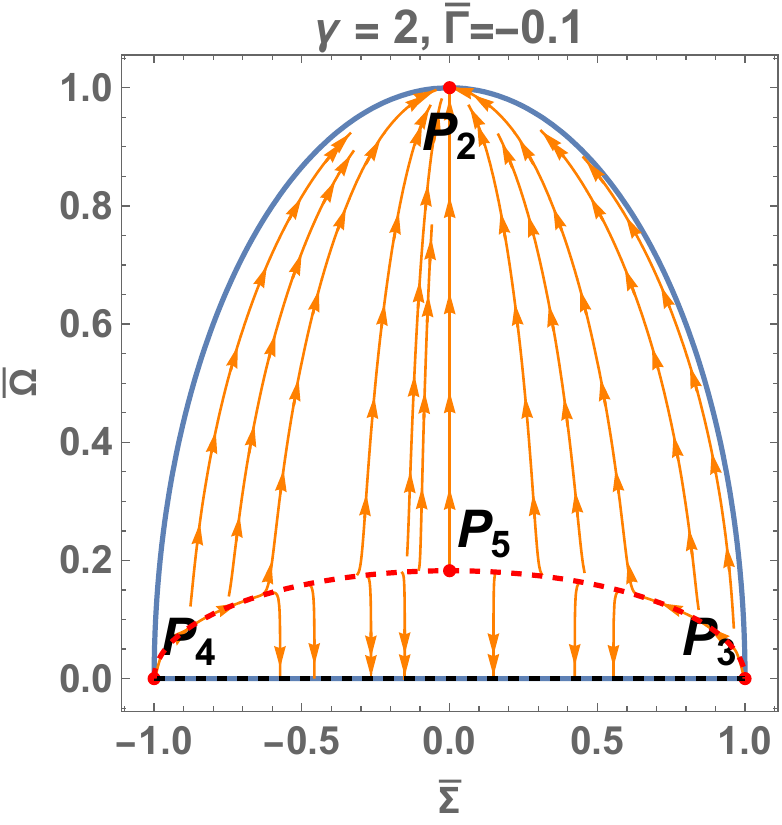}
    \caption{Phase-plane diagrams for system \eqref{int-3-2D-a}-\eqref{int-3-2D-b} setting $\bar{\Gamma}=-0.1$ for different values of $\gamma$. The red-dashed line is the unstable manifold of $P_5$, defined by $\bar{\Omega} = \sqrt{\frac{\bar{\Gamma}}{3(1-\gamma)}}\sqrt{1-\bar{\Sigma}^2}$.}
    \label{Fig5}
\end{figure}
The phase plane diagrams of the system~\eqref{int-3-2D-a}--\eqref{int-3-2D-b}, depicted in FIG.~\ref{Fig5} for different values of \( \gamma \) and \( \bar{\Gamma} = -0.1 \), provide a detailed view of the system's dynamics under these conditions. Notably, the red dashed line represents the unstable manifold of \( P_5 \), defined by \( \bar{\Omega} = \sqrt{\frac{\bar{\Gamma}}{3(1 - \gamma)}} \sqrt{1 - \bar{\Sigma}^2} \), which describes trajectories that initially appear attractive but eventually move away from this point, influencing the overall dynamics of the system. These diagrams highlight how the parameters \( \gamma \) and \( \bar{\Gamma} \) affect the flow curves and the global behaviour of the solutions. Furthermore, analysing these trajectories in relation to the unstable manifold provides key insights into the system’s most sensitive regions and their evolution.

\subsection{Interaction 4: $Q\left(\dot{\phi},\rho_{\phi }, \rho _{m}\right)=\frac{\Gamma}{H_0} H \rho_\phi$}
\label{int:4}
\begin{figure}[h!]
    \centering
    \includegraphics[scale=0.4]{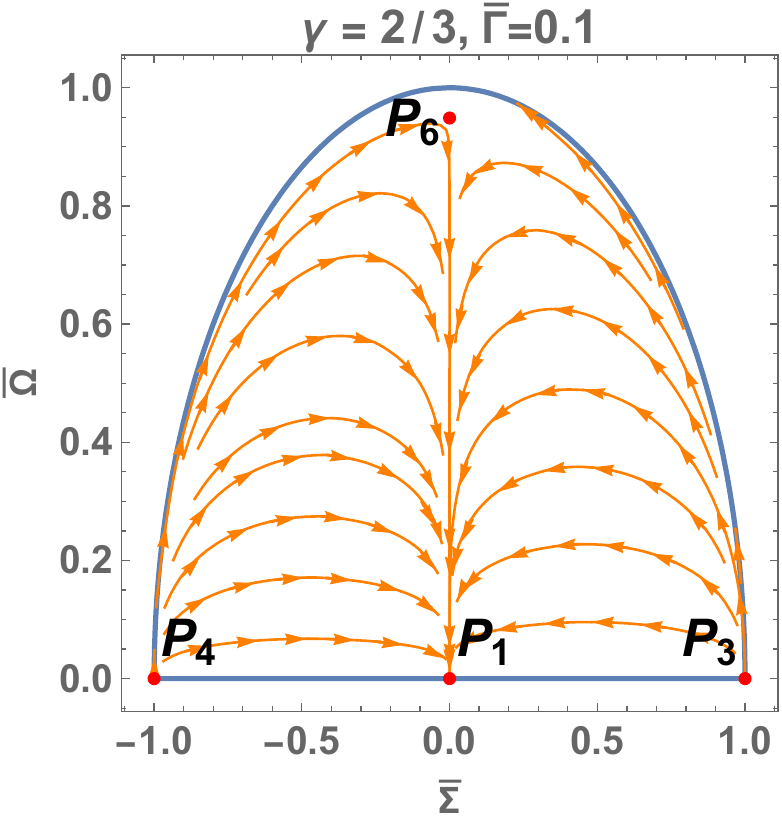}
    \includegraphics[scale=0.4]{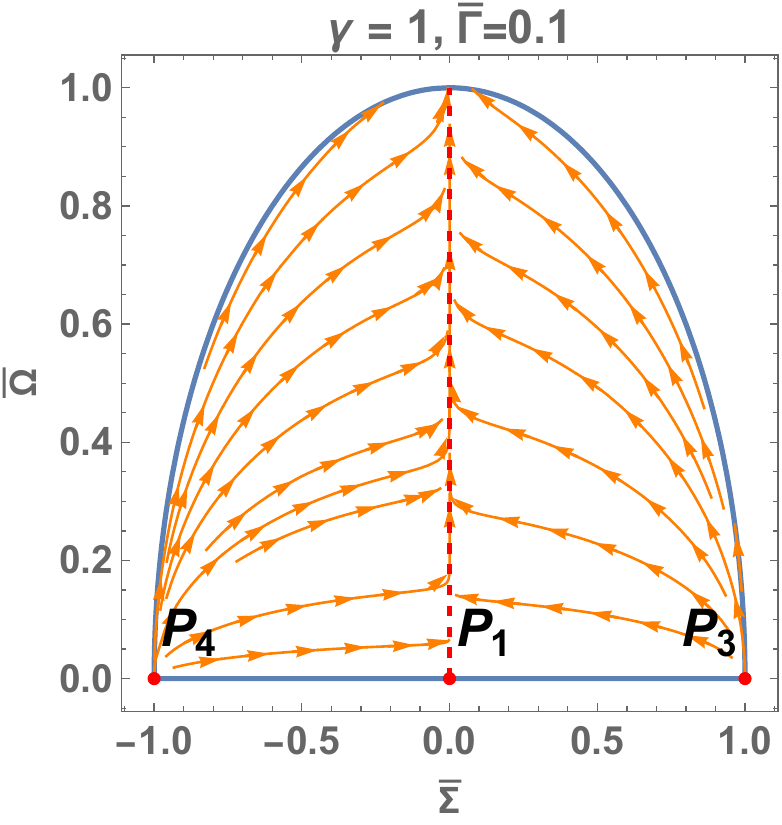}
    \includegraphics[scale=0.4]{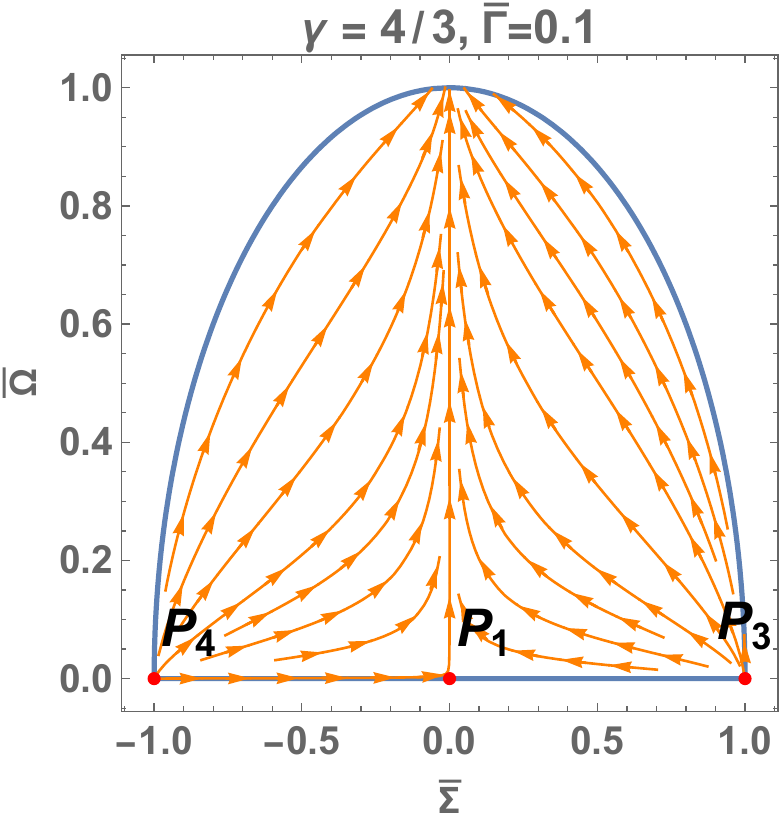}
    \includegraphics[scale=0.4]{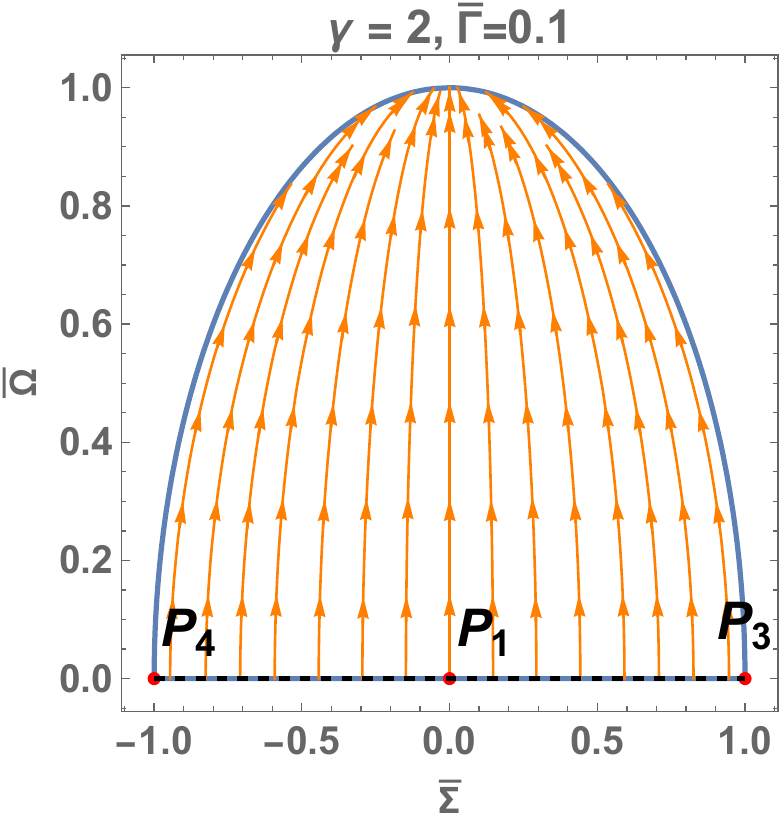}
    \caption{Phase-plane diagrams for system \eqref{int-4-2D-a}-\eqref{int-4-2D-b} setting $\bar{\Gamma}=0.1$ for different values of $\gamma$.}
    \label{fig:int4-2D-Gamma>0}
\end{figure}

\begin{figure}[h!]
    \centering
    \includegraphics[scale=0.4]{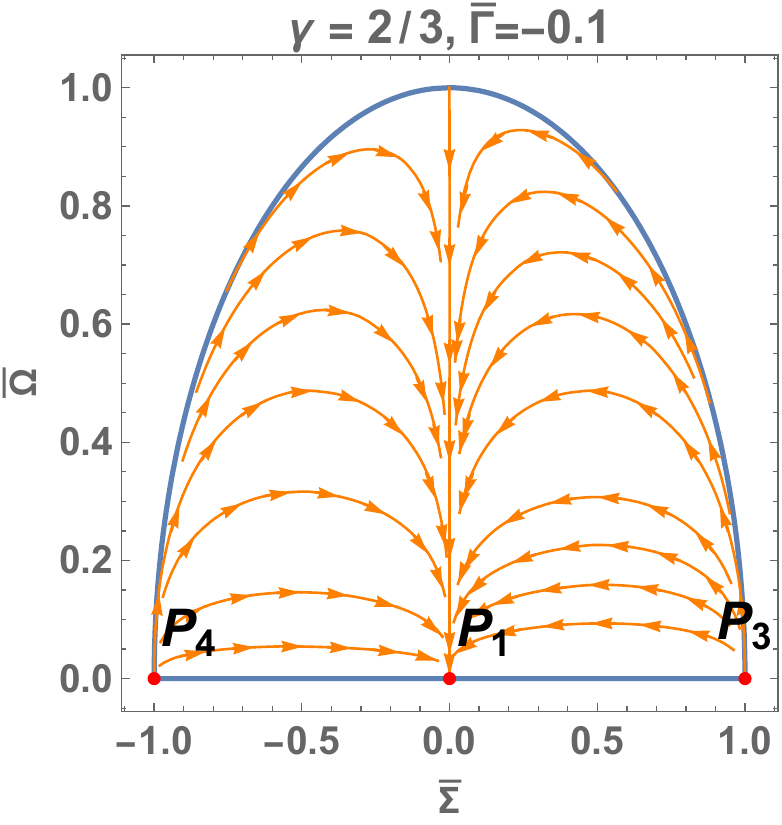}
    \includegraphics[scale=0.4]{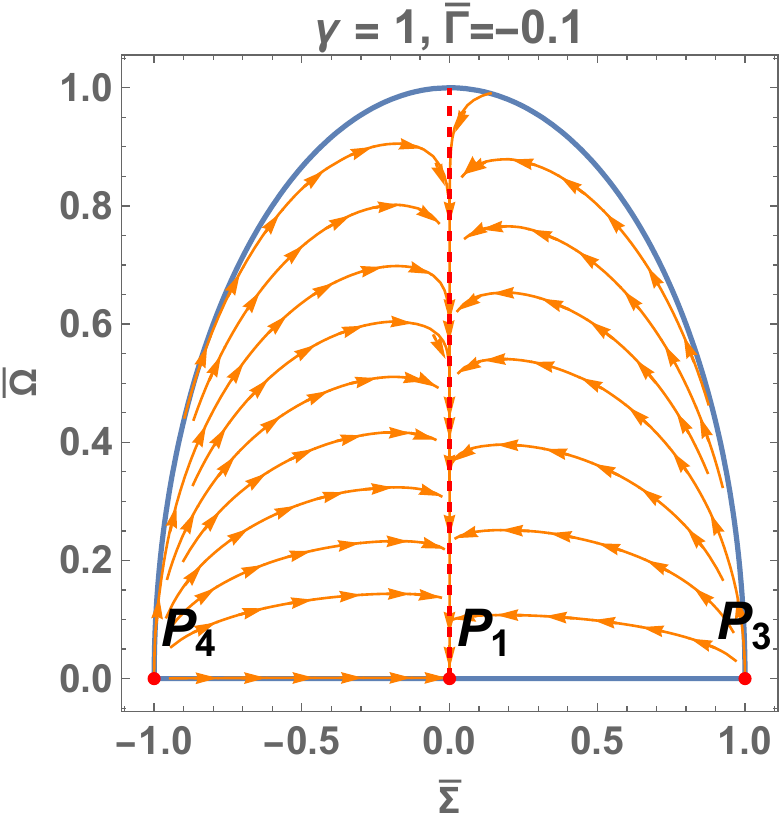}
    \includegraphics[scale=0.4]{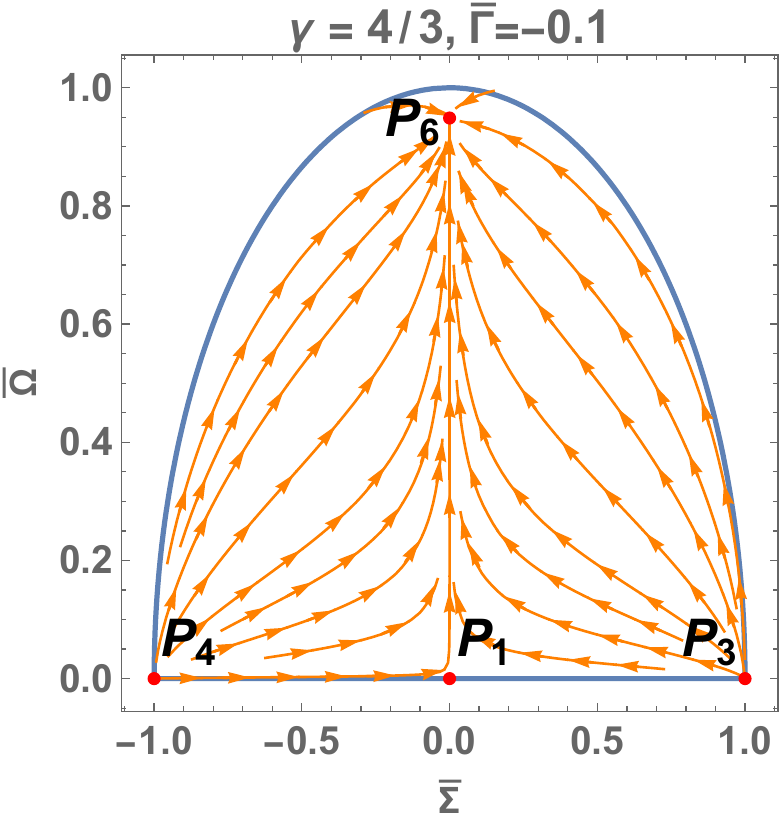}
    \includegraphics[scale=0.4]{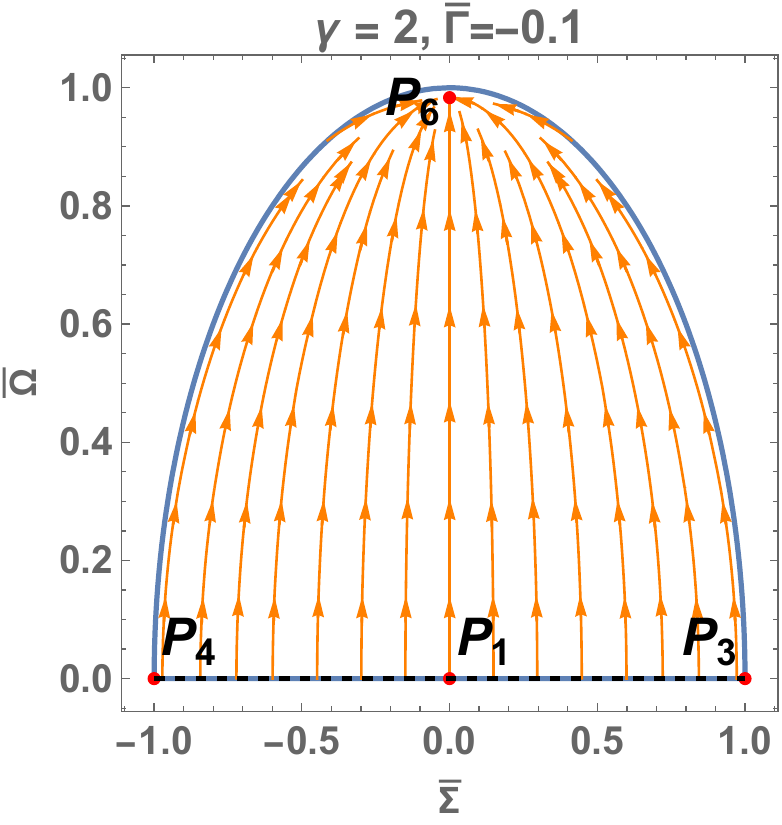}
    \caption{Phase-plane diagrams for system \eqref{int-4-2D-a}-\eqref{int-4-2D-b} setting $\bar{\Gamma}=-0.1$ for different values of $\gamma$.}
    \label{fig:int4-2D-Gamma<0}
\end{figure}

Proceeding as before, setting $\alpha=0,$ $\beta=0,$ $\delta=0$ in system \eqref{gen-redu-1}-\eqref{gen-redu-4}, and expanding in a Taylor series around $H=0$ the truncated system is
\begin{align}
 &\dot{H}=-\frac{3}{2} \left(2 \Sigma ^2+\Omega ^2-\gamma  \left(\Sigma ^2+\Omega ^2-1\right)+\Omega ^2 \cos (2
   (\varphi -t \omega ))\right) H^2+O\left(H^3\right),   \label{comp-int-4-a} \\
    &\dot{\Omega}=\frac{1}{2} \Omega  \left(3 \left(2 \Sigma ^2+\Omega
   ^2-\gamma  \left(\Sigma ^2+\Omega ^2-1\right)+\left(\Omega ^2-1\right) \cos (2 (\varphi -t \omega
   ))-1\right)+\bar{\Gamma}\right) H \nonumber \\&-\frac{\left(\omega ^2-1\right)^3 \Omega ^3 \cos (\varphi -t \omega ) \sin
   ^3(\varphi -t \omega ) H^2}{\omega ^3}+O\left(H^3\right),
\\
    &\dot{\Sigma}=\frac{3}{2} \Sigma  \left(2 \Sigma ^2+\Omega
   ^2-\gamma  \left(\Sigma ^2+\Omega ^2-1\right)+\Omega ^2 \cos (2 (\varphi -t \omega ))-2\right)
   H+O\left(H^3\right),\\
   &\dot{\varphi}=\frac{1}{2} \left(6 \cos ^2(\varphi -t \omega )-\bar{\Gamma}\right) \tan (\varphi -t \omega
   ) H+\frac{\left(\omega ^2-1\right)^3 \Omega ^2 \sin ^4(\varphi -t \omega ) H^2}{\omega
   ^3}+O\left(H^3\right).  \label{comp-int-4-d}
\end{align}
The rescaling $\Gamma =H_0 \bar{\Gamma}$ was used. The averaged system is
\begin{align}
 &\dot{H}=\frac{3}{2} H^2 \left(\gamma  \left(\bar{\Sigma} ^2+\bar{\Omega}^2-1\right)-2 \bar{\Sigma} ^2-\bar{\Omega}^2\right), \label{1-prom-int4} \\
 &\dot{\bar{\Omega}}=\frac{1}{2} H \bar{\Omega} \left(\bar{\Gamma}-3 \gamma  \left(\bar{\Sigma} ^2+\bar{\Omega}^2-1\right)+6 \bar{\Sigma}^2+3 \bar{\Omega}^2-3\right), \label{2-prom-int4} \\
  &\dot{\bar{\Sigma}}=\frac{3}{2} H \bar{\Sigma}  \left(-\gamma  \left(\bar{\Sigma} ^2+\bar{\Omega}^2-1\right)+2 \bar{\Sigma}^2+\bar{\Omega}^2-2\right), \label{3-prom-int4}\\
&\dot{\bar{\varphi}}=0. \label{4-prom-int4}
\end{align}
The guiding system is
\begin{align}
\label{int-4-2D-a}
    &\bar{\Sigma}'=\frac{3}{2}  \bar{\Sigma}  \left(-\gamma  \left(\bar{\Sigma} ^2+\bar{\Omega} ^2-1\right)+2 \bar{\Sigma}
   ^2+\bar{\Omega} ^2-2\right),\\
   \label{int-4-2D-b}&\bar{\Omega}'=\frac{1}{2}  \bar{\Omega}  \left(\bar{\Gamma}-3 \gamma  \left(\bar{\Sigma} ^2+\bar{\Omega} ^2-1\right)+6 \bar{\Sigma}
   ^2+3 \bar{\Omega} ^2-3\right).
\end{align}
The prime derivative means $f'= \frac{1}{H}\dot{f}$ and the system is defined in the compact phase-space 
\begin{equation}
    \left\{(\bar{\Sigma},\bar{\Omega})\in \mathbb{R}^2| \;  \bar{\Omega}^2 + \bar{\Sigma}^2 \leq 1, \bar{\Omega}\geq 0, -1\leq \bar{\Sigma} \leq 1 \right\}.
\end{equation}
The equilibrium points for system \eqref{int-4-2D-a}-\eqref{int-4-2D-b} in the coordinates $(\bar{\Sigma},\bar{\Omega})$ are
\begin{enumerate}
    \item $P_1=(0,0)$ with eigenvalues $\left\{\frac{3 (\gamma -2)}{2},\frac{1}{2} \left(\bar{\Gamma}+3 \gamma -3\right)\right\}.$ It always exists and is
    \begin{enumerate}
        \item sink for $0\leq \gamma <2$, $ \bar{\Gamma}<-3 (\gamma -1),$
        \item saddle for $0\leq \gamma <2$, $ \bar{\Gamma}>-3 (\gamma -1),$
        \item non-hyperbolic for $\bar{\Gamma}=-3 (\gamma -1)$ or $ \gamma =2.$  \end{enumerate}
    \item $P_{3,4}=(\pm 1,0)$ with eigenvalues $\left\{6-3 \gamma ,\frac{1}{2} \left(\bar{\Gamma}+3\right)\right\}.$ They always exist and are
    \begin{enumerate}
        \item sources for $0\leq \gamma <2$, $ \bar{\Gamma}>-3,$
        \item saddles for $0\leq \gamma <2$, $ \bar{\Gamma}<-3,$
        \item non-hyperbolic for $\bar{\Gamma}=-3$ or $ \gamma =2.$
    \end{enumerate}
    \item  $P_6=\Big(0,\sqrt{1+\frac{\bar{\Gamma}}{3 (\gamma -1)}}\Big),$ with eigenvalues $\left\{\frac{1}{2} \left(-\bar{\Gamma}-3\right),-\bar{\Gamma}-3 \gamma +3\right\}.$ This point verifies $\Omega_m=\frac{\bar{\Gamma }}{3-3 \gamma }$. It exist for $0\leq \gamma <1, 0\leq \bar{\Gamma}<3-3 \gamma$ or $1<\gamma \leq 2, 3-3 \gamma <\bar{\Gamma}\leq 0.$
    The point is
    \begin{enumerate}
        \item sink for $1<\gamma \leq 2$, $-3 (\gamma -1)<\bar{\Gamma
   }\leq0,$ 
        \item saddle for $0\leq \gamma <1$, $ 0\leq \bar{\Gamma}<-3 (\gamma-1).$
    \end{enumerate}
\end{enumerate}
Interaction 4~\eqref{int-4} shows that \( \bar{\Gamma} \) influences both the stability and the existence of \( P_6 \). FIG.~\ref{fig:int4-2D-Gamma>0} and FIG.~\ref{fig:int4-2D-Gamma<0} illustrate the phase plane of the guiding system for \( \bar{\Gamma} = 0.1 \) and \( \bar{\Gamma} = -0.1 \), respectively.
\FloatBarrier
\subsection{Interaction 5: $Q\left(\dot{\phi},\rho_{\phi }, \rho _{m}\right)=\frac{\Gamma}{H_0} H \left(\rho_{m}+\rho_\phi\right)$}
\label{int:5}
\begin{figure}[H]
    \centering
    \includegraphics[scale=0.4]{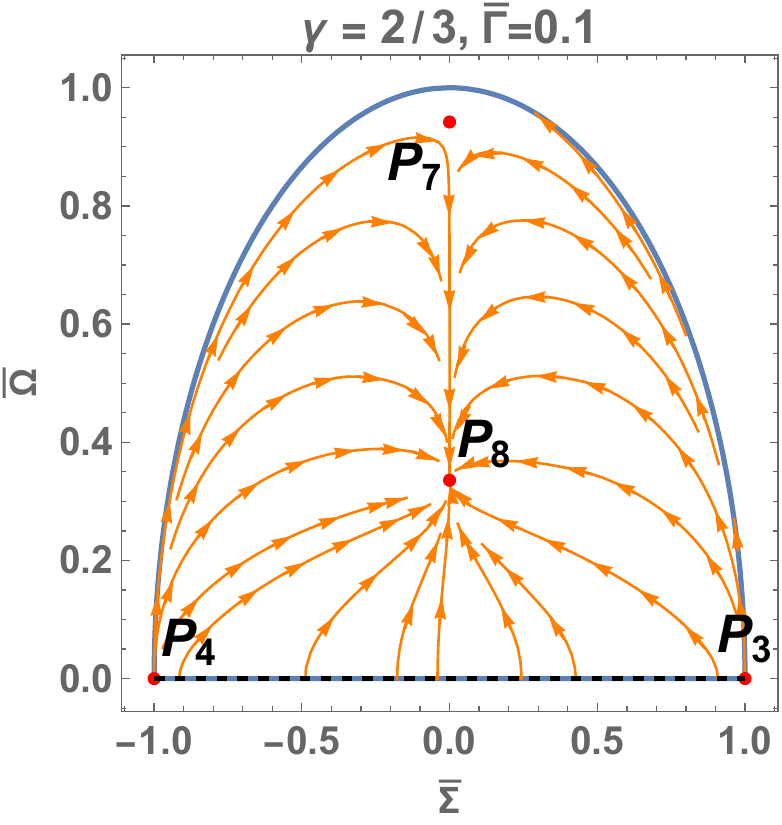}
    \includegraphics[scale=0.4]{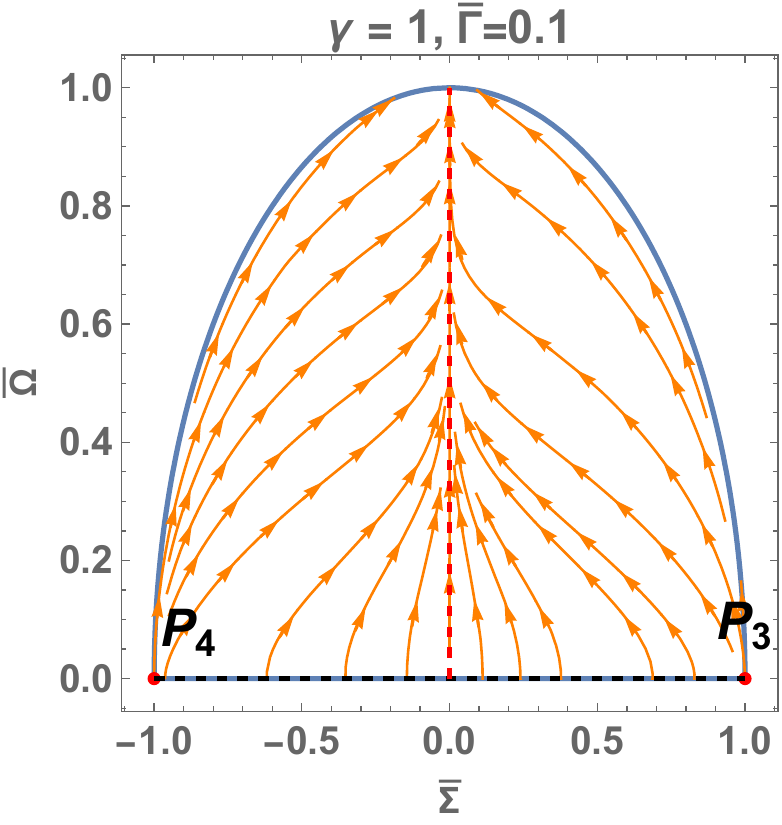}
    \includegraphics[scale=0.4]{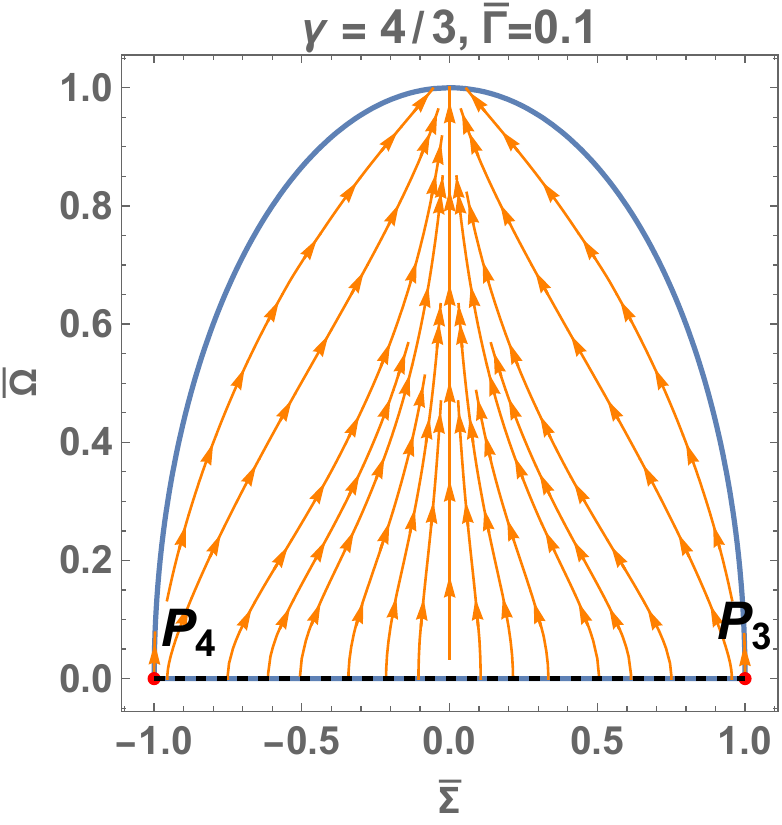}
    \includegraphics[scale=0.4]{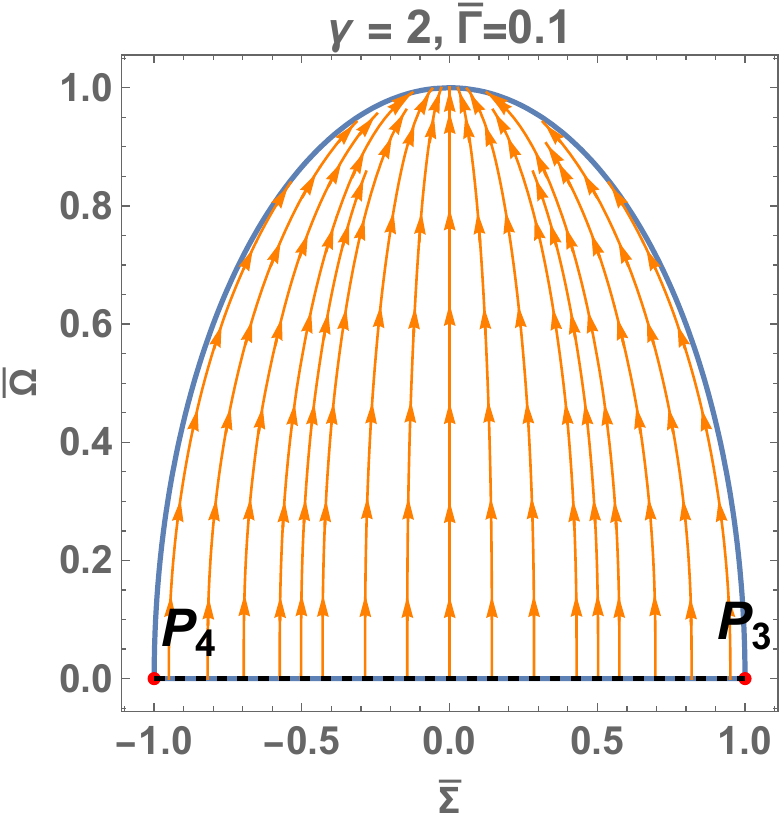}
    \caption{Phase-plane diagrams for system \eqref{int-5-2D-a}-\eqref{int-5-2D-b} setting $\bar{\Gamma}=0.1$ for different values of $\gamma$.}
     \label{fig:int5-2D-Gamma>0}
\end{figure}

\begin{figure}[H]
    \centering
    \includegraphics[scale=0.4]{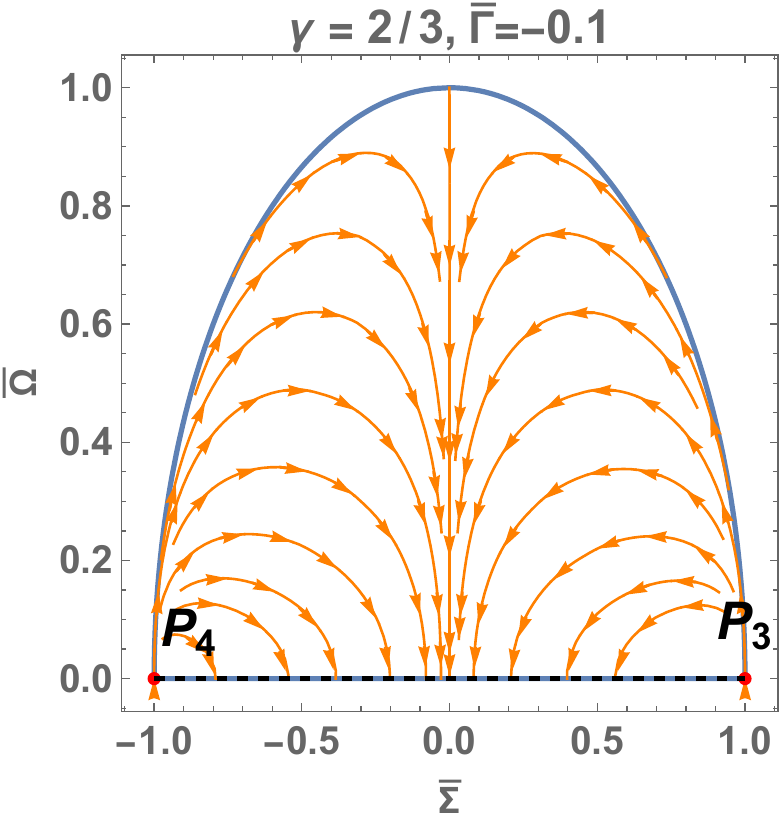}
    \includegraphics[scale=0.4]{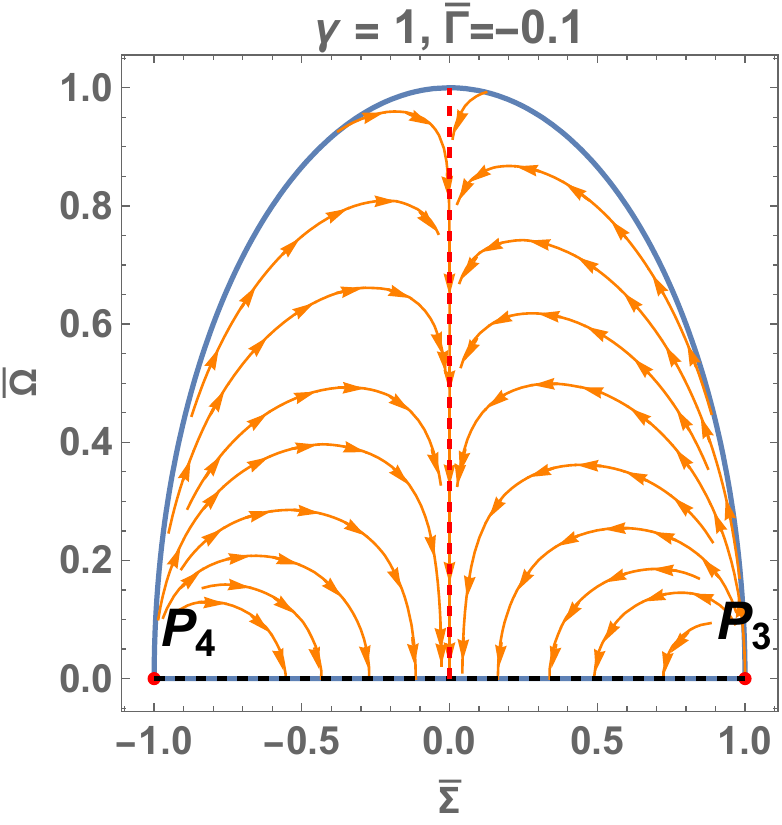}
    \includegraphics[scale=0.4]{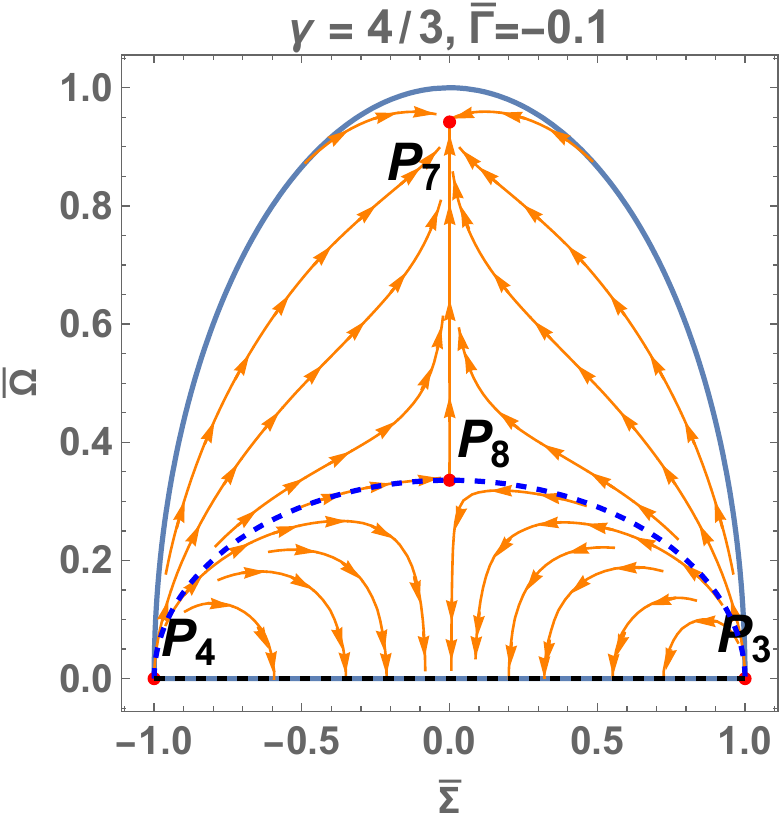}
    \includegraphics[scale=0.4]{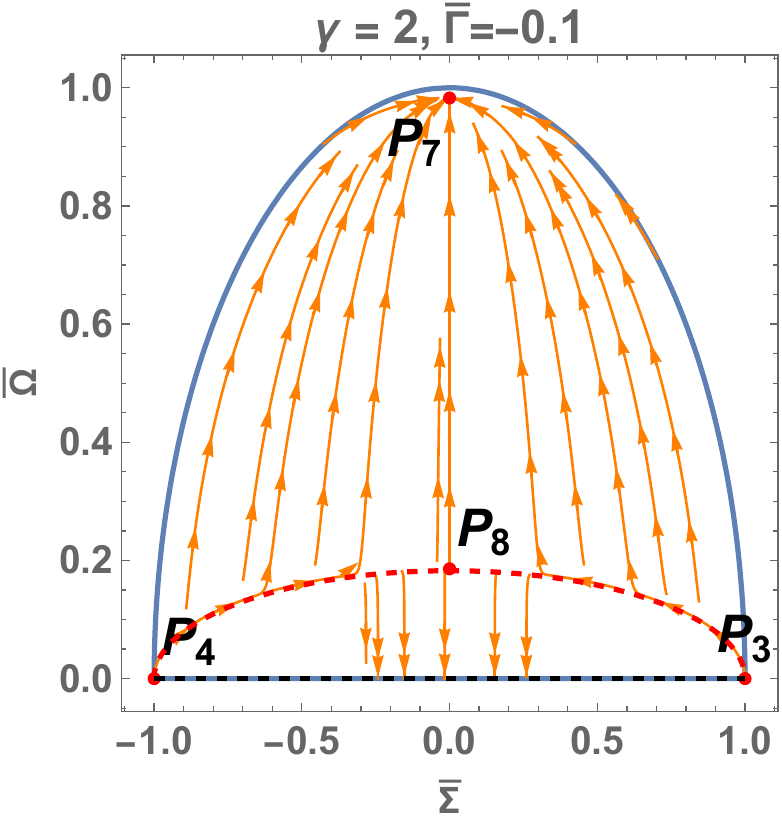}
    \caption{Phase-plane diagrams for system \eqref{int-5-2D-a}-\eqref{int-5-2D-b} setting $\bar{\Gamma}=-0.1$ for different values of $\gamma$. The red-dashed parabola is the unstable manifold of $P_8$, defined by $\bar{\Omega} = \frac{\sqrt{\bar{\Gamma} \left(\bar{\Sigma} ^2-1\right)}}{\sqrt{3}}$. The blue-dashed parabola is the unstable manifold of $P_8$, defined by $\bar{\Omega} = \frac{\sqrt{1-\sqrt{1+4 \bar{\Gamma}}} \sqrt{1-\bar{\Sigma} ^2}}{\sqrt{2}}$. }
    \label{fig:int5-2D-Gamma<0}
\end{figure}

Setting $\alpha=0,$ $\beta=1,$ $\delta=0$ in system \eqref{gen-redu-1}-\eqref{gen-redu-4}, expanding in a Taylor series around $H=0$ and using the rescaling $\Gamma =H_0 \bar{\Gamma}$ the truncated system is
\begin{small}
    \begin{align}
    \label{comp-int-5-a}&\dot{H}=-\frac{3}{2} \left(2 \Sigma ^2+\Omega ^2-\gamma  \left(\Sigma ^2+\Omega ^2-1\right)+\Omega ^2 \cos (2
   (\varphi -t \omega ))\right) H^2+O\left(H^3\right),\\
    &\dot{\Omega}=\left(\frac{3}{2} \Omega  \left(2 \Sigma ^2+\Omega
   ^2-\gamma  \left(\Sigma ^2+\Omega ^2-1\right)+\left(\Omega ^2-1\right) \cos (2 (\varphi -t \omega
   ))-1\right)-\frac{\left(\Sigma ^2-1\right) \bar{\Gamma}}{2 \Omega }\right) H\\ \nonumber
   &-\frac{\left(\omega
   ^2-1\right)^3 \Omega ^3 \cos (\varphi -t \omega ) \sin ^3(\varphi -t \omega ) H^2}{\omega
   ^3}+O\left(H^3\right),\\
    &\dot{\Sigma}=\frac{3}{2} \Sigma  \left(2 \Sigma ^2+\Omega ^2-\gamma  \left(\Sigma ^2+\Omega
   ^2-1\right)+\Omega ^2 \cos (2 (\varphi -t \omega ))-2\right) H+O\left(H^3\right),\\
   \label{comp-int-5-d} &\dot{\varphi}=\frac{\left(6 \Omega ^2 \cos
   ^2(\varphi -t \omega )+\left(\Sigma ^2-1\right) \bar{\Gamma}\right) \tan (\varphi -t \omega ) H}{2 \Omega
   ^2}+\frac{\left(\omega ^2-1\right)^3 \Omega ^2 \sin ^4(\varphi -t \omega ) H^2}{\omega
   ^3}+O\left(H^3\right).
\end{align}
\end{small}
The averaged system is
\begin{align}
   &\dot{H}=\frac{3}{2} H^2 \left(\gamma  \left(\bar{\Sigma}^2+\bar{\Omega}^2-1\right)-2 \bar{\Sigma}^2-\bar{\Omega}
   ^2\right),  \label{1-prom-int5}\\
  &\dot{\bar{\Omega}}=-\frac{H \left(\left(\bar{\Sigma}^2-1\right) \bar{\Gamma}+3 \bar{\Omega}^2 \left(\gamma  \left(\bar{\Sigma}^2+\bar{\Omega}^2-1\right)-2 \bar{\Sigma}^2-\bar{\Omega}^2+1\right)\right)}{2 \bar{\Omega}},  \label{2-prom-int5} \\
 &\dot{\bar{\Sigma}}=\frac{3}{2} H \bar{\Sigma} 
   \left(-\gamma  \left(\bar{\Sigma}^2+\bar{\Omega}^2-1\right)+2 \bar{\Sigma}^2+\bar{\Omega}^2-2\right),  \label{3-prom-int5}\\
 &\dot{\bar{\varphi}}=0.    \label{4-prom-int5}
\end{align}
And the guiding system with the new time derivative defined as $f'= \frac{1}{H}\dot{f}$ is
\begin{align}
\label{int-5-2D-a}
    &\bar{\Sigma}'=\frac{3}{2}  \bar{\Sigma}
   \left(-\gamma  \left(\bar{\Sigma}^2+\bar{\Omega}^2-1\right)+2 \bar{\Sigma}^2+\bar{\Omega}^2-2\right),\\
   \label{int-5-2D-b}&\bar{\Omega}'=-\frac{ \left(\left(\bar{\Sigma}^2-1\right) \bar{\Gamma}+3 \bar{\Omega}^2 \left(\gamma  \left(\bar{\Sigma}
   ^2+\bar{\Omega}^2-1\right)-2 \bar{\Sigma}^2-\bar{\Omega}^2+1\right)\right)}{2 \bar{\Omega}}.
\end{align}
As before, the phase-space is the compact set
\begin{equation}
    \left\{(\bar{\Sigma},\bar{\Omega})\in \mathbb{R}^2| \;  \bar{\Omega}^2 + \bar{\Sigma}^2 \leq 1, \bar{\Omega}\geq 0, -1\leq \bar{\Sigma} \leq 1 \right\}.
\end{equation}
System \eqref{int-5-2D-a}-\eqref{int-5-2D-b} also has a singularity in the line $\bar{\Omega}=0$ \cite{Paliathanasis:2024jxo,Leon:2025sfd,Papagiannopoulos:2025zku}. The equilibrium points of the system \eqref{int-5-2D-a}–\eqref{int-5-2D-b} in the coordinates $(\bar{\Sigma}, \bar{\Omega})$ are
\begin{enumerate}
    \item $P_{3,4}=(\pm 1,0),$ with eigenvalues $\{0,0\}.$ They belong to the singular line $\bar{\Omega}=0$ and are non-hyperbolic.
    \item $P_7=\left(0,\frac{\sqrt{\sqrt{3+\frac{12 \bar{\Gamma}}{\gamma
   -1}+9}}}{\sqrt{6}}\right),$ with eigenvalues 
   $$\left\{\frac{1}{2} \left(3 \gamma -\sqrt{3}
   \sqrt{\gamma -1} \sqrt{3 \gamma +4
   \bar{\Gamma}-3}-9\right),-2 \sqrt{3} \sqrt{\gamma -1}
   \sqrt{3 \gamma +4 \bar{\Gamma}-3}\right\}.$$
   It has $\Omega_m=\frac{1}{6} \left(3-\frac{\sqrt{9 \gamma +12 \bar{\Gamma}-9}}{\sqrt{\gamma -1}}\right)$ and it exists exists for $0\leq \gamma <1, 0\leq \bar{\Gamma}\leq
   \frac{1}{4} (3-3 \gamma )$ or $1<\gamma \leq 2, \frac{1}{4} (3-3
   \gamma )\leq \bar{\Gamma}\leq 0$. The point is
   \begin{enumerate}
       \item sink for $1<\gamma \leq 2, \frac{1}{4} (3-3 \gamma
   )<\bar{\Gamma}\leq 0,$
   \item saddle for $0\leq \gamma <1, 0\leq \bar{\Gamma}<\frac{1}{4} (3-3
   \gamma ),$
   \item non-hyperbolic for $\gamma\neq 1, \bar{\Gamma}=\frac{1}{4} (3-3 \gamma
   )$.
   \end{enumerate}
    \item $P_8=\left(0,\frac{\sqrt{3-\sqrt{\frac{12 \bar{\Gamma}}{\gamma
   -1}+9}}}{\sqrt{6}}\right)$ with eigenvalues $$\left\{\frac{1}{2} \left(3 \gamma +\sqrt{3}
   \sqrt{\gamma -1} \sqrt{3 \gamma +4
   \bar{\Gamma}-3}-9\right),2 \sqrt{3} \sqrt{\gamma -1}
   \sqrt{3 \gamma +4 \bar{\Gamma}-3}\right\}.$$ It has $\Omega_m=\frac{1}{6} \left(\frac{\sqrt{9 \gamma +12 \bar{\Gamma}-9}}{\sqrt{\gamma -1}}+3\right)$, it exists for $0\leq \gamma <1, 0\leq \bar{\Gamma}\leq
   \frac{1}{4} (3-3 \gamma )$ or $1<\gamma \leq 2, \frac{1}{4} (3-3
   \gamma )\leq \bar{\Gamma}\leq 0$. The point is a
   \begin{enumerate}
       \item sink for $0\leq \gamma <1, 0\leq \bar{\Gamma}<\frac{1}{4} (3-3
   \gamma )$,
        \item saddle for $1<\gamma <2, \frac{1}{4} (3-3 \gamma
   )<\bar{\Gamma}\leq 0$ or $\gamma =2,
   -\frac{3}{4}<\bar{\Gamma}<0$,
   \item non-hyperbolic for $\gamma\neq 1, \bar{\Gamma}=\frac{1}{4} (3-3 \gamma
   )$ or $\gamma =2, \bar{\Gamma}=0$.
   \end{enumerate}
\end{enumerate}
Considering Interaction 5~\eqref{int-5}, it is clear that the parameter \( \bar{\Gamma} \) influences the stability and the existence conditions of points \( P_{7,8} \). Moreover, in the full system~\eqref{comp-int-5-a}--\eqref{comp-int-5-d}, the presence of the tangent function modifies the behaviour of the numerical solutions. FIG.~\ref{fig:int5-2D-Gamma>0} shows the phase plane of the guiding system~\eqref{int-3-2D-a}--\eqref{int-3-2D-b} for different values of \( \gamma \) and setting \( \bar{\Gamma} = 0.1 \). On the other hand, FIG.~\ref{fig:int5-2D-Gamma<0} shows the phase plane of the guiding system~\eqref{int-5-2D-a}--\eqref{int-5-2D-b} for different values of \( \gamma \) with \( \bar{\Gamma} = -0.1 \). As shown in Figure~\ref{fig:int5-2D-Gamma<0}, for \( \bar{\Gamma} = -0.1 \), the set \( \bar{\Omega} = 0 \) is an attracting set; however, this set is a singularity of the guiding system~\eqref{int-5-2D-a}--\eqref{int-5-2D-b}. Therefore, this interaction does not define a physically acceptable behaviour for the value \( \bar{\Gamma} = -0.1 \).
\subsection{Interaction 6: $Q\left(\dot{\phi},\rho_{\phi }, \rho _{m}\right)=\frac{\Gamma  H_0 \rho_{\phi }  \dot{\phi}^2}{H}$}
\begin{figure}[H]
    \centering
    \includegraphics[scale=0.4]{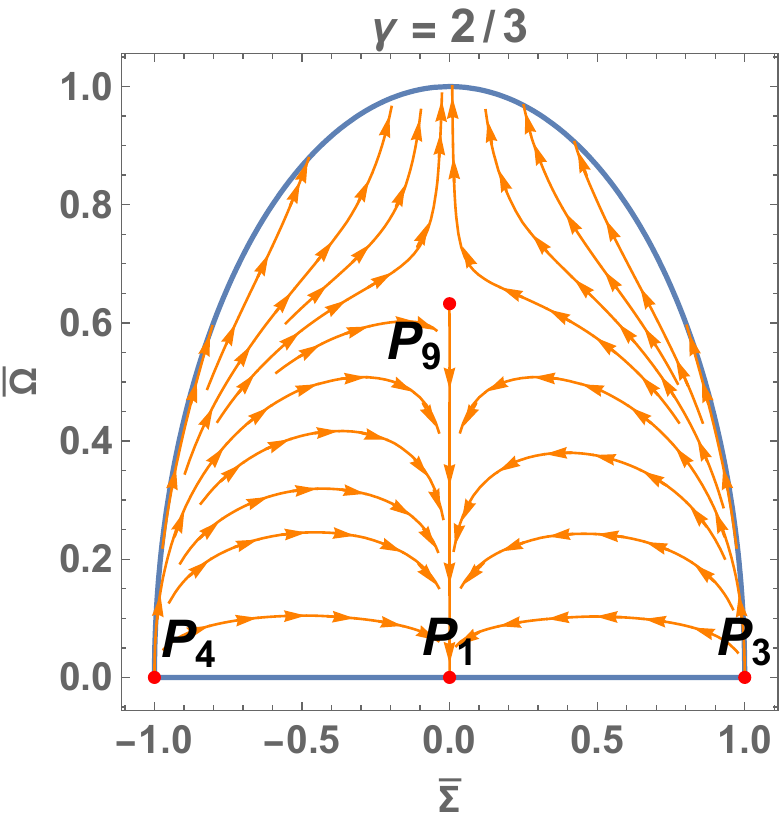}
    \includegraphics[scale=0.4]{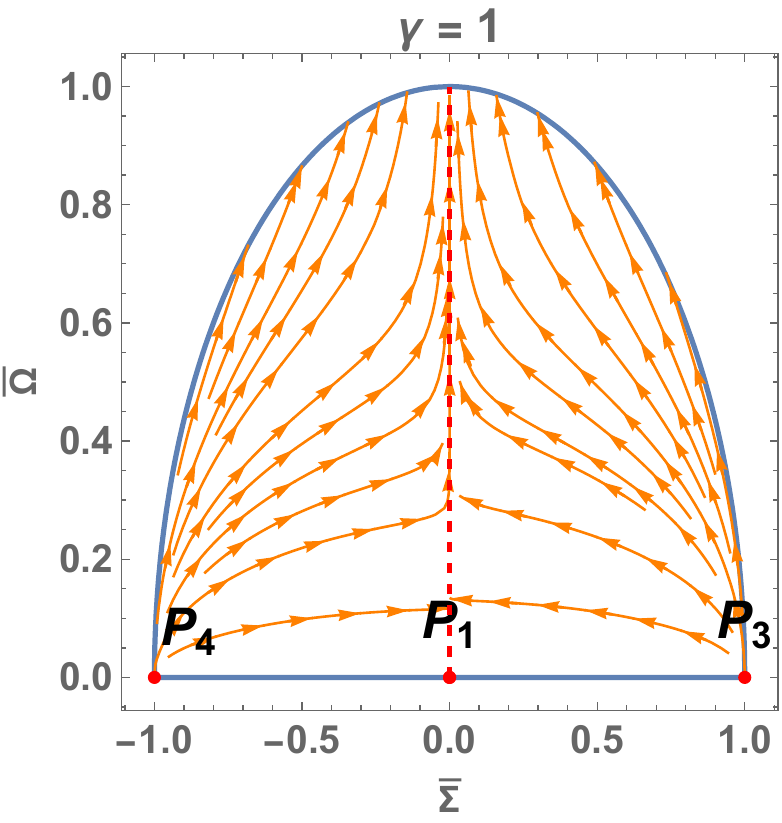}
    \includegraphics[scale=0.4]{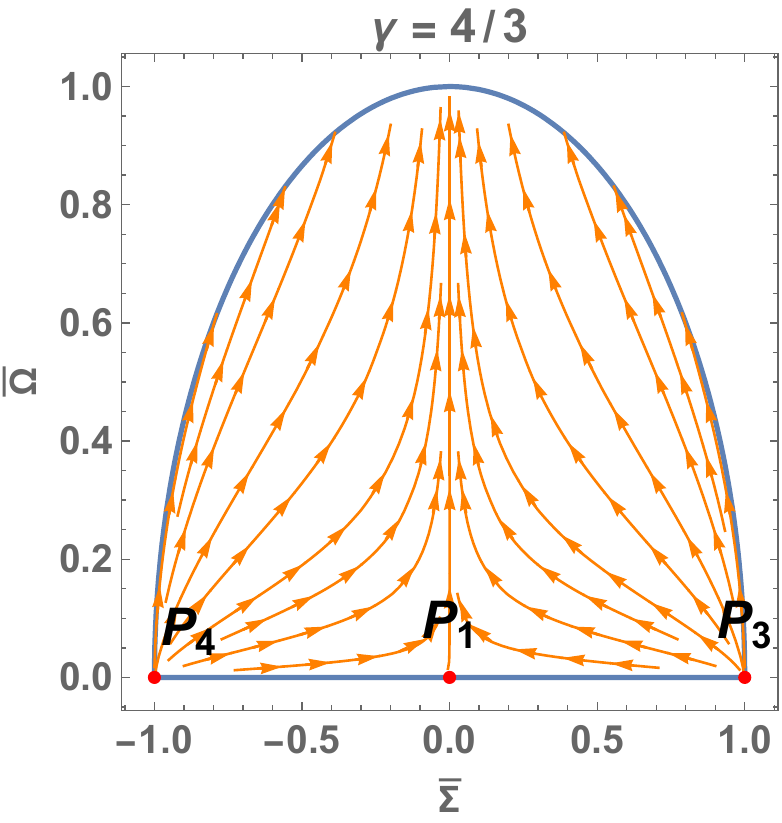}
    \includegraphics[scale=0.4]{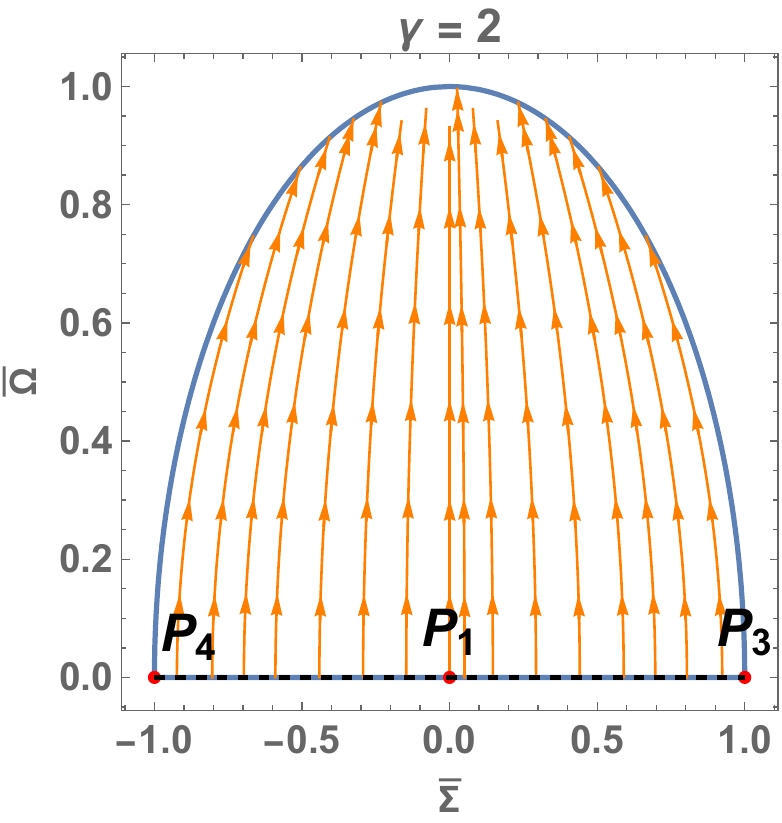}
    \caption{Phase-plane diagrams for system \eqref{int-6-2D-a}-\eqref{int-6-2D-b} setting $m=0.5$ for different values of $\gamma$.}
     \label{fig:int6-2D-m>0}
\end{figure}

\begin{figure}[H]
    \centering
    \includegraphics[scale=0.4]{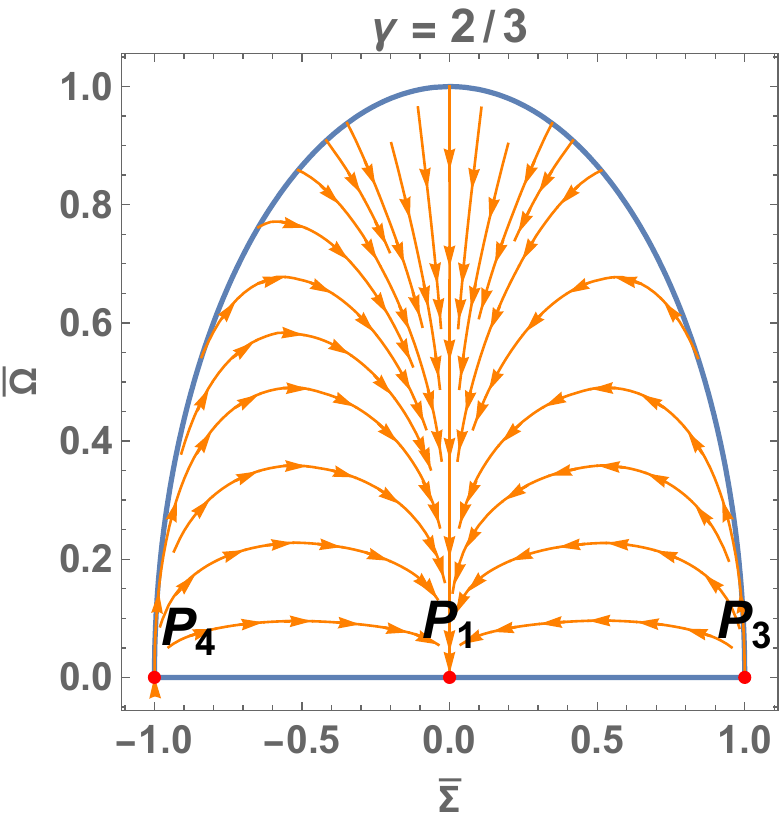}
    \includegraphics[scale=0.4]{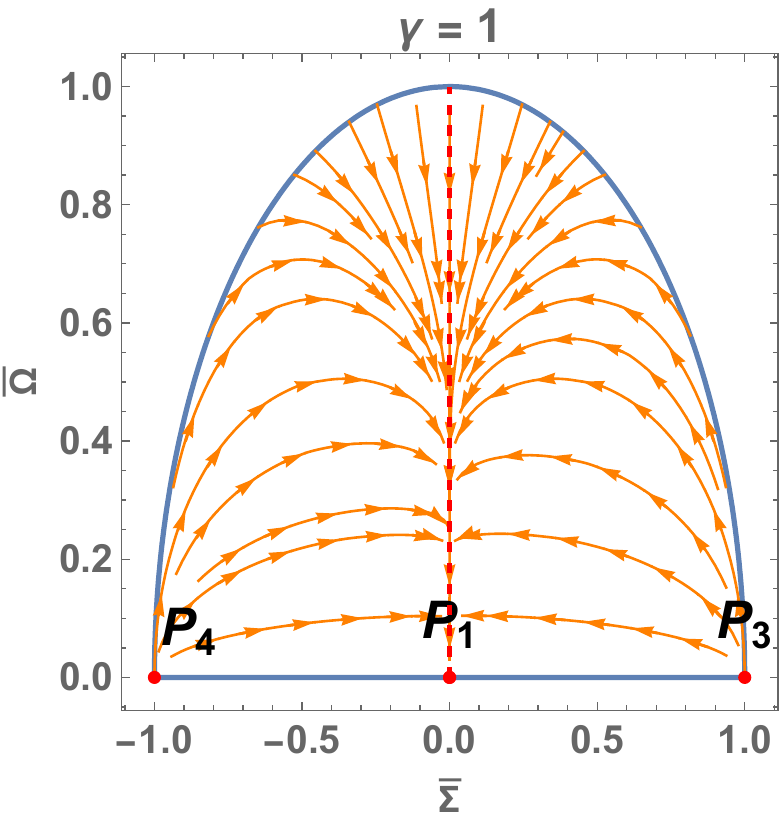}
    \includegraphics[scale=0.4]{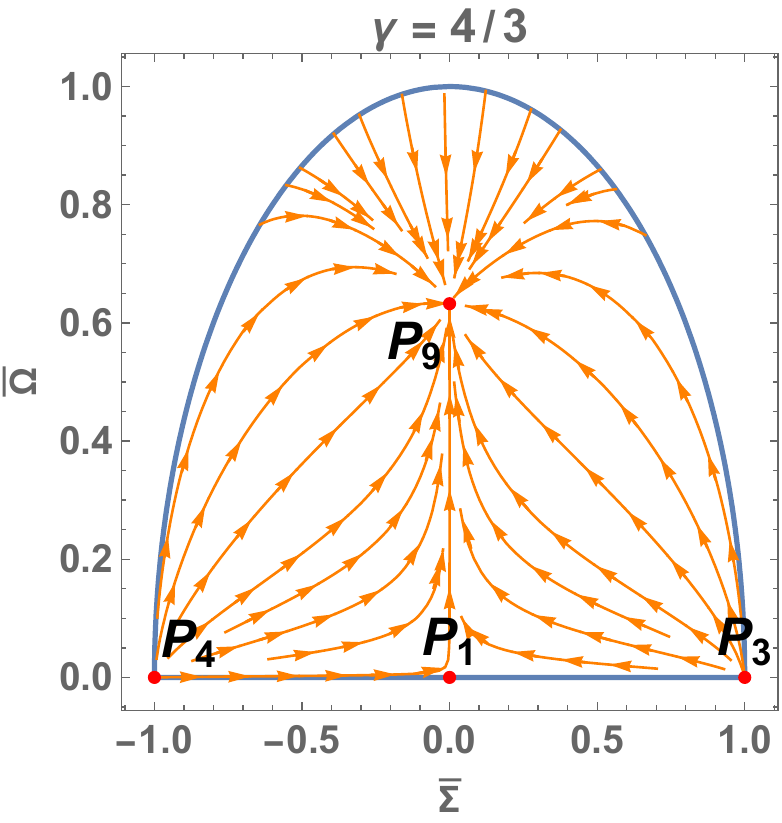}
    \includegraphics[scale=0.4]{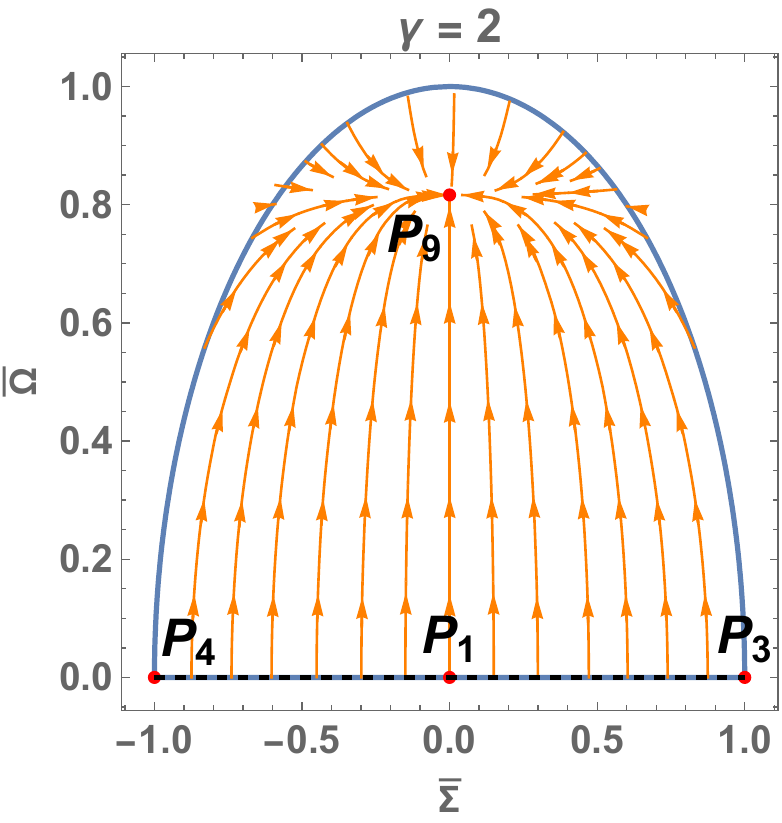}
    \caption{Phase-plane diagrams for system \eqref{int-6-2D-a}-\eqref{int-6-2D-b} setting $m=-0.5$ for different values of $\gamma$.}
 \label{fig:int6-2D-m<0}\end{figure}

Setting $\alpha=0,$ $\beta=0,$ $\delta=2$ in system \eqref{gen-redu-1}-\eqref{gen-redu-4}, expanding in a Taylor series around $H=0$ and using the rescaling $\Gamma =\frac{m}{H_0} $ the truncated system is
\begin{small}
    \begin{align}
    \label{comp-int-6-a}&\dot{H}=-\frac{3}{2} \left(2 \Sigma ^2+\Omega ^2-\gamma  \left(\Sigma
   ^2+\Omega ^2-1\right)+\Omega ^2 \cos (2 (\varphi -t \omega ))\right)
   H^2+O\left(H^3\right),\\
    &\dot{\Omega}=\frac{3}{2} \Omega  \left(-\gamma  \Sigma ^2+2
   \Sigma ^2+(m-\gamma +1) \Omega ^2+\gamma +\left((m+1) \Omega ^2-1\right)
   \cos (2 (\varphi -t \omega ))-1\right) H\\ \nonumber &-\frac{\left(\omega ^2-1\right)^3
   \Omega ^3 \cos (\varphi -t \omega ) \sin ^3(\varphi -t \omega ) H^2}{\omega
   ^3}+O\left(H^3\right),\\
    &\dot{\Sigma}=\frac{3}{2} \Sigma  \left(2 \Sigma ^2+\Omega
   ^2-\gamma  \left(\Sigma ^2+\Omega ^2-1\right)+\Omega ^2 \cos (2 (\varphi -t
   \omega ))-2\right) H+O\left(H^3\right),\\
   \label{comp-int-6-d} &\dot{\varphi}=-\frac{3}{2} \left(\left(m \Omega
   ^2-1\right) \sin (2 (\varphi -t \omega ))\right) H+\frac{\left(\omega
   ^2-1\right)^3 \Omega ^2 \sin ^4(\varphi -t \omega ) H^2}{\omega
   ^3}+O\left(H^3\right).
\end{align}
\end{small}
The averaged system is
\begin{align}
   &\dot{H}=\frac{3}{2} H^2 \left(\gamma  \left(\bar{\Sigma} ^2+\bar{\Omega} ^2-1\right)-2
   \bar{\Sigma} ^2-\bar{\Omega} ^2\right),  \label{1-prom-int6}\\
  &\dot{\bar{\Omega}}=\frac{3}{2} H \bar{\Omega}  \left(-\gamma  \bar{\Sigma}
   ^2+\gamma +\bar{\Omega} ^2 (-\gamma +m+1)+2 \bar{\Sigma} ^2-1\right),  \label{2-prom-int6} \\
 &\dot{\bar{\Sigma}}=\frac{3}{2} H
   \bar{\Sigma}  \left(-\gamma  \left(\bar{\Sigma} ^2+\bar{\Omega} ^2-1\right)+2 \bar{\Sigma}
   ^2+\bar{\Omega} ^2-2\right),  \label{3-prom-int6}\\
 &\dot{\bar{\varphi}}=0.    \label{4-prom-int6}
\end{align}
And the guiding system with the new time derivative defined as $f'= \frac{1}{H}\dot{f}$ is
\begin{align}
\label{int-6-2D-a}
    &\bar{\Sigma}'=\frac{3}{2} \bar{\Sigma}  \left(-\gamma  \left(\bar{\Sigma} ^2+\bar{\Omega}
   ^2-1\right)+2 \bar{\Sigma} ^2+\bar{\Omega} ^2-2\right),\\
   \label{int-6-2D-b}&\bar{\Omega}'=\frac{3}{2} \bar{\Omega} 
   \left(-\gamma  \bar{\Sigma} ^2+\gamma +\bar{\Omega} ^2 (-\gamma +m+1)+2 \bar{\Sigma}
   ^2-1\right).
\end{align}
As before, the phase-space is the compact set
\begin{equation}
    \left\{(\bar{\Sigma},\bar{\Omega})\in \mathbb{R}^2| \;  \bar{\Omega}^2 + \bar{\Sigma}^2 \leq 1, \bar{\Omega}\geq 0, -1\leq \bar{\Sigma} \leq 1 \right\}.
\end{equation}
The equilibrium points of system \eqref{int-6-2D-a}-\eqref{int-6-2D-b} in the coordinates $(\bar{\Sigma},\bar{\Omega})$ are
\begin{enumerate}
    \item $P_1=(0, 0)$ with eigenvalues $\lbrace \frac{3 (\gamma -2)}{2},\frac{3 (\gamma -1)}{2}\rbrace$. It always exists and it verifies $\Omega_m=1$. It describes a matter-dominated flat FLRW solution. The point is 
    \begin{enumerate}
        \item sink for $0\leq \gamma <1$, 
        \item saddle for $1<\gamma \leq 2$,
        \item non-hyperbolic for $\gamma=1,2$.
        \end{enumerate}
    \item $P_{3,4}=(\pm 1,0)$ with eigenvalues $\lbrace \frac{3}{2},-3 (\gamma -2) \rbrace$. They always exist and both represent an anisotropic Bianchi I vacuum solution. They are
    \begin{enumerate}
        \item sources for $0\leq \gamma <2$, 
        \item non-hyperbolic for $\gamma=2$. 
    \end{enumerate}
    \item $P_9=\left(0,\sqrt{\frac{1-\gamma}{1+m-\gamma}}\right),$ with eigenvalues $\left\{3-3 \gamma ,\frac{3 (\gamma +(\gamma -2) m-1)}{2 (-\gamma
   +m+1)}\right\}$. It exist for 
    \begin{enumerate}
        \item $m<0, 1\leq \gamma \leq 2$ or
        \item $ m=0, 0\leq \gamma <1$ or
        \item $m=0,  1<\gamma
   \leq 2$ or
        \item $m>0, 0\leq \gamma \leq 1.$
        Additionally, we verify that $\Omega_m=\frac{m}{1+m-\gamma}.$
    \end{enumerate}
     The point is a
        \begin{enumerate}
            \item sink for $1<\gamma\leq 2,$ $m>0$ or $1<\gamma\leq 2,$ $m<0$,
            \item saddle for $0\leq \gamma<1, m\geq 0,$
            \item non-hyperbolic for $\gamma=1, m\neq 0.$
        \end{enumerate}
\end{enumerate}
Considering Interaction 6~\eqref{int-6}, it is clear that the parameter \( m \) influences the stability and the existence conditions of \( P_{9} \). FIG.~\ref{fig:int6-2D-m>0} shows the phase plane of the guiding system~\eqref{int-6-2D-a}--\eqref{int-6-2D-b} for different values of \( \gamma \) and setting \( m = 0.1 \). On the other hand, FIG.~\ref{fig:int6-2D-m<0} shows the phase plane of the guiding system~\eqref{int-6-2D-a}--\eqref{int-6-2D-b} for different values of \( \gamma \) with \( m = -0.1 \).

\subsection{Interaction 7: $Q\left(\dot{\phi},\rho_{\phi }, \rho _{m}\right)=\frac{\Gamma  H_0 \rho_{m } \dot{\phi}^2}{H}$}
\begin{figure}[H]
    \centering
    \includegraphics[scale=0.4]{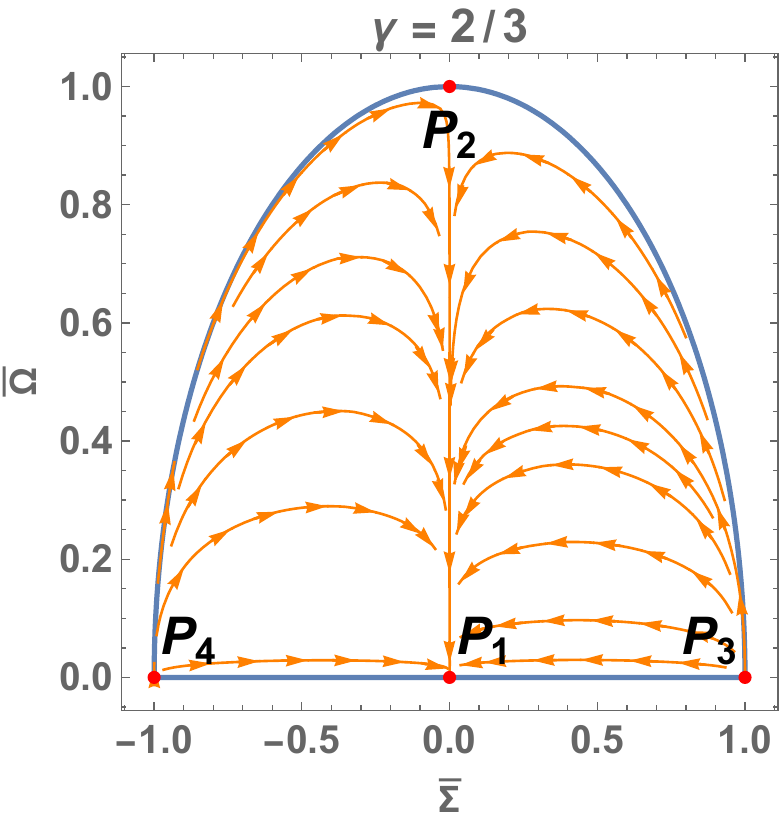}
    \includegraphics[scale=0.4]{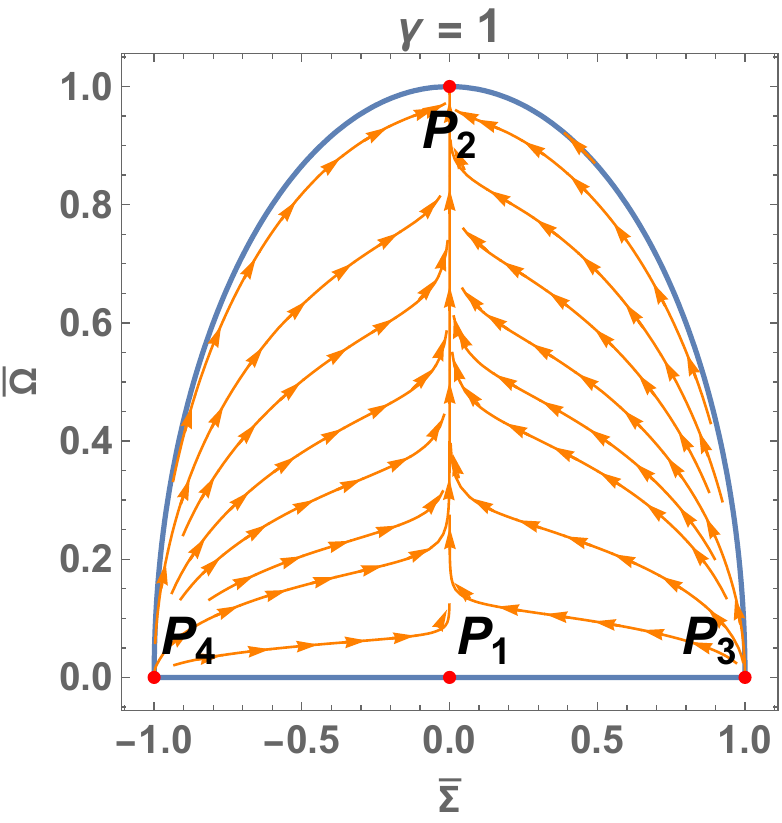}
    \includegraphics[scale=0.4]{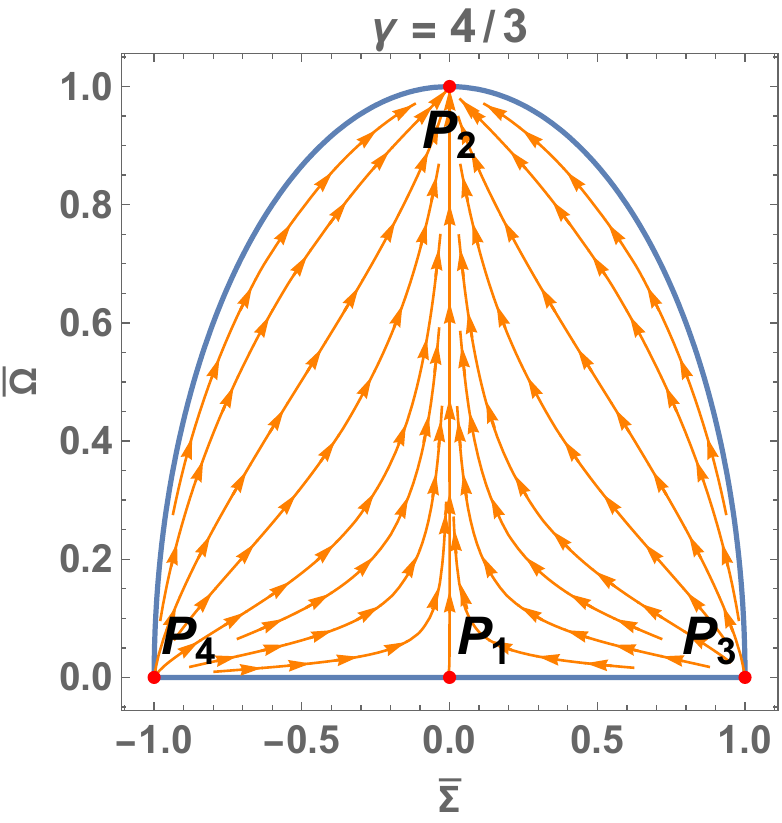}
    \includegraphics[scale=0.4]{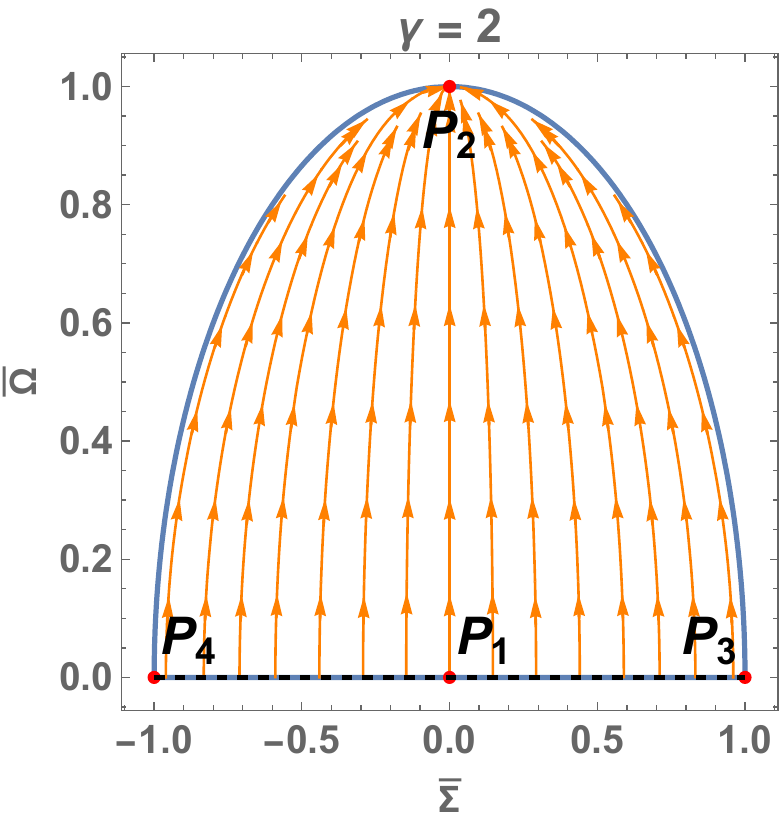}
    \caption{Phase-plane diagrams for system \eqref{int-6-2D-a}-\eqref{int-6-2D-b} setting $m=0.1$ for different values of $\gamma$.}
     \label{fig:int7-2D-m>0}
\end{figure}

\begin{figure}[H]
    \centering
    \includegraphics[scale=0.4]{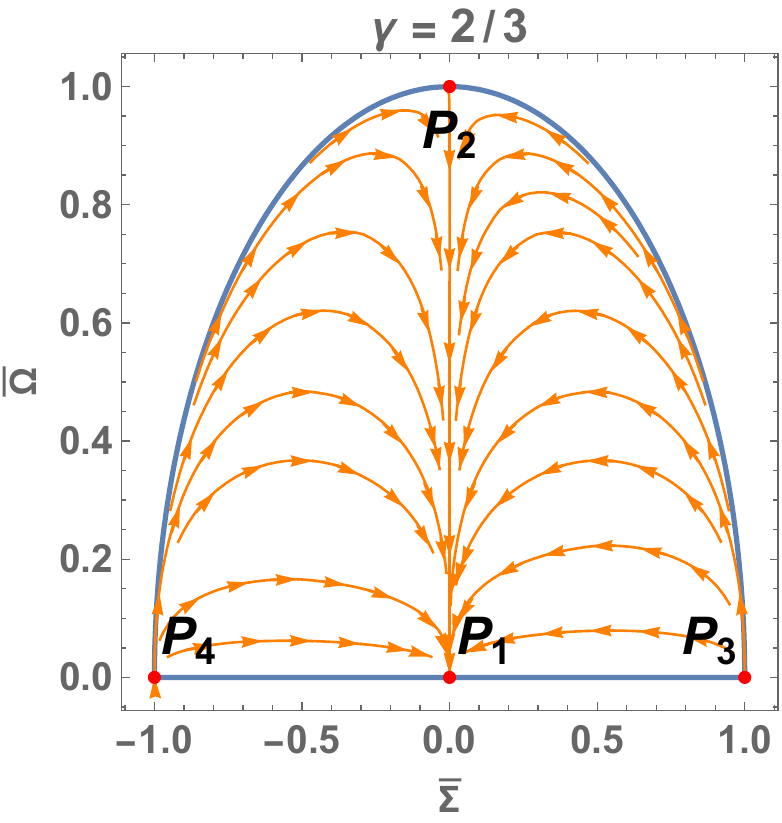}
    \includegraphics[scale=0.4]{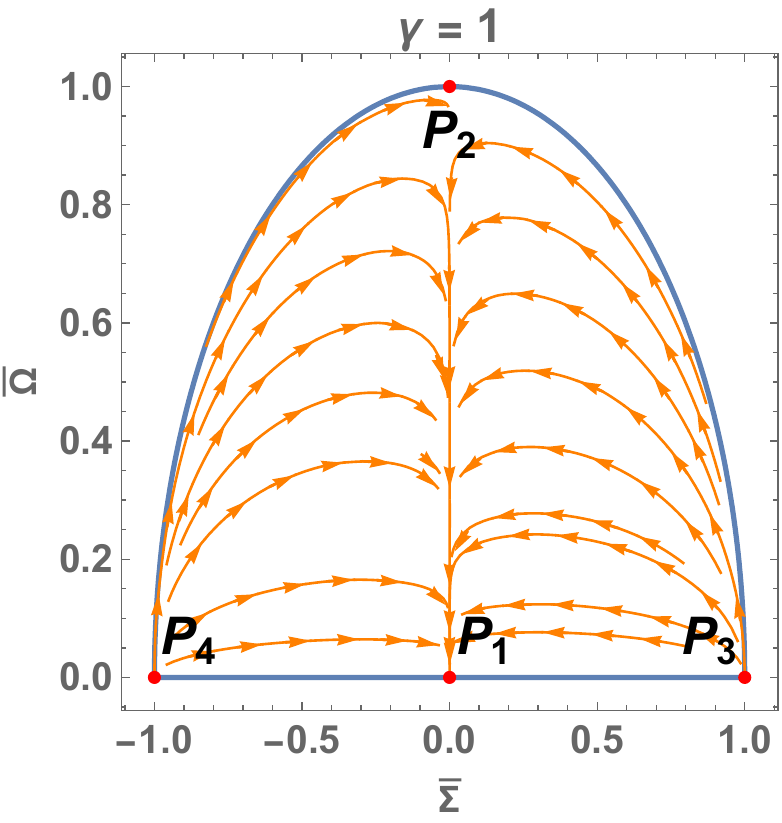}
    \includegraphics[scale=0.4]{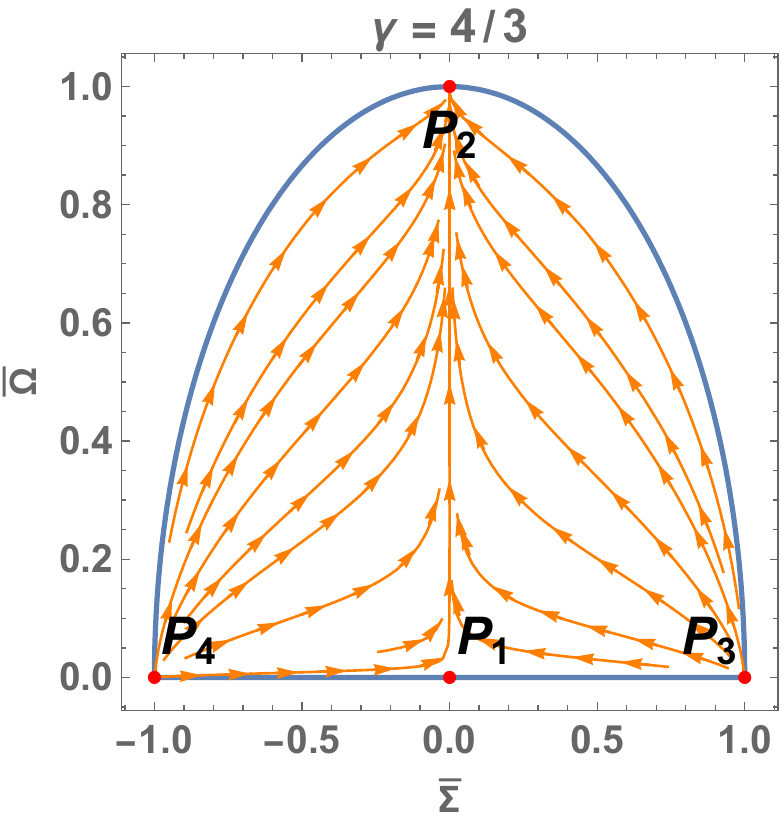}
    \includegraphics[scale=0.4]{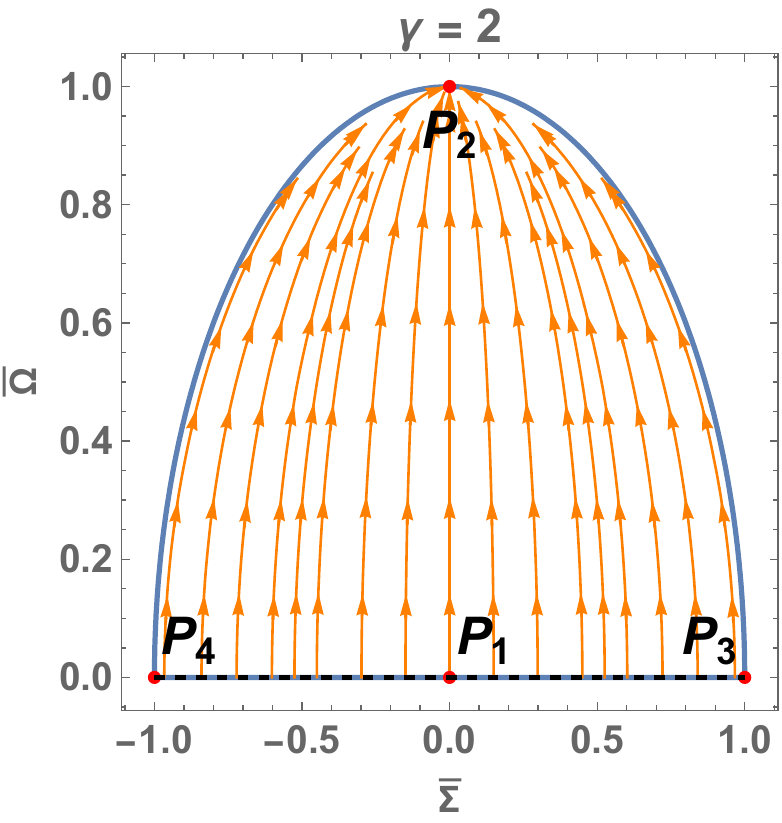}
    \caption{Phase-plane diagrams for system \eqref{int-6-2D-a}-\eqref{int-6-2D-b} setting $m=-0.1$ for different values of $\gamma$.}
 \label{fig:int7-2D-m<0}\end{figure}

\begin{figure}[H]
    \centering
    \includegraphics[scale=0.4]{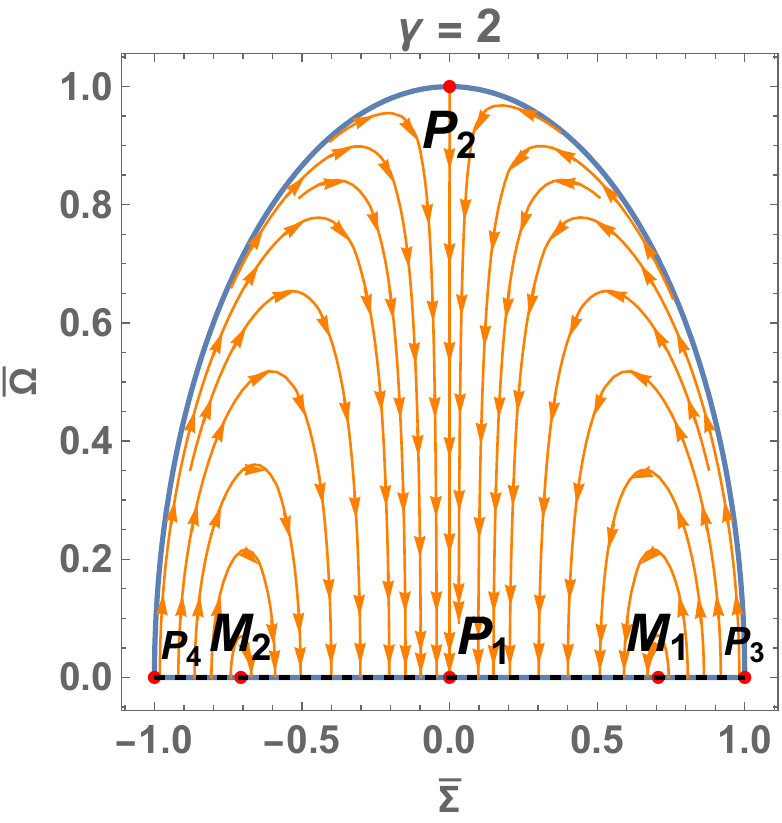}
    \includegraphics[scale=0.4]{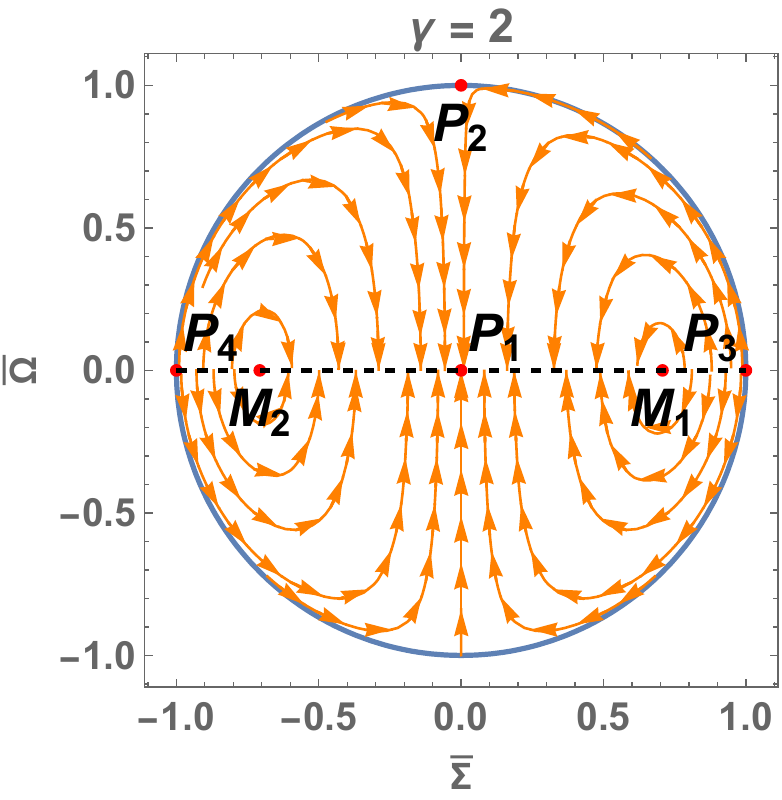}
    \caption{Phase-space diagrams for the special case $\gamma=2, m\leq -1$ setting $m=-2.$}
    \label{fig:special}
\end{figure}

Setting $\alpha=1,$ $\beta=0,$ $\delta=2$ in system \eqref{gen-redu-1}-\eqref{gen-redu-4}, expanding in a Taylor series around $H=0$ and using the rescaling $\Gamma =\frac{m}{H_0} $ the truncated system is
\begin{small}
    \begin{align}
    \label{comp-int-7-a}&\dot{H}=-\frac{3}{2} \left(2 \Sigma ^2+\Omega
   ^2-\gamma  \left(\Sigma ^2+\Omega
   ^2-1\right)+\Omega ^2 \cos (2 (\varphi -t \omega
   ))\right) H^2+O\left(H^3\right),\\
    &\dot{\Omega}=\frac{3}{2}
   \Omega  \left(-m \Sigma ^2-\gamma  \Sigma
   ^2+2 \Sigma ^2-(m+\gamma -1) \Omega
   ^2+m+\gamma +\left(\Omega ^2-m \left(\Sigma
   ^2+\Omega ^2-1\right)-1\right) \cos (2 (\varphi
   -t \omega ))-1\right) H\\ \nonumber &-\frac{\left(\omega
   ^2-1\right)^3 \Omega ^3 \cos (\varphi -t \omega
   ) \sin ^3(\varphi -t \omega ) H^2}{\omega
   ^3}+O\left(H^3\right),\\
    &\dot{\Sigma}=\frac{3}{2} \Sigma 
   \left(2 \Sigma ^2+\Omega ^2-\gamma 
   \left(\Sigma ^2+\Omega ^2-1\right)+\Omega ^2
   \cos (2 (\varphi -t \omega ))-2\right)
   H+O\left(H^3\right),\\
   \label{comp-int-7-d} &\dot{\varphi}=\frac{3}{2} \left(m
   \left(\Sigma ^2+\Omega ^2-1\right)+1\right)
   \sin (2 (\varphi -t \omega ))
   H+\frac{\left(\omega ^2-1\right)^3 \Omega ^2
   \sin ^4(\varphi -t \omega ) H^2}{\omega
   ^3}+O\left(H^3\right).
\end{align}
\end{small}
The averaged system is
\begin{align}
   &\dot{H}=\frac{3}{2} H^2 \left(\gamma 
   \left(\bar{\Sigma} ^2+\bar{\Omega} ^2-1\right)-2 \bar{\Sigma}
   ^2-\bar{\Omega} ^2\right),  \label{1-prom-int7}\\
  &\dot{\bar{\Omega}}=\frac{3}{2} H \bar{\Omega} 
   \left(-\gamma  \bar{\Sigma} ^2+\gamma -\bar{\Omega} ^2
   (\gamma +m-1)-m \bar{\Sigma} ^2+m+2 \bar{\Sigma}
   ^2-1\right),  \label{2-prom-int7} \\
 &\dot{\bar{\Sigma}}=\frac{3}{2} H \bar{\Sigma} 
   \left(-\gamma  \left(\bar{\Sigma} ^2+\bar{\Omega}
   ^2-1\right)+2 \bar{\Sigma} ^2+\bar{\Omega}
   ^2-2\right),  \label{3-prom-int7}\\
 &\dot{\bar{\varphi}}=0.    \label{4-prom-int7}
\end{align}
And the guiding system with the new time derivative defined as $f'= \frac{1}{H}\dot{f}$ is
\begin{align}
\label{int-7-2D-a}
    &\bar{\Sigma}'=\frac{3}{2} \bar{\Sigma}  \left(-\gamma 
   \left(\bar{\Sigma} ^2+\bar{\Omega} ^2-1\right)+2 \bar{\Sigma}
   ^2+\bar{\Omega} ^2-2\right),\\
   \label{int-7-2D-b}&\bar{\Omega}'=\frac{3}{2} \bar{\Omega} 
   \left(-\gamma  \bar{\Sigma} ^2+\gamma -\bar{\Omega} ^2
   (\gamma +m-1)-m \bar{\Sigma} ^2+m+2 \Sigma
   ^2-1\right).
\end{align}
As before, the phase-space is the compact set
\begin{equation}
    \left\{(\bar{\Sigma},\bar{\Omega})\in \mathbb{R}^2| \;  \bar{\Omega}^2 + \bar{\Sigma}^2 \leq 1, \bar{\Omega}\geq 0, -1\leq \bar{\Sigma} \leq 1 \right\}.
\end{equation}
The equilibrium points of system \eqref{int-7-2D-a}-\eqref{int-7-2D-b} in the coordinates $(\bar{\Sigma},\bar{\Omega})$ are
\begin{enumerate}
    \item $P_1=(0, 0)$ with eigenvalues $\left\{\frac{3 (\gamma -2)}{2},\frac{3}{2}
   (\gamma +m-1)\right\}$. It always exists and it verifies $\Omega_m=1$. It describes a matter-dominated flat FLRW solution. The point is a 
    \begin{enumerate}
        \item sink for $m\leq -1, 0\leq \gamma <2$ or $
   -1<m<1, 0\leq \gamma <1-m$, 
        \item saddle for $-1<m\leq 1, 1-m<\gamma <2 $ or $ m>1,
   0\leq \gamma <2$,
        \item non-hyperbolic for $\gamma =2$ or $  (-1<m\leq 1, \gamma +m=1)$.
        \end{enumerate}
        \item $P_2=(0,1)$ with eigenvalues $\left\{-\frac{3}{2},-3 (\gamma +m-1)\right\}$. It always exists and represents a scalar field dominated solution. The point always exists and is a
        \begin{enumerate}
            \item sink for $-1<m\leq 1, 1-m<\gamma \leq 2$ or $
   m>1, 0\leq \gamma \leq 2$,
   \item saddle for $m<-1, 0\leq \gamma \leq 2$ or $-1\leq
   m<1, 0\leq \gamma <1-m$, 
   \item non-hyperbolic for 
   $m=1-\gamma.$
        \end{enumerate}
    \item $P_{3,4}=(\pm 1,0)$ with eigenvalues $\lbrace \frac{3}{2},-3 (\gamma -2) \rbrace$. They always exist and both represent an anisotropic Bianchi I vacuum solution. They are
    \begin{enumerate}
        \item sources for $0\leq \gamma <2$, 
        \item non-hyperbolic for $\gamma=2$. 
    \end{enumerate}
    \item A special case arises when $\gamma=2.$
The points $M_{1,2}=\left(\pm \sqrt{\frac{\gamma
   +m-1}{m}},\sqrt{\frac{2-\gamma }{m}}\right)$ only exist for $\gamma=2$ and $m\leq -1.$ But in this case, the points are $M_{1,2}=\left(\pm\sqrt{\frac{m+1}{m}},0\right)$ and they belong to the set of equilibrium points defined by the line $\bar{\Omega}=0.$ The eigenvalues for this normally hyperbolic set are $\left\{-\frac{3}{2} \left(m \left(\Sigma
   ^2-1\right)-1\right),0\right\},$ we will study the stability considering only the real part of the non-zero eigenvalue. The set is a
   \begin{enumerate}
       \item source for $-1\leq \bar{\Sigma}
   <-\sqrt{\frac{m+1}{m}}$ or $
   \sqrt{\frac{m+1}{m}}<\bar{\Sigma} \leq 1$, 
   \item sink for $-\sqrt{\frac{m+1}{m}}<\bar{\Sigma}
   <\sqrt{\frac{m+1}{m}},$
   \item non-hyperbolic for \begin{enumerate}
       \item $m=1, \bar{\Sigma}=0,$
       \item $m<-1, \bar{\Sigma}=\sqrt{\frac{m+1}{m}},$
       \item $m<-1, \bar{\Sigma}=\sqrt{\frac{m+1}{m}}.$
   \end{enumerate}
   \end{enumerate}
\end{enumerate}
Considering Interaction 7~\eqref{int-7}, it is clear that the parameter \( m \) influences the stability and the existence conditions of the equilibrium points, in particular, the stability of $P_1$ depends on $m$ which was not the case for the previous interactions. FIG.~\ref{fig:int7-2D-m>0} shows the phase plane of the guiding system~\eqref{int-7-2D-a}--\eqref{int-6-2D-b} for different values of \( \gamma \) and setting \( m = 0.1 \). On the other hand, FIG.~\ref{fig:int7-2D-m<0} shows the phase plane of the guiding system~\eqref{int-7-2D-a}--\eqref{int-6-2D-b} for different values of \( \gamma \) with \( m = -0.1 \). The special case $\gamma=2, m\leq -1$ is depicted separately in FIG. \ref{fig:special} in which we observe that the points $M_{1,2}$ act as bifurcation points for the stability of the line $\bar{\Omega}=0$ and behave almost like a centre when the restriction $\bar{\Omega}\geq 0$ from the compact set is ignored.
\FloatBarrier
\subsection{Interaction 8: $Q\left(\dot{\phi},\rho_{\phi }, \rho _{m}\right)=\frac{\Gamma  \rho_{m } \rho_{\phi }  \dot{\phi}}{\rho_{m }+\rho_{\phi }}$}
\begin{figure}[H]
    \centering
    \includegraphics[scale=0.4]{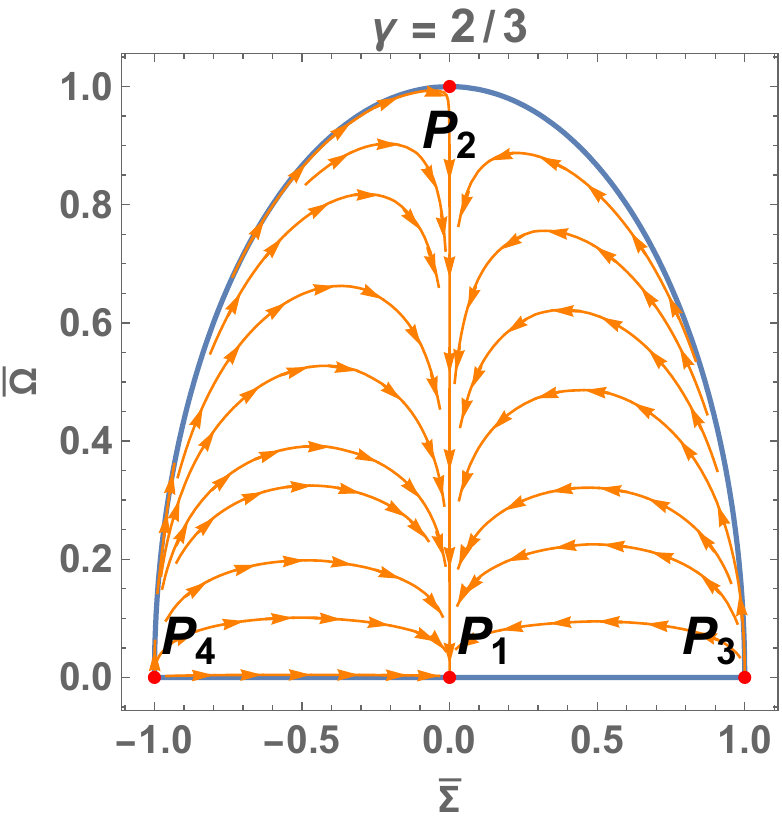}
    \includegraphics[scale=0.4]{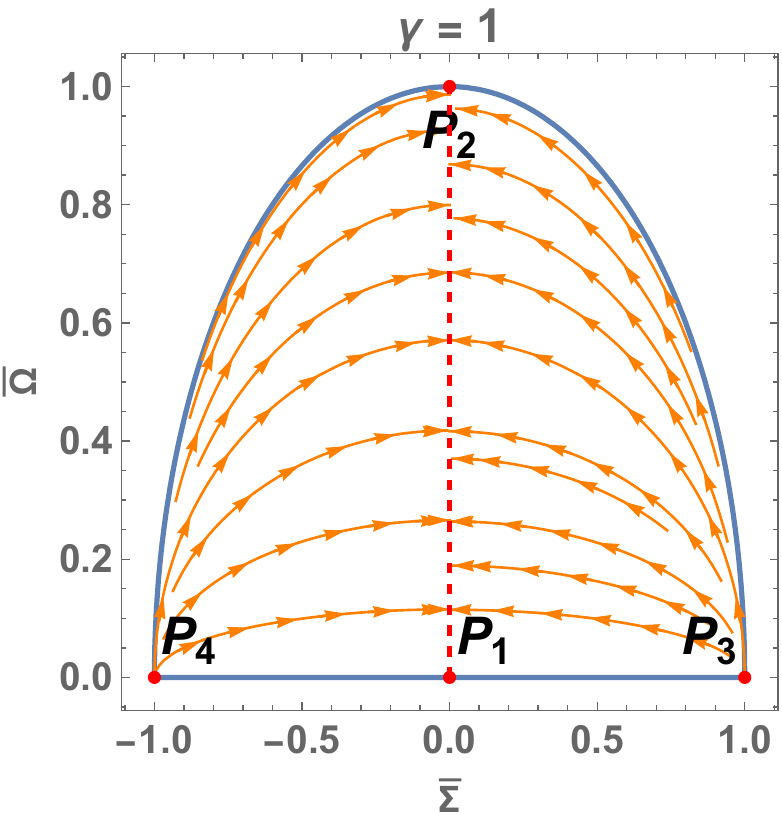}
    \includegraphics[scale=0.4]{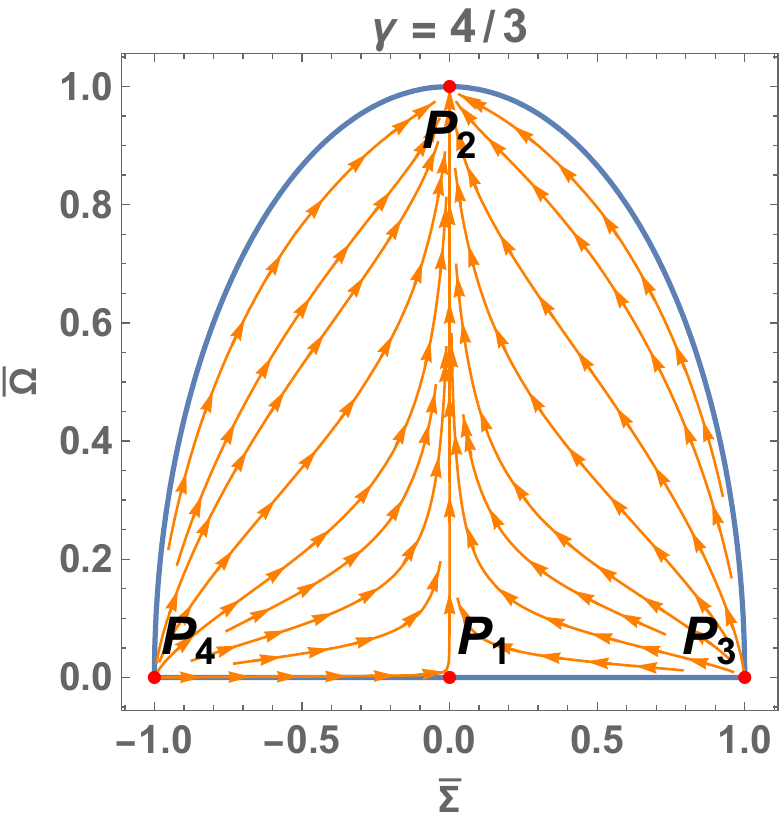}
    \includegraphics[scale=0.4]{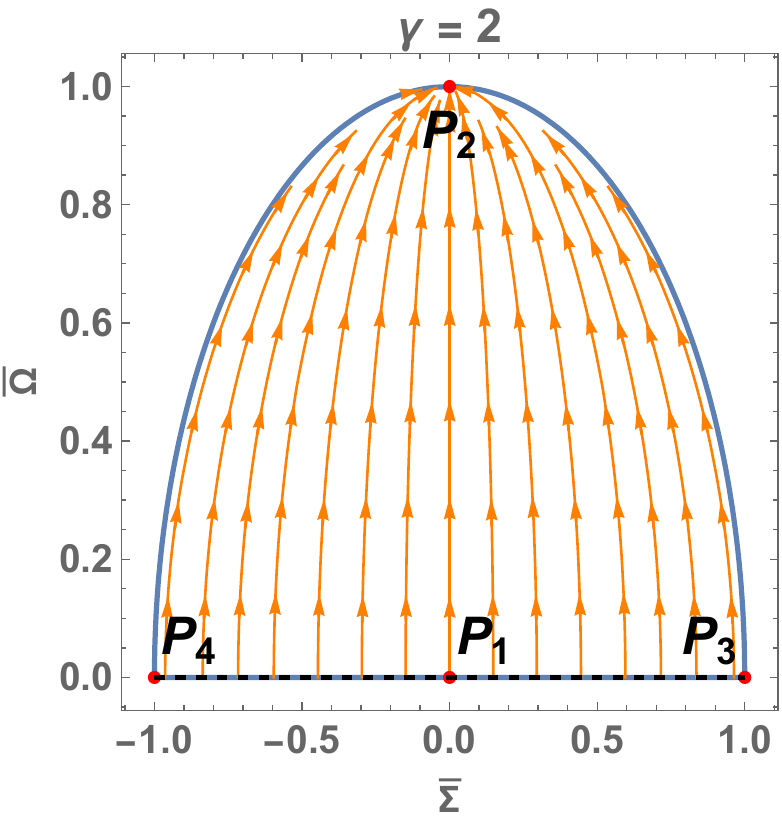}
    \caption{Phase-plane diagrams for system \eqref{int-6-2D-a}-\eqref{int-6-2D-b} for different values of $\gamma$.}
     \label{fig:int8-2D}
\end{figure}

Setting $\alpha=1,$ $\beta=-1,$ $\delta=1$ in system \eqref{gen-redu-1}-\eqref{gen-redu-4}, expanding in a Taylor series around $H=0$, the truncated system is
\begin{small}
    \begin{align}
    \label{comp-int-8-a}&\dot{H}=-\frac{3}{2} \left(2 \Sigma ^2+\Omega
   ^2-\gamma  \left(\Sigma ^2+\Omega
   ^2-1\right)+\Omega ^2 \cos (2 (\varphi -t \omega
   ))\right) H^2+O\left(H^3\right),\\
    &\dot{\Omega}=\frac{1}{2}
   \Omega  \left(6 \Sigma ^2+6 \left(\Omega
   ^2-1\right) \cos ^2(\varphi -t \omega )-3 \gamma
    \left(\Sigma ^2+\Omega
   ^2-1\right)+\frac{\sqrt{6} \Gamma  \Omega 
   \left(\Sigma ^2+\Omega ^2-1\right) \cos (\varphi
   -t \omega )}{\Sigma ^2-1}\right)
   H\\ \nonumber &-\frac{\left(\omega ^2-1\right)^3 \Omega ^3
   \cos (\varphi -t \omega ) \sin ^3(\varphi -t \omega
   ) H^2}{\omega
   ^3}+O\left(H^3\right),\\
    &\dot{\Sigma}=\frac{3}{2} \Sigma 
   \left(2 \Sigma ^2+\Omega ^2-\gamma 
   \left(\Sigma ^2+\Omega ^2-1\right)+\Omega ^2
   \cos (2 (\varphi -t \omega ))-2\right)
   H+O\left(H^3\right),\\
   \label{comp-int-8-d} &\dot{\varphi}=\frac{1}{2} \left(6 \cos
   (\varphi -t \omega )-\frac{\sqrt{6} \Gamma 
   \Omega  \left(\Sigma ^2+\Omega
   ^2-1\right)}{\Sigma ^2-1}\right) \sin (\varphi
   -t \omega ) H+\frac{\left(\omega
   ^2-1\right)^3 \Omega ^2 \sin ^4(\varphi -t
   \omega ) H^2}{\omega
   ^3}+O\left(H^3\right).
\end{align}
\end{small}
The averaged system is
\begin{align}
   &\dot{H}=\frac{3}{2} H^2 \left(\gamma 
   \left(\bar{\Sigma} ^2+\bar{\Omega} ^2-1\right)-2 \bar{\Sigma}
   ^2-\bar{\Omega} ^2\right),  \label{1-prom-int8}\\
  &\dot{\bar{\Omega}}=\frac{3}{2} H \bar{\Omega} 
   \left(-\gamma  \left(\bar{\Sigma} ^2+\bar{\Omega}
   ^2-1\right)+2 \bar{\Sigma} ^2+\bar{\Omega}
   ^2-1\right),  \label{2-prom-int8} \\
 &\dot{\bar{\Sigma}}=\frac{3}{2} H \bar{\Sigma} 
   \left(-\gamma  \left(\bar{\Sigma} ^2+\bar{\Omega}
   ^2-1\right)+2 \bar{\Sigma} ^2+\bar{\Omega}
   ^2-2\right),  \label{3-prom-int8}\\
 &\dot{\bar{\varphi}}=0.    \label{4-prom-int8}
\end{align}
And the guiding system with the new time derivative defined as $f'= \frac{1}{H}\dot{f}$ is
\begin{align}
\label{int-8-2D-a}
    &\bar{\Sigma}'=\frac{3}{2} \bar{\Sigma}  \left(-\gamma 
   \left(\bar{\Sigma} ^2+\bar{\Omega} ^2-1\right)+2 \bar{\Sigma}
   ^2+\bar{\Omega} ^2-2\right),\\
   \label{int-8-2D-b}&\bar{\Omega}'=\frac{3}{2} \bar{\Omega} 
   \left(-\gamma  \left(\bar{\Sigma} ^2+\bar{\Omega}
   ^2-1\right)+2 \bar{\Sigma} ^2+\bar{\Omega}
   ^2-1\right).
\end{align}
As before, the phase-space is the compact set
\begin{equation}
    \left\{(\bar{\Sigma},\bar{\Omega})\in \mathbb{R}^2| \;  \bar{\Omega}^2 + \bar{\Sigma}^2 \leq 1, \bar{\Omega}\geq 0, -1\leq \bar{\Sigma} \leq 1 \right\}.
\end{equation}
The equilibrium points of system \eqref{int-8-2D-a}-\eqref{int-8-2D-b} in the coordinates $(\bar{\Sigma},\bar{\Omega})$ are
\begin{enumerate}
    \item $P_1=(0, 0)$ with eigenvalues $\lbrace \frac{3 (\gamma -2)}{2},\frac{3 (\gamma -1)}{2}\rbrace$. It always exists and it verifies $\Omega_m=1$. It describes a matter-dominated flat FLRW solution. The point is 
    \begin{enumerate}
        \item sink for $0\leq \gamma <1$, 
        \item saddle for $1<\gamma \leq 2$,
        \item non-hyperbolic for $\gamma=1,2$.
    \end{enumerate}
    \item $P_2=(0,1)$ with eigenvalues $\lbrace -\frac{3}{2},-3 (\gamma -1) \rbrace$. It always exists and represents a scalar field dominated flat FLRW solution. The point is 
    \begin{enumerate}
        \item saddle for $0\leq \gamma <1$, 
        \item sink for $1<\gamma \leq 2$, 
        \item non-hyperbolic for $\gamma=1$.
    \end{enumerate}
    \item $P_{3,4}=(\pm 1,0)$ with eigenvalues $\lbrace \frac{3}{2},-3 (\gamma -2) \rbrace$. They always exist and both represent an anisotropic Bianchi I vacuum solution. They are
    \begin{enumerate}
        \item sources for $0\leq \gamma <2$, 
        \item non-hyperbolic for $\gamma=2$. 
    \end{enumerate}
\end{enumerate}
We note that the equilibrium points for interaction $8$ \eqref{int-8} are the same as those obtained from interaction 1 \eqref{int-1}. In FIG. \ref{fig:int8-2D} we present the dynamics on the phase-plane for system \eqref{int-8-2D-a}-\eqref{int-8-2D-b}.
\subsection{Interaction 9: $Q\left(\dot{\phi},\rho_{\phi }, \rho _{m}\right)=\frac{\Gamma  H_0 \rho_{m } \rho_{\phi }  \dot{\phi}^2}{H
   (\rho_{m }+\rho_{\phi } )}$}
\begin{figure}[H]
    \centering
    \includegraphics[scale=0.4]{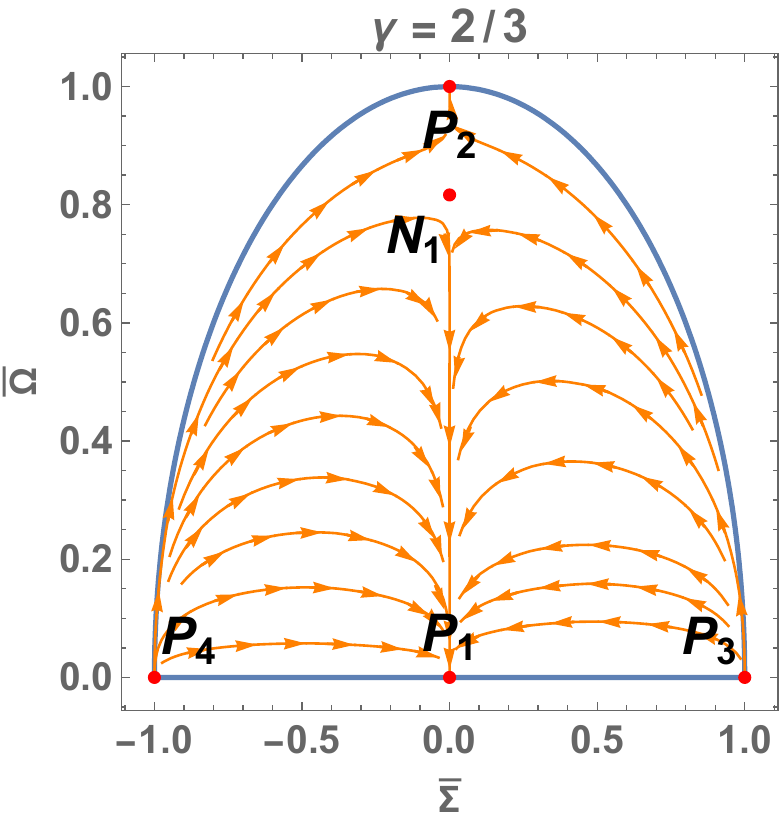}
    \includegraphics[scale=0.4]{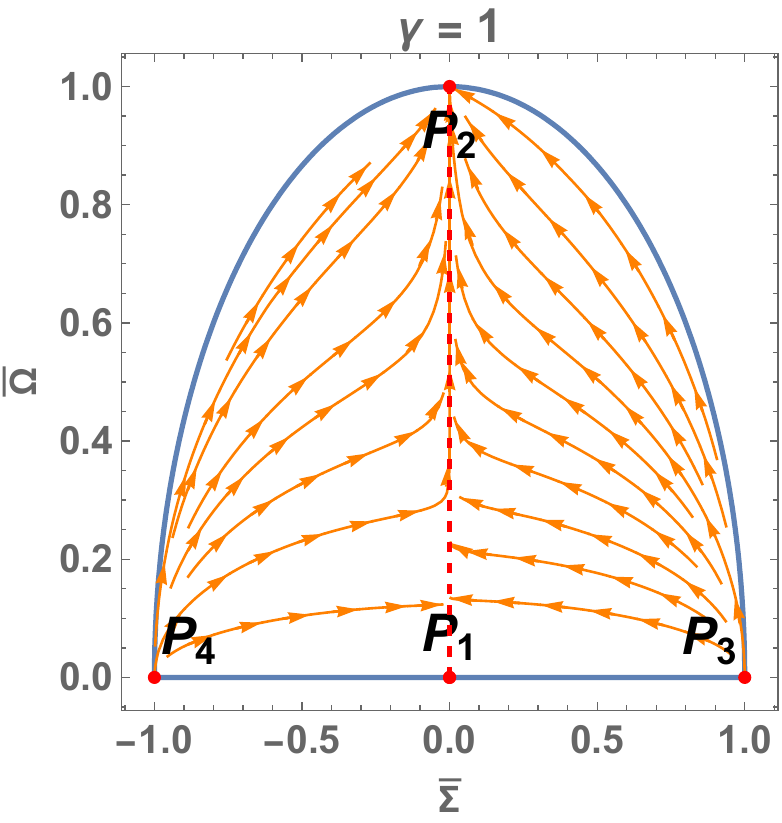}
    \includegraphics[scale=0.4]{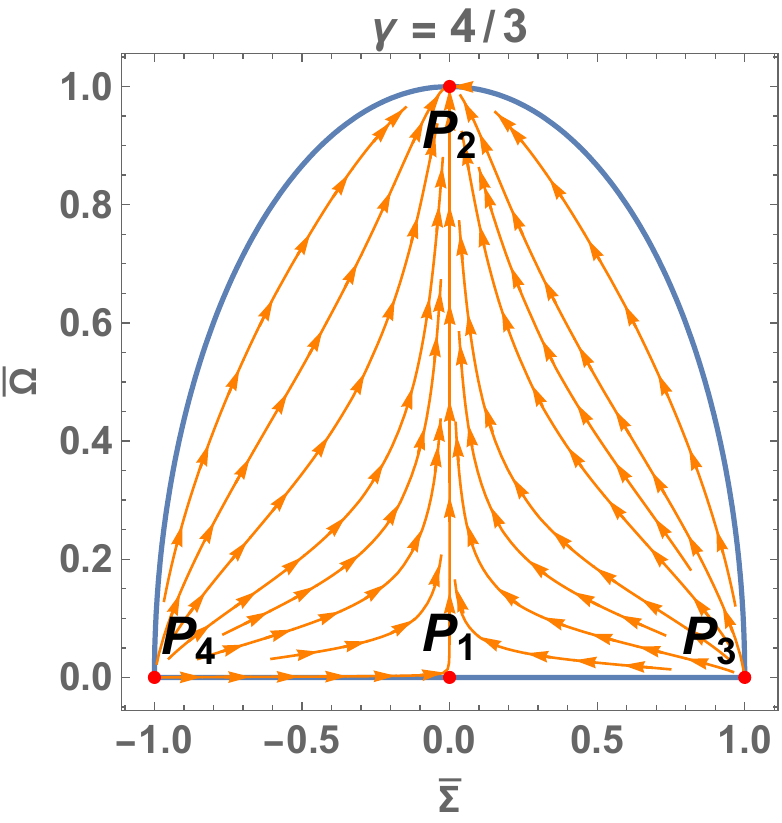}
    \includegraphics[scale=0.4]{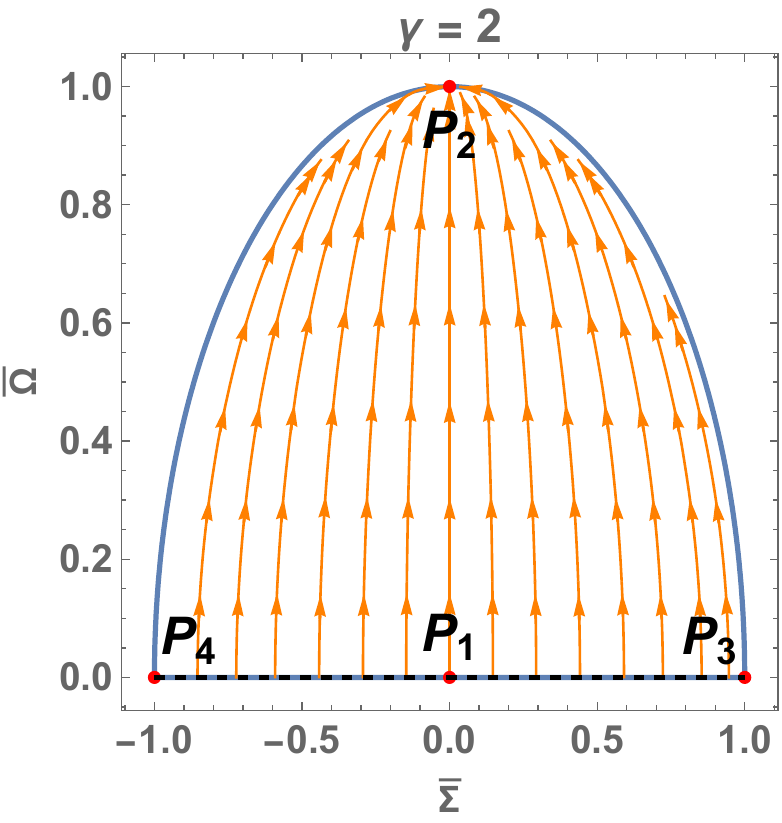}
    \caption{Phase-plane diagrams for system \eqref{int-9-2D-a}-\eqref{int-9-2D-b} setting $m=0.5$ for different values of $\gamma$.}
     \label{fig:int9-2D-m>0}
\end{figure}

\begin{figure}[H]
    \centering
    \includegraphics[scale=0.4]{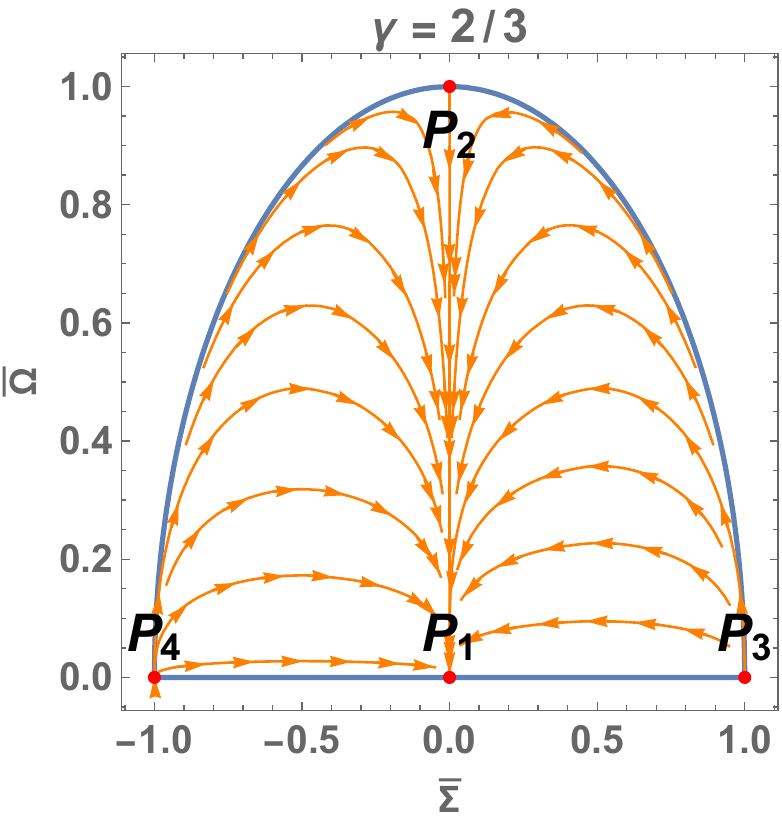}
    \includegraphics[scale=0.4]{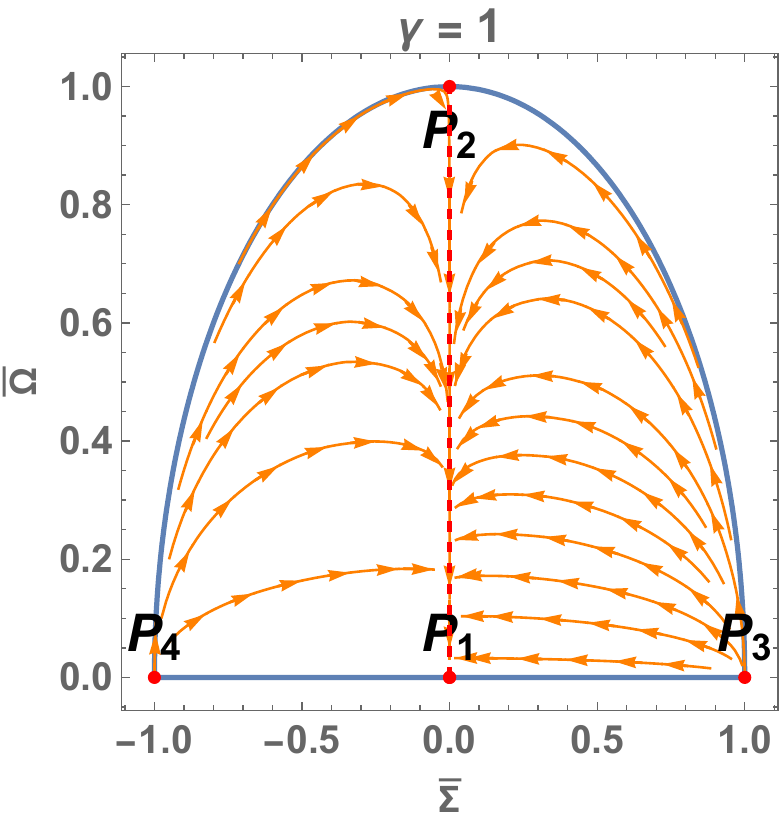}
    \includegraphics[scale=0.4]{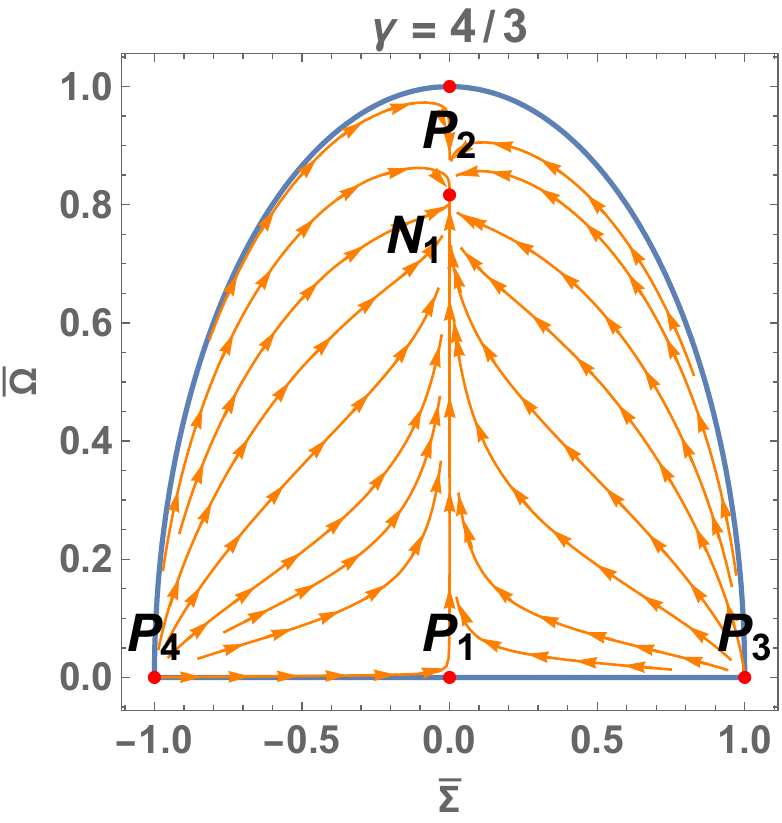}
    \includegraphics[scale=0.4]{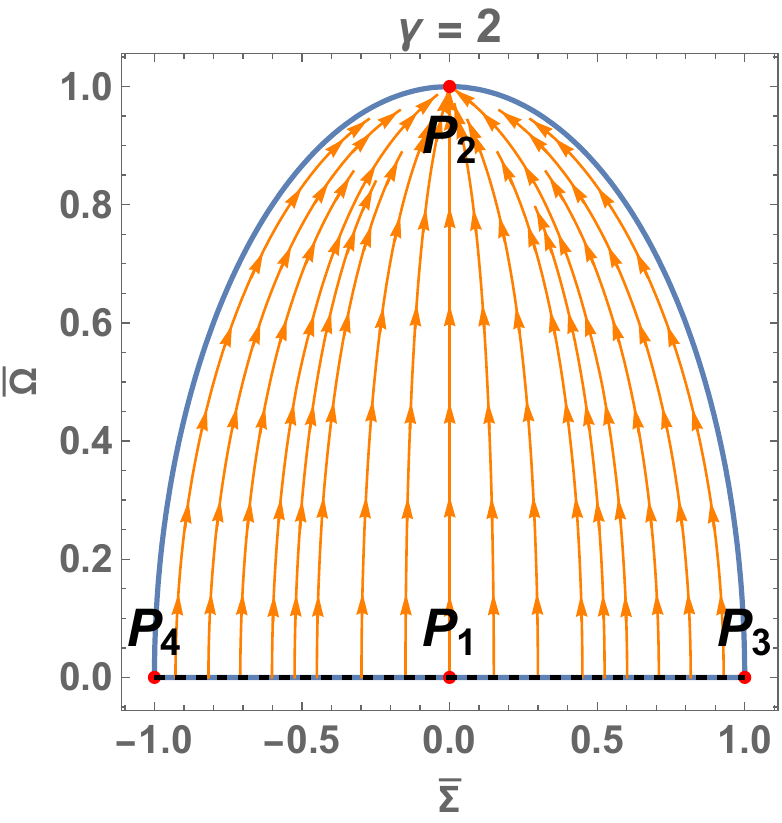}
    \caption{Phase-plane diagrams for system \eqref{int-9-2D-a}-\eqref{int-9-2D-b} setting $m=-0.5$ for different values of $\gamma$.}
 \label{fig:int9-2D-m<0}\end{figure}

   Finally, setting $\alpha=1,$ $\beta=-1,$ $\delta=2$ in system \eqref{gen-redu-1}-\eqref{gen-redu-4}, expanding in a Taylor series around $H=0$ and using the rescaling $\Gamma =\frac{m}{H_0} $ the truncated system is
\begin{small}
    \begin{align}
    \label{comp-int-9-a}&\dot{H}=-\frac{3}{2} \left(2 \Sigma ^2+\Omega
   ^2-\gamma  \left(\Sigma ^2+\Omega
   ^2-1\right)+\Omega ^2 \cos (2 (\varphi -t \omega
   ))\right) H^2+O\left(H^3\right),\\
    &\dot{\Omega}=\frac{3}{2} \Omega  \left(-\left((\gamma
   -2) \Sigma ^2\right)+\gamma
   -1\right)-\frac{\left(\omega
   ^2-1\right)^3 \Omega ^3 \cos (\varphi -t \omega
   ) \sin ^3(\varphi -t \omega ) H^2}{\omega
   ^3}\\ \nonumber &+\frac{3 \Omega  \left(\left(\Sigma
   ^2-1\right) \Omega ^2 (-\gamma
   +m+1)+\left((m+1) \left(\Sigma ^2-1\right)
   \Omega ^2+m \Omega ^4-\Sigma ^2+1\right)
   \cos (2 (\varphi -t \omega ))+m \Omega
   ^4\right)}{2 \left(\Sigma
   ^2-1\right)}+O\left(H^3\right),\\
    &\dot{\Sigma}=\frac{3}{2} \Sigma 
   \left(2 \Sigma ^2+\Omega ^2-\gamma 
   \left(\Sigma ^2+\Omega ^2-1\right)+\Omega ^2
   \cos (2 (\varphi -t \omega ))-2\right)
   H+O\left(H^3\right),\\
   \label{comp-int-9-d} &\dot{\varphi}=-\frac{3 \left(\left(m
   \Omega ^4+m \left(\Sigma ^2-1\right) \Omega
   ^2-\Sigma ^2+1\right) \sin (2 (\varphi -t \omega
   ))\right) H}{2 \left(\Sigma
   ^2-1\right)}+\frac{\left(\omega
   ^2-1\right)^3 \Omega ^2 \sin ^4(\varphi -t
   \omega ) H^2}{\omega
   ^3}+O\left(H^3\right).
\end{align}
\end{small}
The averaged system is
\begin{align}
   &\dot{H}=\frac{3}{2} H^2 \left(\gamma 
   \left(\bar{\Sigma} ^2+\bar{\Omega} ^2-1\right)-2 \bar{\Sigma}
   ^2-\bar{\Omega} ^2\right),  \label{1-prom-int9}\\
  &\dot{\bar{\Omega}}=\frac{3}{2} H \bar{\Omega} 
   \left(-(\gamma -2) \bar{\Sigma} ^2+\gamma +\bar{\Omega}
   ^2 (-\gamma +m+1)+\frac{m \bar{\Omega} ^4}{\bar{\Sigma}
   ^2-1}-1\right),  \label{2-prom-int9} \\
 &\dot{\bar{\Sigma}}=\frac{3}{2} H \bar{\Sigma} 
   \left(-\gamma  \left(\bar{\Sigma} ^2+\bar{\Omega}
   ^2-1\right)+2 \bar{\Sigma} ^2+\bar{\Omega}
   ^2-2\right),  \label{3-prom-int9}\\
 &\dot{\bar{\varphi}}=0.    \label{4-prom-int9}
\end{align}
And the guiding system with the new time derivative defined as $f'= \frac{1}{H}\dot{f}$ is
\begin{align}
\label{int-9-2D-a}
    &\bar{\Sigma}'=\frac{3}{2} \bar{\Sigma}  \left(-\gamma 
   \left(\bar{\Sigma} ^2+\bar{\Omega} ^2-1\right)+2 \bar{\Sigma}
   ^2+\bar{\Omega} ^2-2\right),\\
   \label{int-9-2D-b}&\bar{\Omega}'=\frac{3}{2} \bar{\Omega} 
   \left(-(\gamma -2) \bar{\Sigma} ^2+\gamma +\bar{\Omega}
   ^2 (-\gamma +m+1)+\frac{m \bar{\Omega} ^4}{\bar{\Sigma}
   ^2-1}-1\right).
\end{align}
As before, the phase-space is the compact set
\begin{equation}
    \left\{(\bar{\Sigma},\bar{\Omega})\in \mathbb{R}^2| \;  \bar{\Omega}^2 + \bar{\Sigma}^2 \leq 1, \bar{\Omega}\geq 0, -1\leq \bar{\Sigma} \leq 1 \right\}.
\end{equation}
We observe that there is a singularity in the denominators for the values $\Sigma=\pm 1$,  therefore, it is necessary to properly analyse the equilibrium points involving that coordinate \cite{Paliathanasis:2024jxo,Leon:2025sfd,Papagiannopoulos:2025zku}. The equilibrium points for system . The equilibrium points of system \eqref{int-9-2D-a}-\eqref{int-9-2D-b} in the coordinates $(\bar{\Sigma},\bar{\Omega})$ are
\begin{enumerate}
    \item $P_1=(0,0)$, with eigenvalues $\lbrace \frac{3 (\gamma -2)}{2},\frac{3 (\gamma -1)}{2}\rbrace$. It always exists and it verifies $\Omega_m=1$. It describes a matter-dominated flat FLRW solution. The point is 
    \begin{enumerate}
        \item sink for $0\leq \gamma <1$, 
        \item saddle for $1<\gamma \leq 2$,
        \item non-hyperbolic for $\gamma=1,2$.
    \end{enumerate}
    \item $P_2=(0,1)$, with eigenvalues $\left\{-\frac{3}{2},-3 (\gamma +m-1)\right\}$ . It always exists and represents a scalar field dominated solution. The point always exists and is a
        \begin{enumerate}
            \item sink for $-1<m\leq 1, 1-m<\gamma \leq 2$ or $
   m>1, 0\leq \gamma \leq 2$,
   \item saddle for $m<-1, 0\leq \gamma \leq 2$ or $-1\leq
   m<1, 0\leq \gamma <1-m$, 
   \item non-hyperbolic for 
   $m=1-\gamma.$
        \end{enumerate}
    \item $P_{3,4}=(\pm 1,0)$. They always exist, to study their stability we substitute $\bar{\Omega}=0$ in the Jacobian matrix and then take the limit as $\Sigma\rightarrow \pm 1.$ The eigenvalues are 
    $\left\{\frac{3}{2} (4-2 \gamma
   ),\frac{3}{2}\right\}.$ They are
   \begin{enumerate}
       \item sources for $0\leq \gamma <2,$
       \item non-hyperbolic for $\gamma=2.$
   \end{enumerate}
    \item $N_1=\left(0,\sqrt{\frac{1-\gamma}{m}}\right)$, with eigenvalues $\left\{\frac{3 \left((\gamma -1)^2+(\gamma -2)
   m\right)}{2 m},-\frac{3 (\gamma -1) (\gamma
   +m-1)}{m}\right\}$. The point exists for $m>0, 0\leq \gamma \leq 1$ or $ m<0,
   1\leq \gamma \leq 2$ and is a
   \begin{enumerate}
       \item sink for $0\leq \gamma <1, \frac{-\gamma ^2+2
   \gamma -1}{\gamma -2}<m<1-\gamma $ or $
   1<\gamma \leq 2, m<1-\gamma $,
   \item saddle for 
   \begin{enumerate}
       \item $0\leq \gamma <1, 0<m<\frac{-\gamma ^2+2
   \gamma -1}{\gamma -2}$ or
       \item $0\leq \gamma <1, m>1-\gamma$ or
       \item $
   1<\gamma \leq 2, 1-\gamma <m<0$,
   \end{enumerate}  
   \item non-hyperbolic for
   \begin{enumerate}
       \item $m\neq 0, \gamma =1$ or
       \item $\gamma +m=1, 1<\gamma \leq 2$ or
       \item $0\leq \gamma
   <1, \frac{(\gamma -1)^2}{\gamma
   -2}+m=0$ or
       \item $0\leq \gamma
   <1,  \gamma +m=1.$
   \end{enumerate}
   \end{enumerate}
\end{enumerate}
Considering interaction $9$ \eqref{int-9}, we highlight the equilibrium point $N_1$ which is not present in any other interaction. In Figures \ref{fig:int9-2D-m>0} and \ref{fig:int9-2D-m<0} we present the dynamics on the phase-plane for system \eqref{int-9-2D-a}-\eqref{int-9-2D-b} considering $m=0.5>0$ and $m-0.5<0$ and different values of the equation of state parameter.

\section{Deceleration parameter for the equilibrium points}
\label{sect-5}
In this section, we investigate the dynamical behaviour of the deceleration parameter. The definition of the deceleration parameter $q$ is as follows
\begin{equation}
\label{dece}
    q:=-1-\frac{\dot{H}(t)}{H^2(t)}.
\end{equation}
For the models under consideration, equation \eqref{dece} becomes
\begin{small}
    \begin{align}
q=&-1+\frac{3}{2} \Bigg[
  \frac{1}{3} \gamma 
  \Bigg(
    \frac{
      \cos \left(
        \frac{\sqrt{6} H (\omega^2 - 1) \Omega \sin(\varphi - t \omega)}
        {\omega}
      \right)
    }{
      H^2 (\omega^2 - 1)
    }
    - \frac{1}{H^2 (\omega^2 - 1)} 
    - 3 \Sigma^2 \notag \\
    &\quad
    - \frac{3 \Omega^2 \left(\omega^2 \cos^2(\varphi - t \omega) + \sin^2(\varphi - t \omega)\right)}{\omega^2} 
    + 3
  \Bigg)
  + 2 \Sigma^2 \notag \\
  &\quad
  + \Omega^2 \cos\left(2 (\varphi - t \omega)\right) 
  + \Omega^2
\Bigg].
\end{align}
\end{small}
To better understand the behaviour near $H = 0$, we expand this expression in a Taylor series about $H = 0$ and retain terms up to first order
\begin{equation}
\label{qseries}
    q_{\text{series}}=-1+\frac{3}{2}  \left(-\gamma  \left(\Sigma
   ^2+\Omega ^2-1\right)+2 \Sigma ^2+\Omega ^2
   \cos (2 (\varphi -t \omega ))+\Omega
   ^2\right)+O\left(H^2\right).
\end{equation}
Averaging over the oscillatory term yields:
\begin{equation}
\label{qaver}
    q_{\text{averaged}}=-1+\frac{3}{2}  \left(-\gamma  \left(\Sigma
   ^2+\Omega ^2-1\right)+2 \Sigma ^2+\Omega
   ^2\right).
\end{equation}
Given that $0\leq \Omega \leq 1$ and $-1\leq \cos (\alpha)\leq 1$ these two quantities verify
\begin{equation}
   |\Delta q|:= |q_{\text{series}}- q_{\text{averaged}}|=\left|\frac{3}{2} \Omega ^2 \cos (2 (\varphi -t \omega
   ))\right|\leq \frac{3}{2}\Omega^2\leq \frac{3}{2}.
\end{equation}
In the following section, we evaluate these quantities at the equilibrium points obtained in Section~\ref{Sect:4}. Since one of the main advantages of the averaging method is that the equilibrium points and the late-time behaviour of the averaged system coincide with those of the original system, we shall assume that $\bar{\Sigma} = \Sigma$ and $\bar{\Omega} = \Omega$ for the purpose of this evaluation.
\subsection{Analysis of $ q_{\text{series}}$ and $q_{\text{averaged}}$}
In this section, we study the value of the deceleration parameter \eqref{qseries} and \eqref{qaver} evaluated in the equilibrium points obtained in Section \ref{Sect:4}. Recall that $0\leq \gamma \leq 2$ and $\Gamma, \bar{\Gamma}, m$ are real parameters.
\begin{enumerate}
    \item For $P_1=(0,0),$ the values are $$q_{\text{series}}=q_{\text{averaged}}=-1+\frac{3 \gamma }{2}.$$ It describes acceleration for $0\leq \gamma <\frac{2}{3}.$
    Additionally, we can obtain an expression for $H$ by solving the following differential equation 
    $$\frac{\dot{H}}{H}=-\frac{3 \gamma }{2}.$$
    Setting the integration constant to zero, the Hubble function is \begin{equation}
    \label{H1eq}
        H(t)=\frac{2}{3 \gamma  t},
    \end{equation}
    which is defined for $\gamma\neq 0,$ is always positive and decreasing as $t\rightarrow \infty$. This is a scaling solution where only the matter source contribute in the cosmic fluid.
    \item For $P_2=(0,1),$ the values are
    \begin{align}
        q_{\text{series}}&=\frac{1}{2}+\frac{3}{2} \cos (2 (\varphi -t
   \omega )),\\
   q_{\text{averaged}}&=\frac{1}{2}.
    \end{align}
    In Figure \ref{comp1} we see that $q_{\text{series}}$ oscillates between deceleration and acceleration. On the other hand, $q_{\text{averaged}}$ is a constant decelerated value.  We verify that for $q_{\text{averaged}}$ the Hubble function is $H(t)=\frac{2}{3 t}.$
    \begin{figure}[H]
        \centering
        \includegraphics[width=0.7\linewidth]{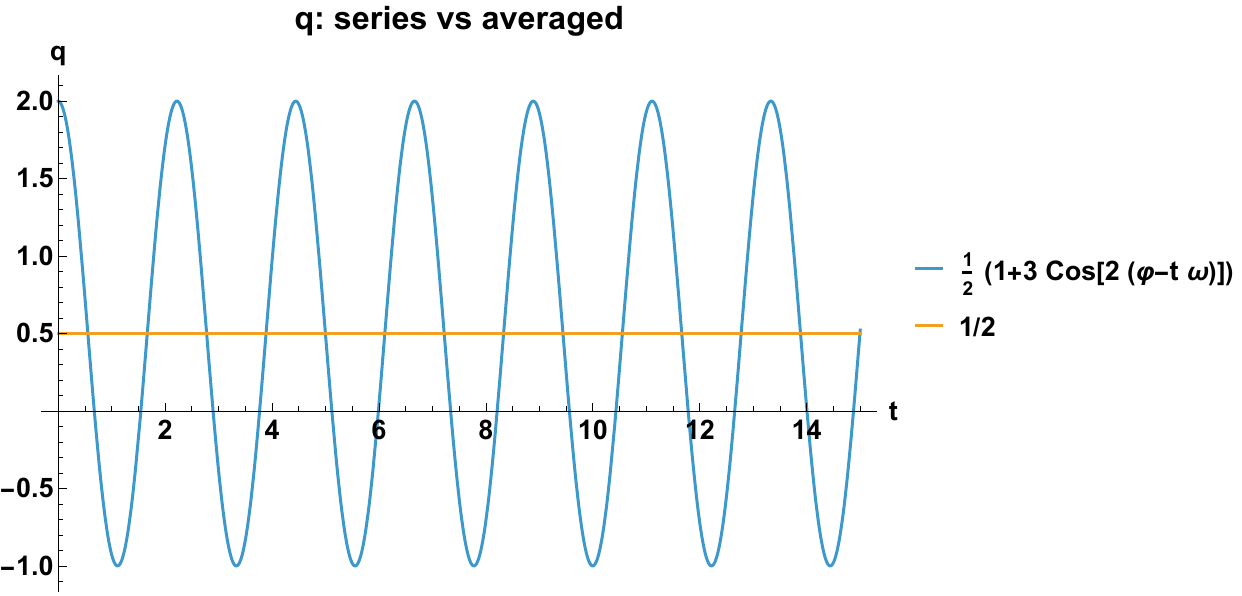}
        \caption{Comparison between $q_{\text{series}}$ and $q_{\text{averaged}}$ for $\omega=\sqrt{2}$ and $\varphi=0.$}
        \label{comp1}
    \end{figure}
    \item For $P_{3,4}=(\pm 1,0),$ the values are $q_{\text{series}}=q_{\text{averaged}}=2,$ they describe deceleration.  The Hubble function is $H(t)=\frac{1}{3 t}.$
    \item For $P_5=(0,\sqrt{\frac{\bar{\Gamma}}{3 (1-\gamma )}}),$ the values are
    \begin{align}
        q_{\text{series}}&=-1+\frac{\bar{\Gamma }}{2}+\frac{3 \gamma }{2}-\frac{\bar{\Gamma } \cos (2 (\varphi -t \omega ))}{2
   (\gamma -1)},\\
   q_{\text{averaged}}&=-1+\frac{\bar{\Gamma }}{2}+\frac{3 \gamma }{2}.
    \end{align}
    We focus on $q_{\text{averaged}},$ it describes acceleration for $\bar{\Gamma }<2-3 \gamma.$ The Hubble function is \begin{equation}
        H(t)=\frac{2}{t \left(\bar{\Gamma }+3 \gamma \right)}.
    \end{equation}
    This function satisfies the condition $\lim_{t\rightarrow \infty}H(t)=0.$ It is positive for $-3 \gamma <\bar{\Gamma }$. This means that in the interval $-3 \gamma <\bar{\Gamma }<2-3 \gamma,$ the Hubble function is positive and $q_{\text{averaged}}$ describes acceleration.
    \item For $P_6=\Big(0,\sqrt{1+\frac{\bar{\Gamma}}{3 (\gamma -1)}}\Big),$ the values are
    \begin{align}
        q_{\text{series}}&=\frac{1}{2}+\frac{\bar{\Gamma } (1-\gamma +\cos (2 (\varphi -t \omega )))}{2 (\gamma
   -1)}+\frac{3}{2} \cos (2 (\varphi -t \omega )),\\
  q_{\text{averaged}} &=\frac{1}{2}-\frac{\bar{\Gamma }}{2}.
    \end{align}
     We focus on $q_{\text{averaged}},$ it describes acceleration for $\bar{\Gamma }>1.$ The Hubble function is 
     \begin{equation}
        H(t)=\frac{2}{t \left(3-\bar{\Gamma }\right)}.
    \end{equation}
    This function satisfies the condition $\lim_{t\rightarrow \infty}H(t)=0.$ It is positive for $\bar{\Gamma }<3$. The interval for $q_{\text{averaged}}<0$ and $H(t)>0$ is $1<\bar{\Gamma }<3.$
    \item For $P_7=\left(0,\frac{\sqrt{\sqrt{3+\frac{12 \bar{\Gamma}}{\gamma
   -1}+9}}}{\sqrt{6}}\right),$ the values are 
   \begin{align}
       q_{\text{series}}&=-\frac{1}{4}+\frac{3 \gamma }{4}-\frac{1}{4} \sqrt{3} \sqrt{\gamma -1} \sqrt{3 \gamma
   +4 \bar{\Gamma}-3}\\ \nonumber 
   &+\frac{\left(3 \sqrt{\gamma -1}+\sqrt{9 \gamma +12 \bar{\Gamma}-9}\right)
   \cos (2 (\varphi -t \omega ))}{4 \sqrt{\gamma -1}},\\
   q_{\text{averaged}}&=-\frac{1}{4}+\frac{3 \gamma }{4}-\frac{1}{4} \sqrt{3} \sqrt{\gamma -1} \sqrt{3 \gamma
   +4 \bar{\Gamma}-3}.
   \end{align}
    We focus on $q_{\text{averaged}}.$ This quantity represents a real number for $1\leq \gamma \leq 2, \bar{\Gamma}\geq \frac{1}{4} (3-3 \gamma )$ and it describes acceleration for $1<\gamma \leq 2,  \bar{\Gamma}>\frac{1}{3 (\gamma -1)}+1.$ The Hubble function is 
     \begin{equation}
        H(t)=\frac{4}{t \left(3 \gamma -\sqrt{3} \sqrt{\gamma -1} \sqrt{3 \gamma +4
   \bar{\Gamma}-3}+3\right)}.
    \end{equation}
    This function satisfies the condition $\lim_{t\rightarrow \infty}H(t)=0.$ It is positive for $1<\gamma \leq 2, \frac{1}{4} (3-3 \gamma )<\bar{\Gamma}<\frac{3 \gamma }{\gamma
   -1}$. The intervals for $q_{\text{averaged}}<0$ and $H(t)>0$ are $1<\gamma \leq 2, \frac{3 \gamma -2}{3 \gamma -3}<\bar{\Gamma}<\frac{3 \gamma
   }{\gamma -1}.$
   \item For $P_8=\left(0,\frac{\sqrt{3-\sqrt{\frac{12 \bar{\Gamma}}{\gamma
   -1}+9}}}{\sqrt{6}}\right),$  the values are 
   \begin{align}
       q_{\text{series}}&=-\frac{1}{4}+\frac{3 \gamma }{4}+\frac{1}{4} \sqrt{3} \sqrt{\gamma -1} \sqrt{3 \gamma
   +4 \bar{\Gamma}-3}\\ \nonumber &+\frac{\left(3 \sqrt{\gamma -1}-\sqrt{9 \gamma +12 \bar{\Gamma}-9}\right)
   \cos (2 (\varphi -t \omega ))}{4 \sqrt{\gamma -1}},\\
   q_{\text{averaged}}&=-\frac{1}{4}+\frac{3 \gamma }{4}+\frac{1}{4} \sqrt{3} \sqrt{\gamma -1} \sqrt{3 \gamma
   +4 \bar{\Gamma}-3}.
   \end{align}
    We focus on $q_{\text{averaged}}.$ This quantity represents a real number for $1\leq \gamma \leq 2, \bar{\Gamma}\geq \frac{1}{4} (3-3 \gamma )$ and it cannot describe acceleration. The Hubble function is 
     \begin{equation}
        H(t)=\frac{4}{t \left(3 \gamma +\sqrt{3} \sqrt{\gamma -1} \sqrt{3 \gamma +4
   \bar{\Gamma}-3}+3\right)}.
    \end{equation}
    This function satisfies the condition $\lim_{t\rightarrow \infty}H(t)=0.$ It is positive for $1<\gamma \leq 2,  \bar{\Gamma}>\frac{1}{4} (3-3 \gamma )$.
    \item For $P_9=\left(0,\frac{\sqrt{1-\gamma }}{\sqrt{m-\gamma +1}}\right),$ the values are
    \begin{align}
       q_{\text{series}} &=\frac{-\gamma +3 \gamma  m-2 m+1}{-2 \gamma +2 m+2}-\frac{3 (\gamma -1)
   \cos (2 (\varphi -t \omega ))}{2 (-\gamma +m+1)},\\
        q_{\text{averaged}}&=\frac{-\gamma +3 \gamma  m-2 m+1}{-2 \gamma +2 m+2}.
    \end{align}
    We focus on $ q_{\text{averaged}}.$ It describes acceleration for \begin{itemize}
        \item $0\leq \gamma <\frac{2}{3}, m<\gamma -1$ or
        \item $0\leq \gamma <\frac{2}{3}, m>\frac{\gamma -1}{3 \gamma -2}$ or
        \item $\gamma =\frac{2}{3}, m<-\frac{1}{3}$ or
        \item $\frac{2}{3}<\gamma <1, \frac{\gamma -1}{3 \gamma -2}<m<\gamma -1$ or
        \item $1<\gamma \leq 2, \frac{\gamma -1}{3 \gamma -2}<m<\gamma -1.$
    \end{itemize}
    The Hubble function is \begin{equation}
        H(t)=\frac{2 (-\gamma +m+1)}{3 t (-\gamma +\gamma  m+1)},
    \end{equation}
    which is positive for \begin{itemize}
        \item $\gamma =0, m>-1$ or
        \item $0<\gamma \leq 2, m<\frac{\gamma -1}{\gamma }$ or
        \item $0<\gamma \leq 2, m>\gamma -1.$
    \end{itemize}
    In this case, $ q_{\text{averaged}}<0$ and $H(t)>0$ occur for the following conditions
    \begin{itemize}
        \item $\gamma =0, m>\frac{1}{2}$ or
        \item $\gamma =\frac{2}{3}, m<-\frac{1}{2}$ or
        \item $0<\gamma <\frac{2}{3}, m>\frac{\gamma -1}{3 \gamma -2}$ or
        \item $0<\gamma <\frac{2}{3}, m<\frac{\gamma -1}{\gamma }$ or
        \item $1<\gamma \leq 2, \frac{\gamma -1}{3 \gamma -2}<m<\frac{\gamma
   -1}{\gamma }$ or
   \item $\frac{2}{3}<\gamma <1, \frac{\gamma -1}{3 \gamma -2}<m<\frac{\gamma
   -1}{\gamma }.$
    \end{itemize}
   \item For $M_{1,2}=\left(\pm\sqrt{\frac{m+1}{m}},0\right),$ the values are $q_{\text{series}}=q_{\text{averaged}}=\frac{-3 \gamma +4 m+6}{2 m}.$ They describe acceleration for $0\leq \gamma <2, \frac{1}{4} (3 \gamma -6)<m<0.$ The Hubble function is 
   \begin{equation}
   H(t)=\frac{2 m}{t (-3 \gamma +6 m+6)},    
   \end{equation}
   which is positive for 
   \begin{itemize}
       \item $m<\frac{\gamma -2}{2}$ or
       \item $m>0.$
   \end{itemize}
   In this case, $q_{\text{averaged}}<0$ and $H(t)>0$ occur for the following conditions $0\leq \gamma <2, \frac{1}{4} (3 \gamma -6)<m<\frac{\gamma -2}{2}.$
    \item For $N_1=\left(0,\sqrt{\frac{1-\gamma}{m}}\right),$ the values are 
    \begin{align}
        q_{\text{series}}&=\frac{3 (\gamma -1)^2+(3 \gamma -2) m}{2 m}-\frac{3 (\gamma -1) \cos (2
   (\varphi -t \omega ))}{2 m},\\
   q_{\text{averaged}}&=\frac{3 (\gamma -1)^2+(3 \gamma -2) m}{2 m}.
    \end{align}
     We focus on $ q_{\text{averaged}}.$ It describes acceleration for
     \begin{itemize}
     \item $\gamma =\frac{2}{3}, m<0$ or
         \item $0\leq \gamma <\frac{2}{3}, m<0$ or
         \item $0\leq \gamma <\frac{2}{3}, m>\frac{-3 \gamma ^2+6 \gamma -3}{3 \gamma
   -2}$ or
   \item $\frac{2}{3}<\gamma <1, \frac{-3 \gamma ^2+6 \gamma -3}{3 \gamma
   -2}<m<0$ or
   \item $1<\gamma \leq 2, \frac{-3 \gamma ^2+6 \gamma -3}{3 \gamma -2}<m<0.$
     \end{itemize}
    The Hubble function is 
    \begin{equation}
        H(t)=\frac{2 m}{3 t (\gamma  (\gamma +m-2)+1)},
    \end{equation}
    which is positive for 
    \begin{itemize}
        \item $\gamma =0, m>0$ or 
        \item $0<\gamma \leq 2, m<\frac{-\gamma ^2+2 \gamma -1}{\gamma }$ or
        \item $0<\gamma \leq 2, m>0.$
    \end{itemize}
    Finally, In this case $ q_{\text{averaged}}<0$ and $H(t)>0$ occur for the following conditions
    \begin{itemize}
        \item $\gamma =0, m>\frac{3}{2}$ or
        \item $\gamma =\frac{2}{3}, m<-\frac{1}{6}$ or
        \item $0<\gamma <\frac{2}{3}, m>\frac{-3 \gamma ^2+6 \gamma -3}{3 \gamma -2}$ or
        \item $0<\gamma <\frac{2}{3}, m<\frac{-\gamma ^2+2 \gamma -1}{\gamma }$ or
        \item $1<\gamma \leq 2, \frac{-3 \gamma ^2+6 \gamma -3}{3 \gamma
   -2}<m<\frac{-\gamma ^2+2 \gamma -1}{\gamma }$ or
   \item $\frac{2}{3}<\gamma <1, \frac{-3 \gamma ^2+6 \gamma -3}{3 \gamma
   -2}<m<\frac{-\gamma ^2+2 \gamma -1}{\gamma }.$
    \end{itemize}
\end{enumerate}
We observe that for every equilibrium point, the Hubble function has the form \begin{equation}
    H(t)=C \frac{1}{t}, \quad C\in \mathbb{R},
\end{equation}
where $C:=f(\gamma,\bar{\Gamma})$. These parameters can be tuned to recover behaviour consistent with current observations. For example in Figures \ref{obs-a} and \ref{obs-b} we show that $H(t)$ for points $P_1$ and $P_6$ respectively can match observational values by tuning the parameters $\gamma$ and $\bar{\Gamma}.$ For Figures \ref{obs-a} and \ref{obs-b} we considered $H_0\approx 67.4 km s^{-1} Mpc^{-1}$ \cite{Planck:2018vyg} (\textbf{left plot}) and $H_0\approx 74 km s^{-1} Mpc^{-1}$ \cite{Riess:2021jrx} (\textbf{right plot}) .

\begin{figure}[H]
    \centering
    \includegraphics[scale=0.5]{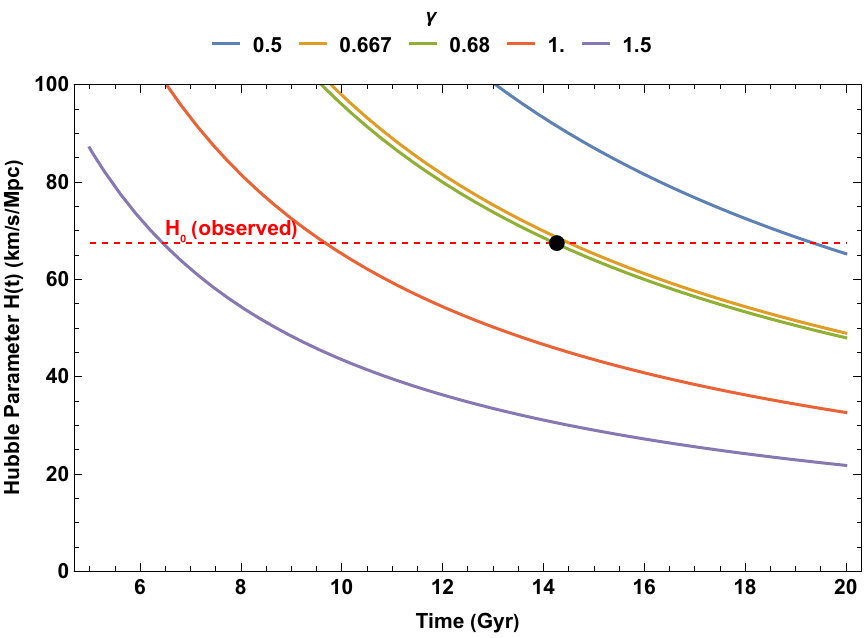}
    \includegraphics[scale=0.5]{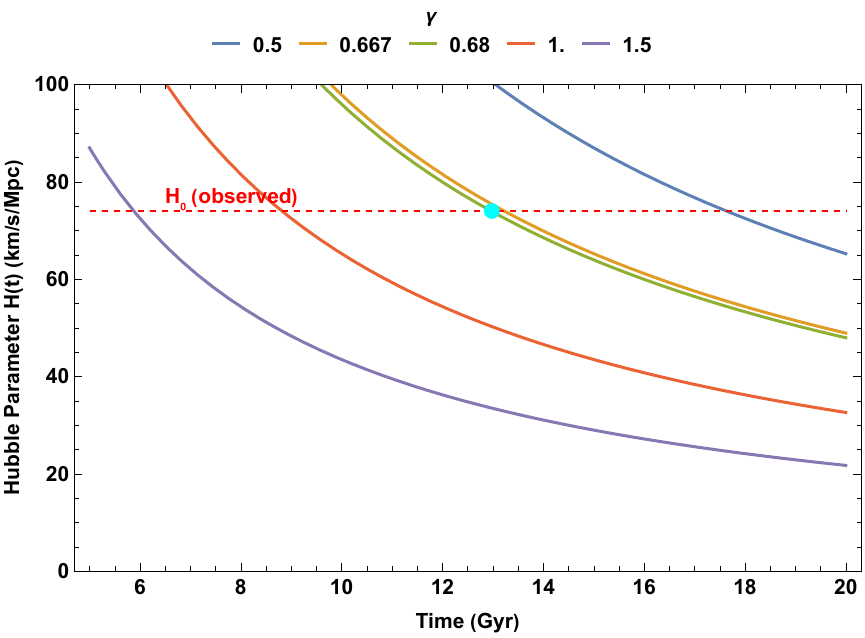}
    \caption{$H(t)=\frac{2}{3 \gamma  t}$ for different values of $\gamma.$ The black point is $(14.27, 67.4)$ and the cyan point is $(12.98, 74).$}
    \label{obs-a}
\end{figure}
\begin{figure}[H]
    \centering
    \includegraphics[scale=0.5]{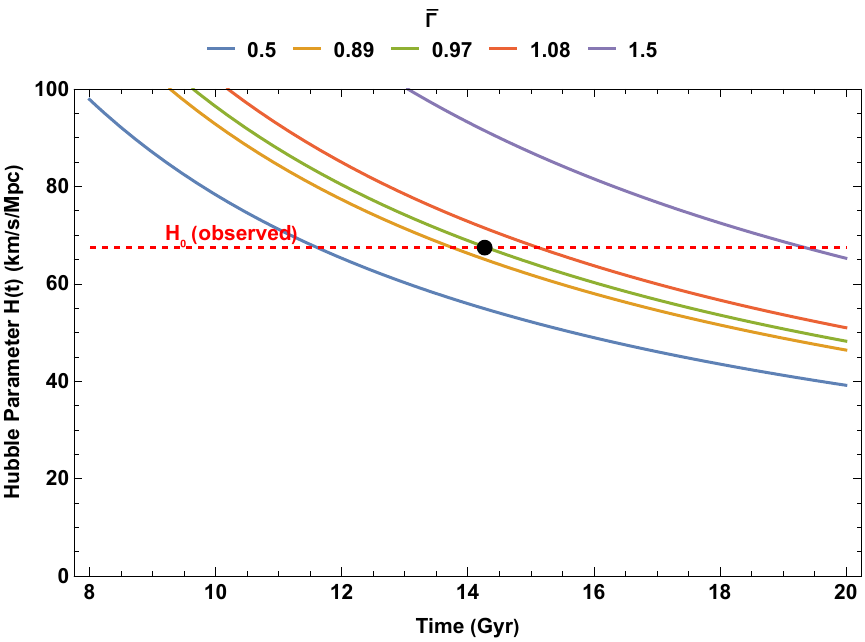}
    \includegraphics[scale=0.5]{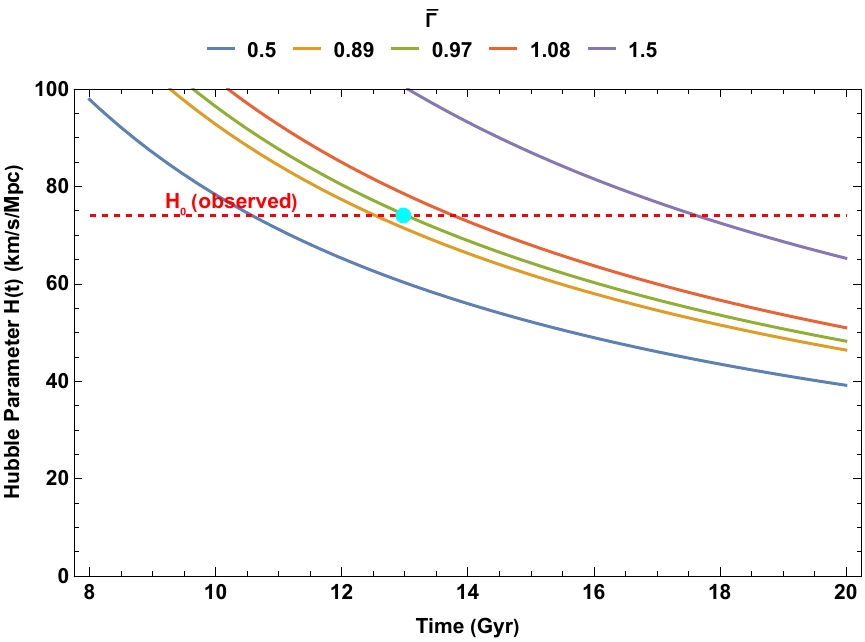}
    \caption{$H(t)=\frac{2}{t \left(3-\bar{\Gamma }\right)}$ for different values of $\bar{\Gamma}.$ The black point is $(14.27, 67.4)$ and the cyan point is $(12.98, 74).$}
    \label{obs-b}
\end{figure}
\section{Numerical simulations}
\label{int:numerics}
In this section, we present the numerical solutions of both the \textit{full} and \textit{averaged} systems corresponding to each of the interaction models under consideration. The computations were performed using \textit{Wolfram Mathematica} \cite{Mathematica}, under a student license. The numerical integration employed the \enquote{StiffnessSwitching} method, supplemented by the \enquote{ExplicitRungeKutta} sub-method. The initial conditions used in the simulations are summarized in Table~\ref{Tab-final}.
    \begin{table}[H]
  \centering    
    \begin{tabular}{|c|c|c|c|c|c|}\hline
   Sol.  &  {$H(0)$} &  {${\bar{\Sigma}(0)}$} &  {$\bar{\Omega}(0)$} &   {$\bar{\Phi}(0)$}  &  {$t(0)$}  \\\hline
              I &  $0.1$ & $0.5$ & $0.5$ & $0$ & $0$\\
        II &  $0.1$ & $0.1$ & $0.9$   & $0$ & $0$\\
        III &  $0.1$ & $0.1$ & $0.1$ & $0$ & $0$\\
        IV &  $0.1$ & $0.9$ & $0.1$ & $0$ & $0$\\
        V &  $0.1$ & $0.5$ & $0.1$ & $0$ & $0$\\
        II$^{*}$& $0.1$&$-0.1$ &$0.35$& $0$& 0\\\hline
    \end{tabular} \caption{\label{Tab-final} Six initial conditions were used for the numerical solution of systems \eqref{gen-redu-1}--\eqref{gen-redu-4} and \eqref{BI-prom-1}--\eqref{BI-prom-4}. For interactions 1, 2, 3, 6, 7, 8 and 9 (\eqref{int-1}, \eqref{int-2}, \eqref{int-3}, \eqref{int-6}, \eqref{int-7}, \eqref{int-8}, and \eqref{int-9}), initial conditions I--V were employed. For interactions 4 \eqref{int-4} and 5 \eqref{int-5}, condition II was replaced by II$^{*}$.}
\end{table}
\subsection{Evolution of $q_{\text{series}}$ and $q_{\text{averaged}}$}
We evaluate the two expressions of the deceleration parameter, given by equations \eqref{qseries} and \eqref{qaver}, using the numerical solutions of both the original and the averaged systems. We illustrate this with representative examples in which the deceleration parameter exhibits a transition from a decelerated expansion phase to an accelerated one.

\begin{figure}[H]
    \centering
    \includegraphics[scale=0.35]{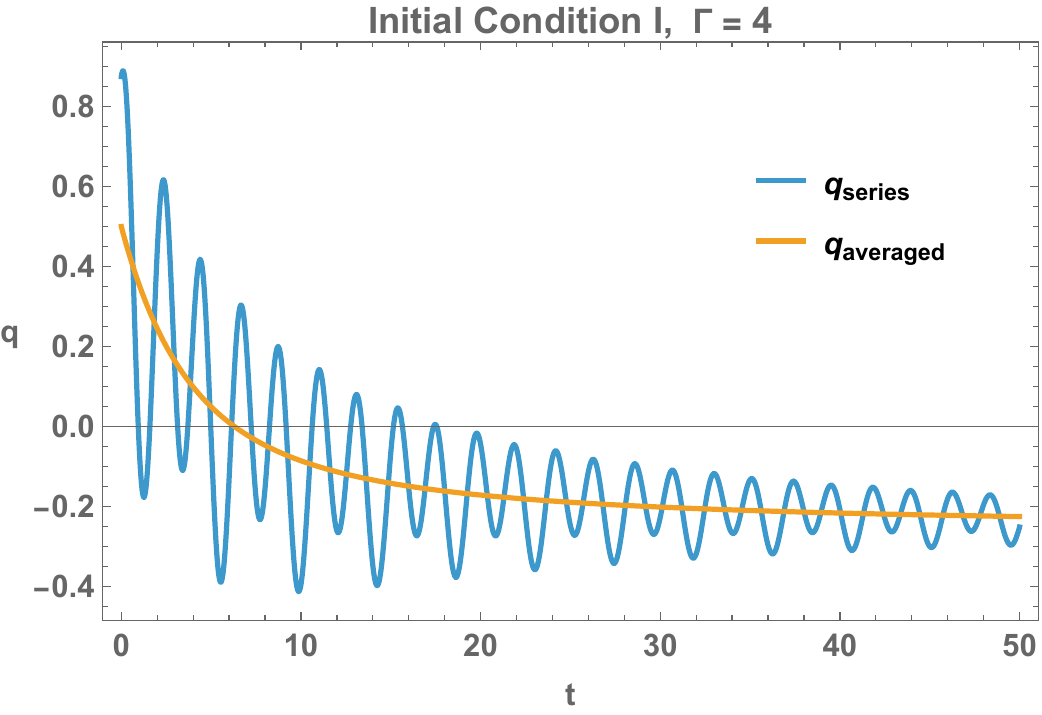}
    \includegraphics[scale=0.35]{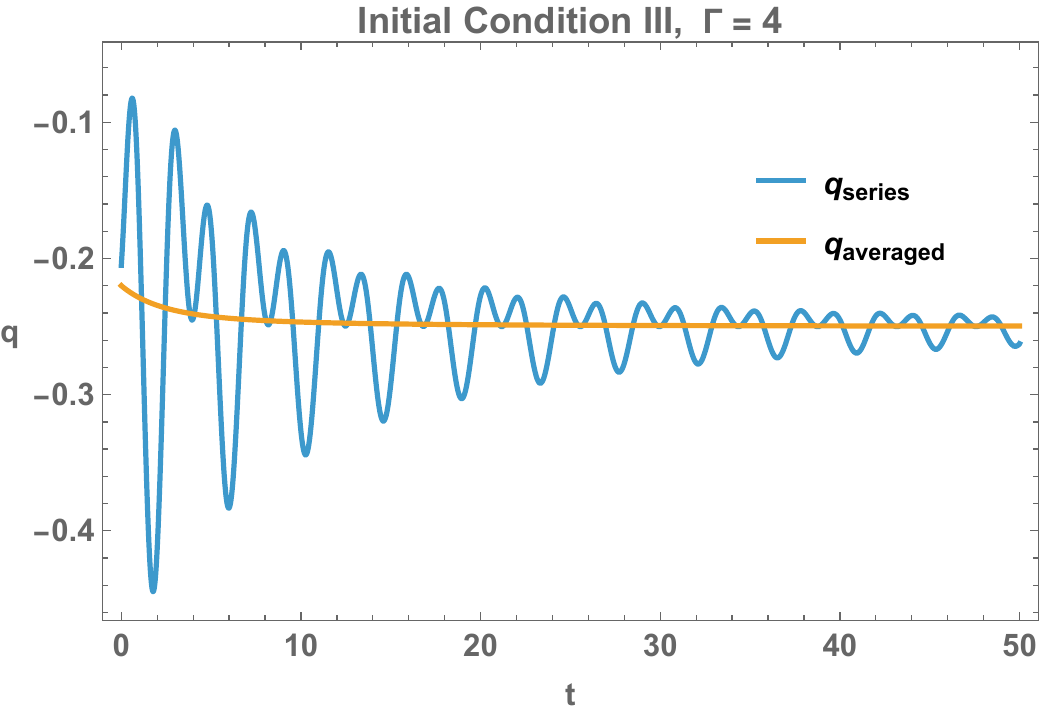}
    \includegraphics[scale=0.35]{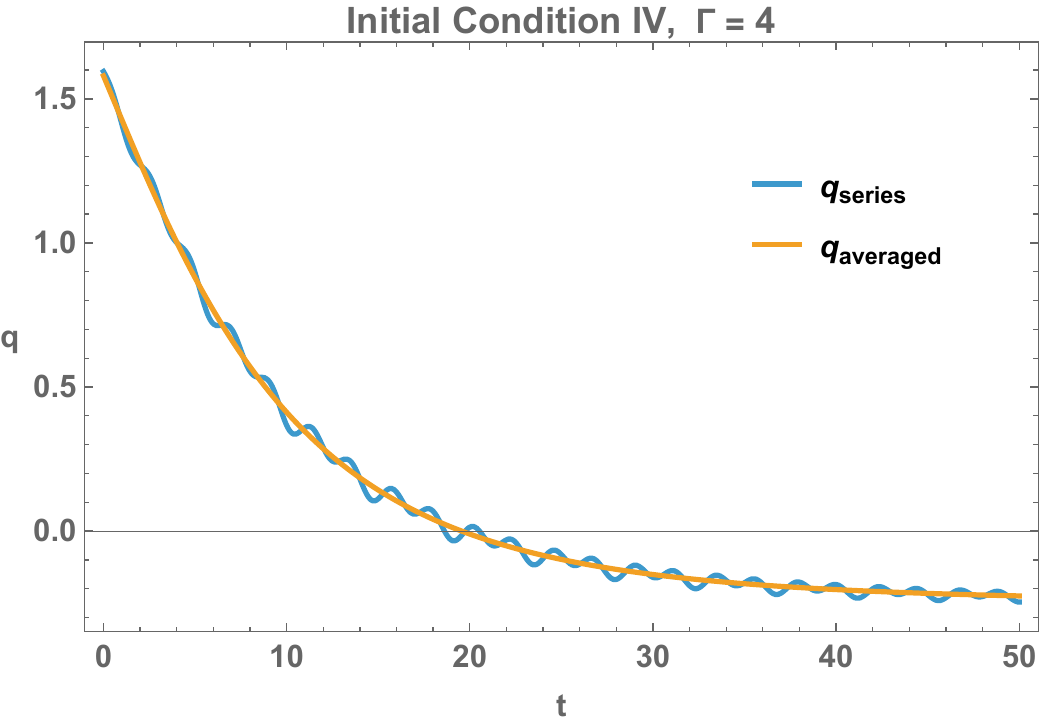}
    \includegraphics[scale=0.35]{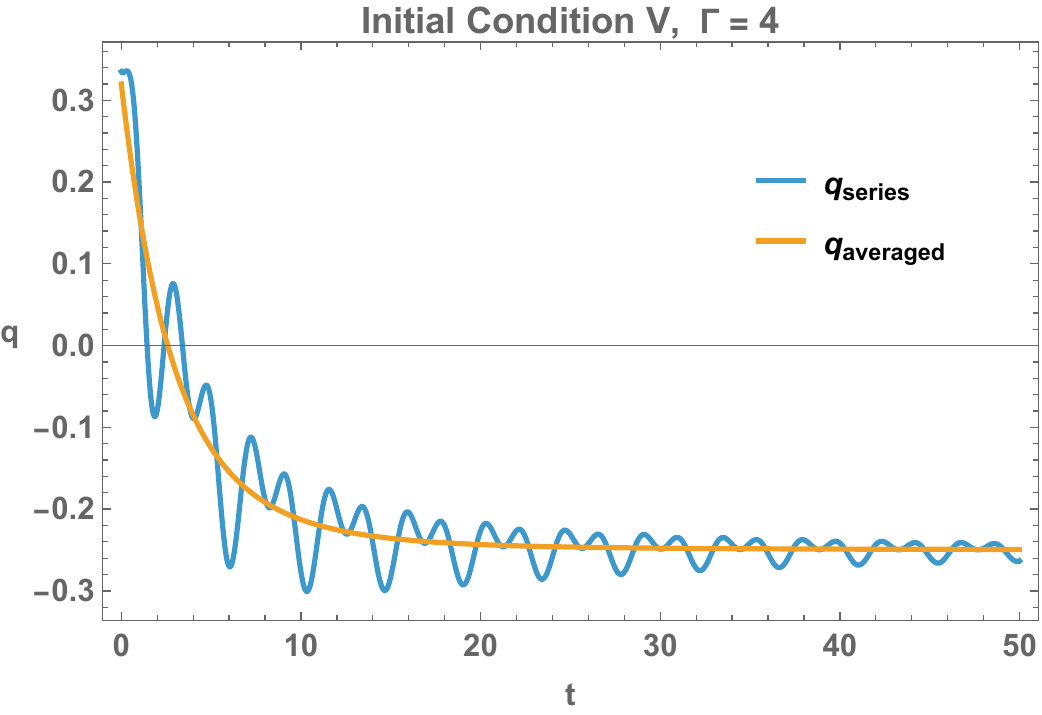}
    \caption{\textbf{Interaction 1 \eqref{int-1}}. Evolution of $q_{\text{series}}$ and $q_{\text{averaged}}$ for some of the initial conditions presented in Table \ref{Tab-final} for $\Gamma=4$. We fixed the value $\gamma=\frac{1}{2}$ to display the transition behaviour from a decelerated epoch to an accelerated one.}
    \label{evo-1-q-a}
\end{figure}

\begin{figure}[H]
    \centering
    \includegraphics[scale=0.35]{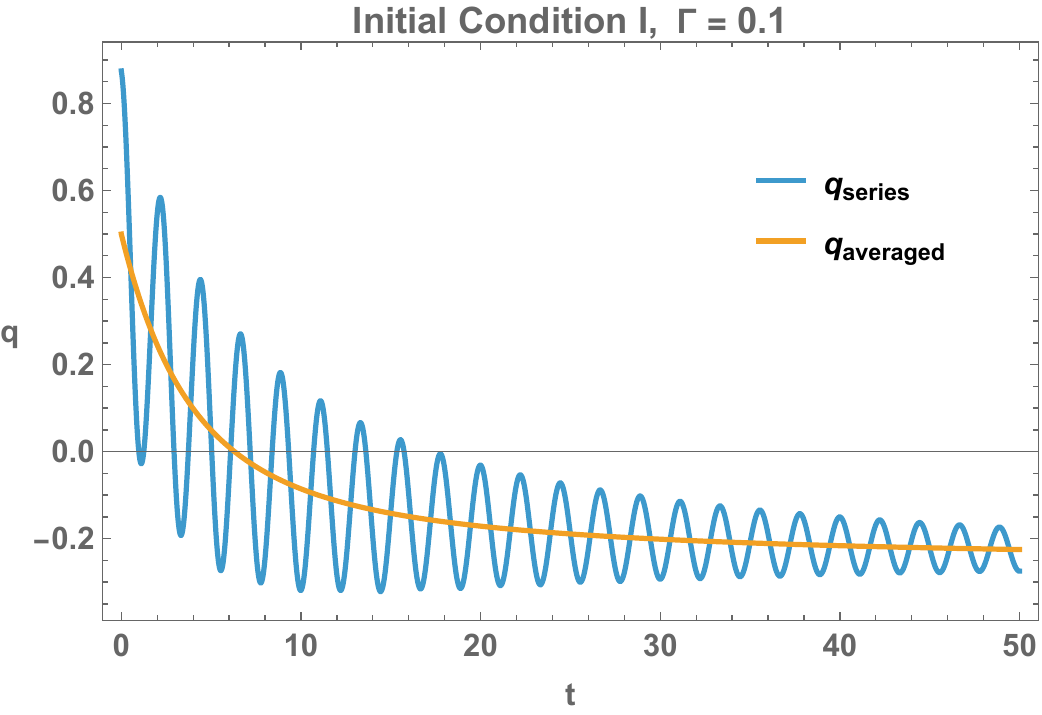}
    \includegraphics[scale=0.35]{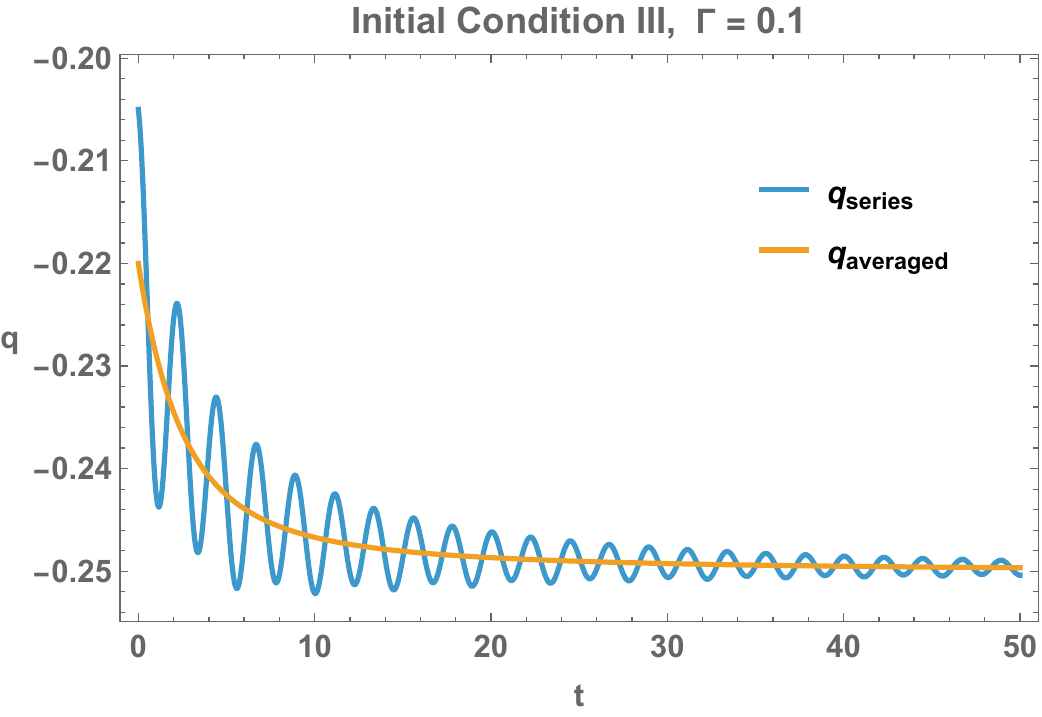}
    \includegraphics[scale=0.35]{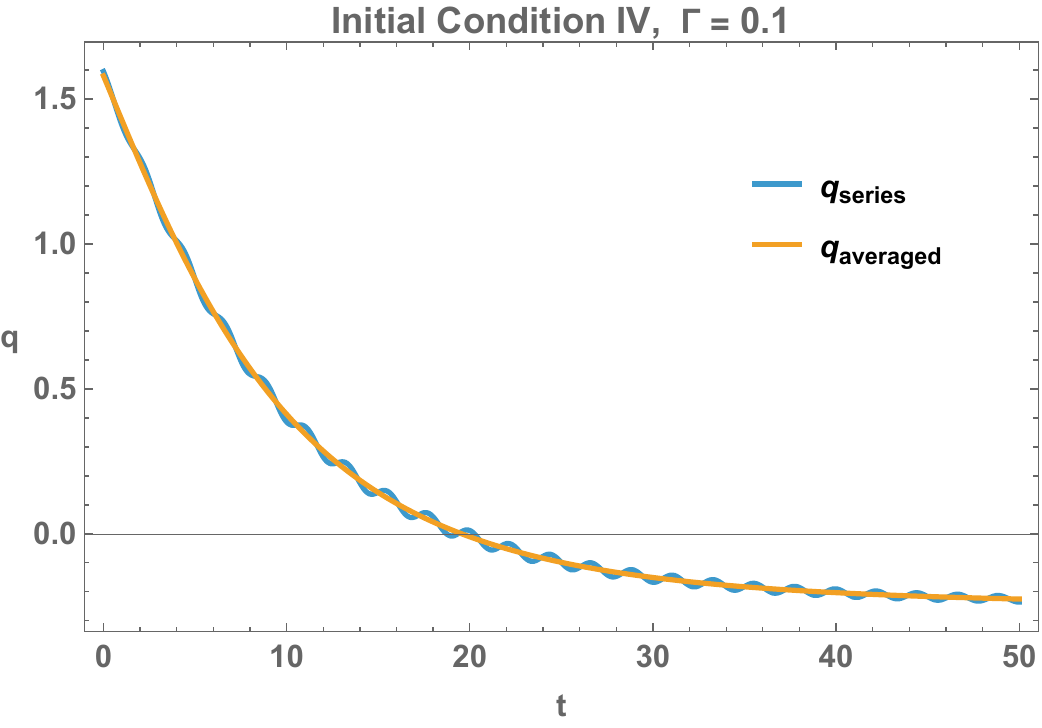}
    \includegraphics[scale=0.35]{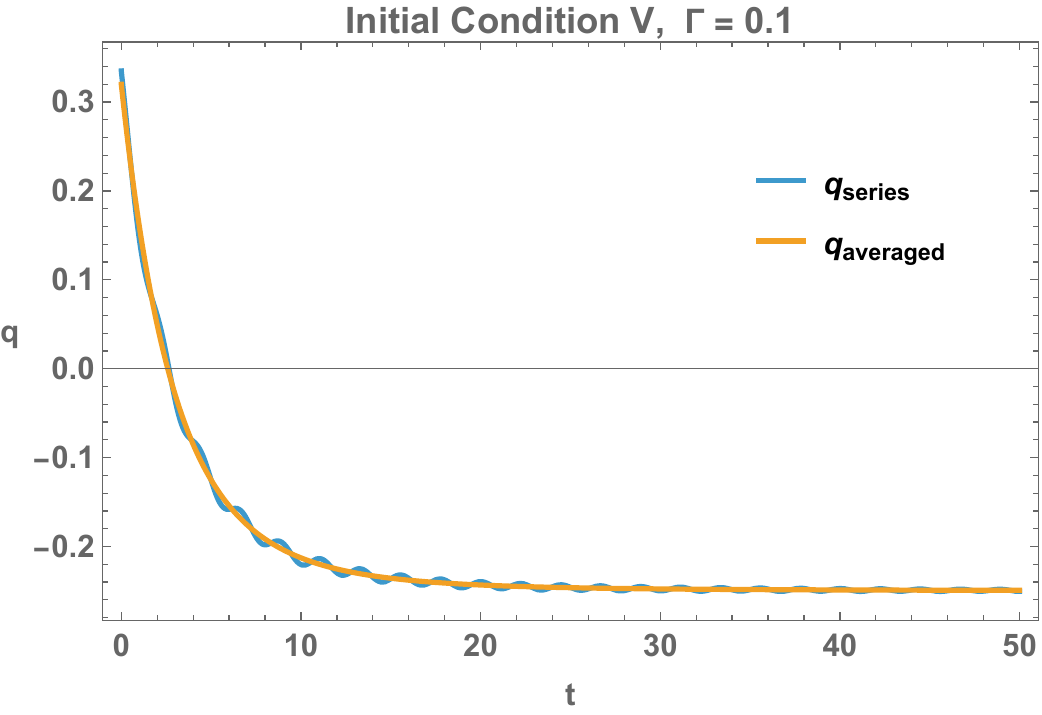}
    \caption{\textbf{Interaction 1 \eqref{int-1}}. Evolution of $q_{\text{series}}$ and $q_{\text{averaged}}$ for some of the initial conditions presented in Table \ref{Tab-final} for $\Gamma=0.1$. We fixed the value $\gamma=\frac{1}{2}$ to display the transition behaviour from a decelerated epoch to an accelerated one.}
    \label{evo-1-q-b}
\end{figure}

For instance, Figures~\ref{evo-1-q-a} and~\ref{evo-1-q-b} show the evolution of both $q_{\text{series}}$ and $q_{\text{averaged}}$ for Interaction 1 \eqref{int-1}, considering different values of the parameter $\Gamma$, while fixing $\gamma = \frac{1}{2} \leq \frac{2}{3}$ to ensure an accelerating expansion.

\begin{figure}[H]
    \centering
    \includegraphics[scale=0.4]{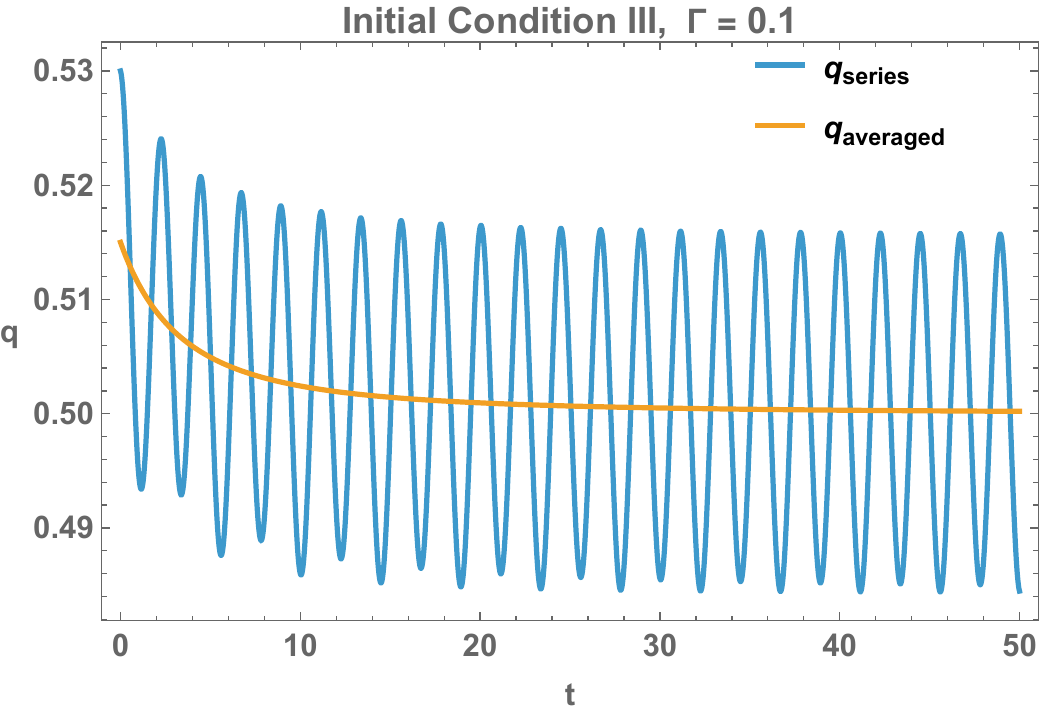}
        \includegraphics[scale=0.4]{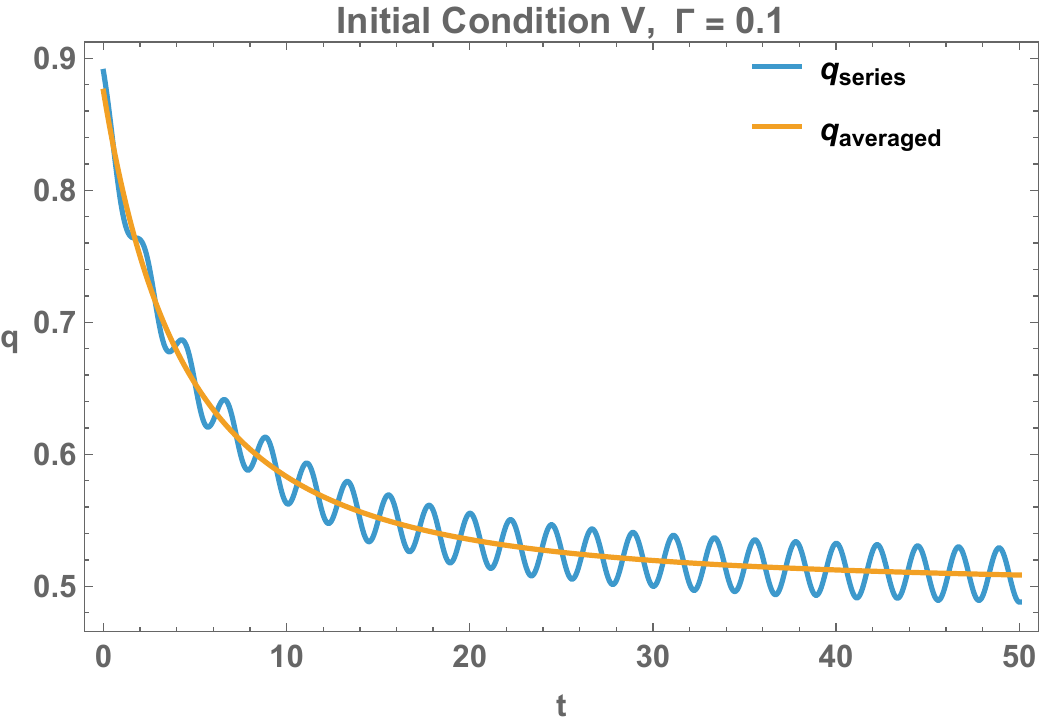}

    \caption{\textbf{Interaction 1 \eqref{int-1}}. Evolution of $q_{\text{series}}$ and $q_{\text{averaged}}$ for some of the initial conditions presented in Table \ref{Tab-final} for $\Gamma=0.1$. We fixed the value $\gamma=1$, this describes a decelerating epoch.}
    \label{evo-1-q-c}
\end{figure}

Additionally, in Figure \ref{evo-1-q-c}, still considering Interaction 1 \eqref{int-1}, we fixed the equation of state parameter value $\gamma=1,$ in this case, the evolution of $q_{\text{series}}$ and $q_{\text{averaged}}$ stays above the positive axis describing deceleration.

\begin{figure}[H]
    \centering
    \includegraphics[scale=0.35]{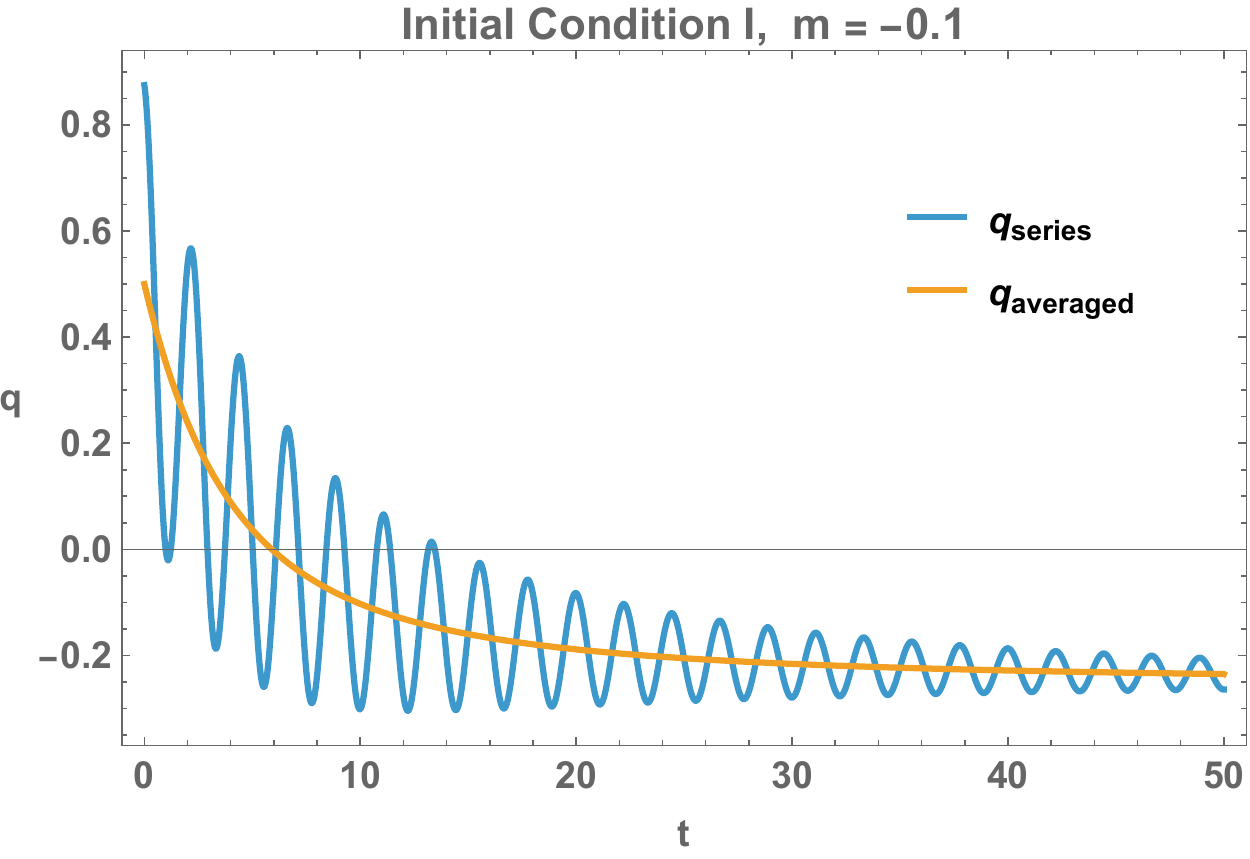}
    \includegraphics[scale=0.35]{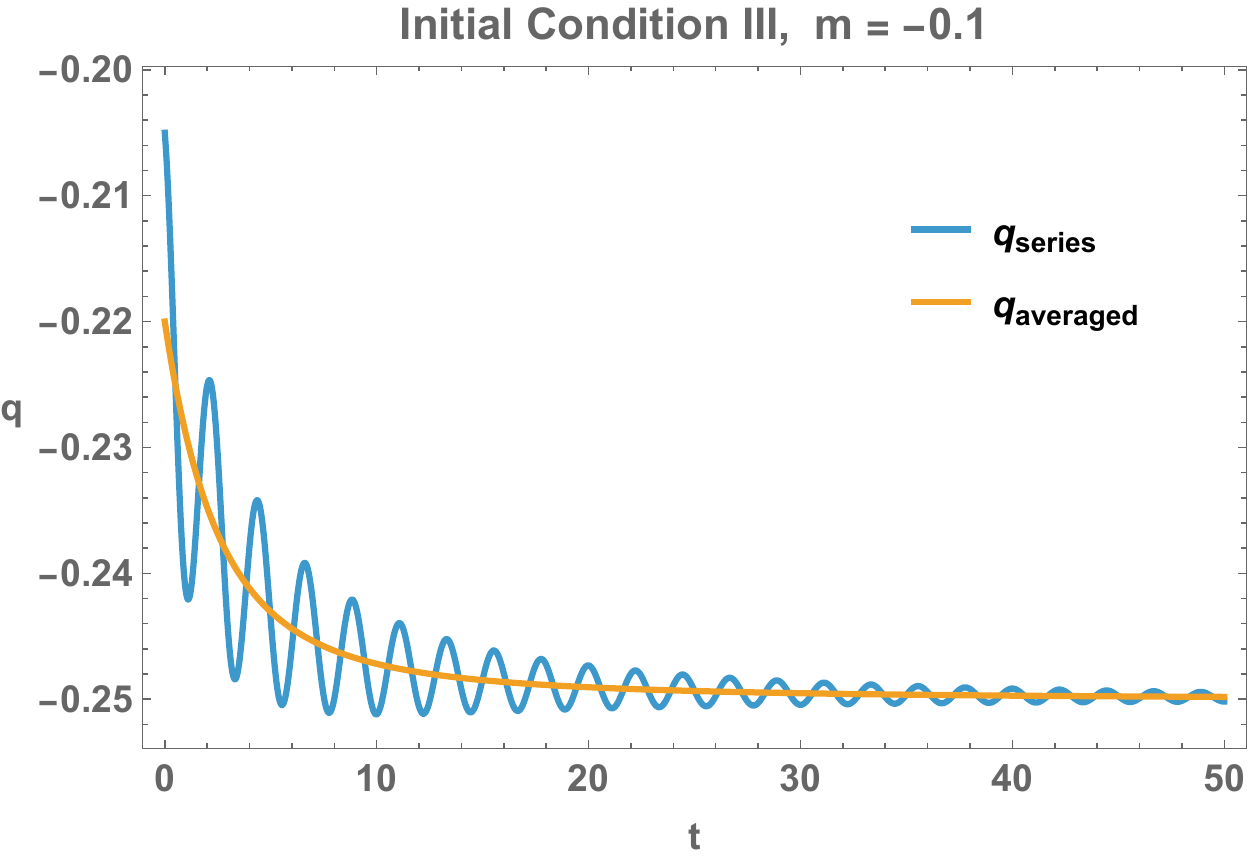}
    \includegraphics[scale=0.35]{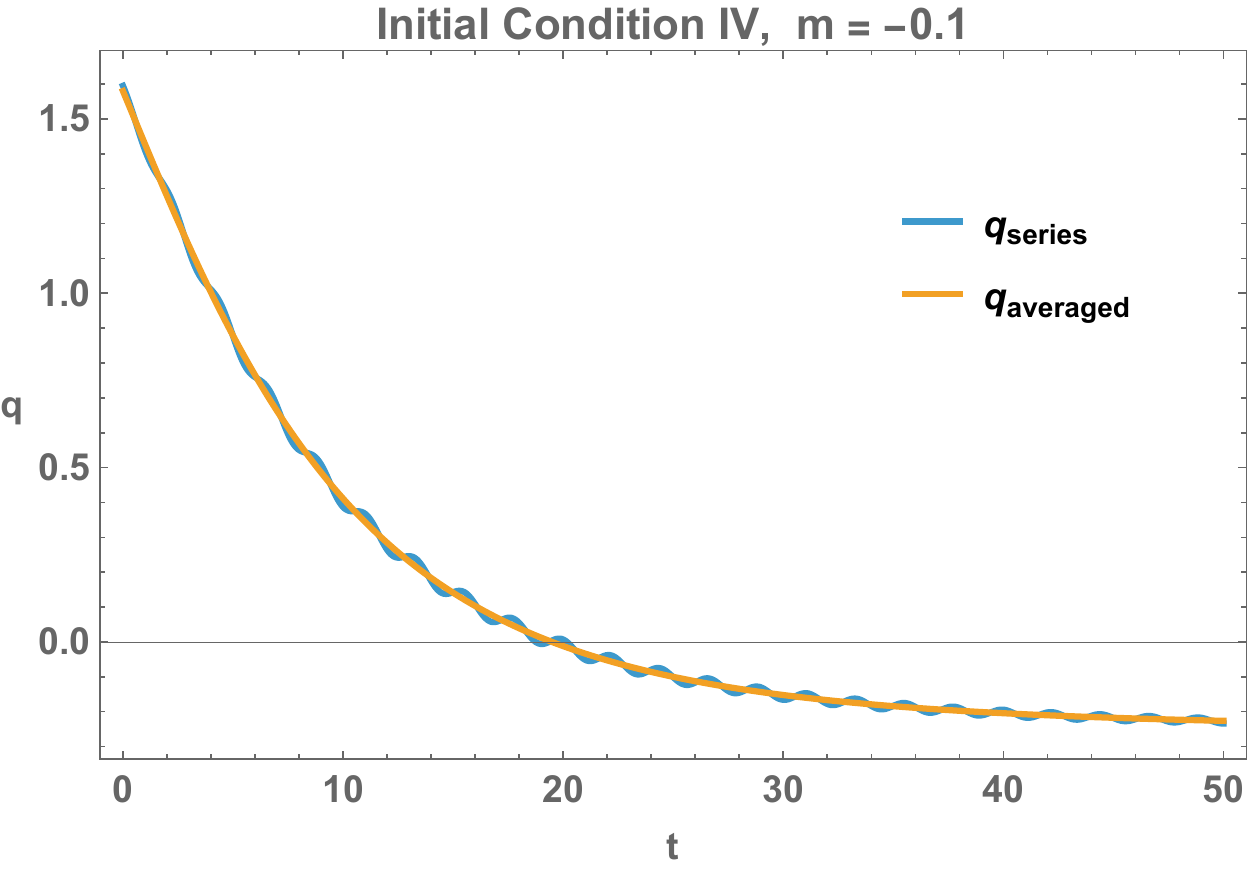}
    \includegraphics[scale=0.35]{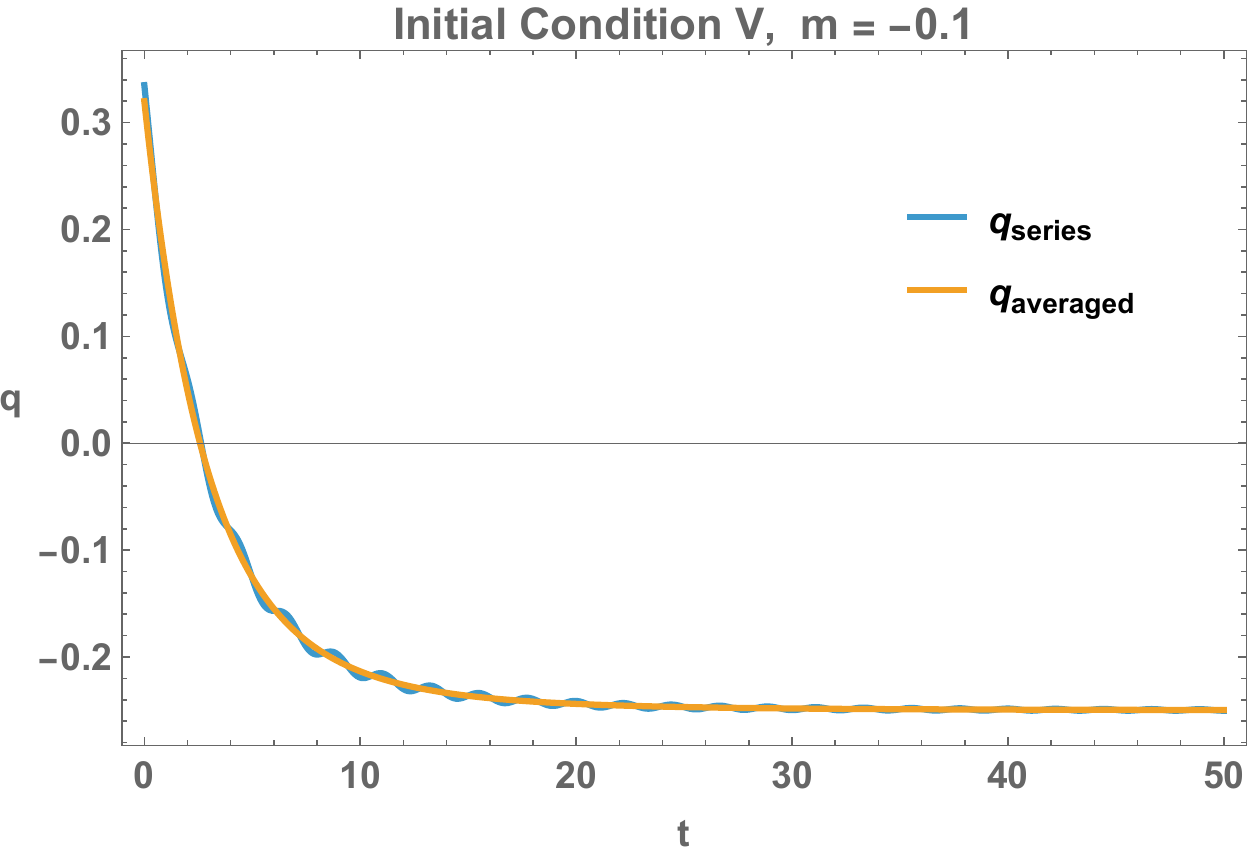}
    \caption{\textbf{Interaction 7 \eqref{int-7}}. Evolution of $q_{\text{series}}$ and $q_{\text{averaged}}$ for some of the initial conditions presented in Table \ref{Tab-final} for $m=-0.1$. We fixed the value $\gamma=\frac{1}{2}$ to display the transition behaviour from a decelerated epoch to an accelerated one.}
    \label{evo-7-q-a}
\end{figure}

\begin{figure}[H]
    \centering
    \includegraphics[scale=0.35]{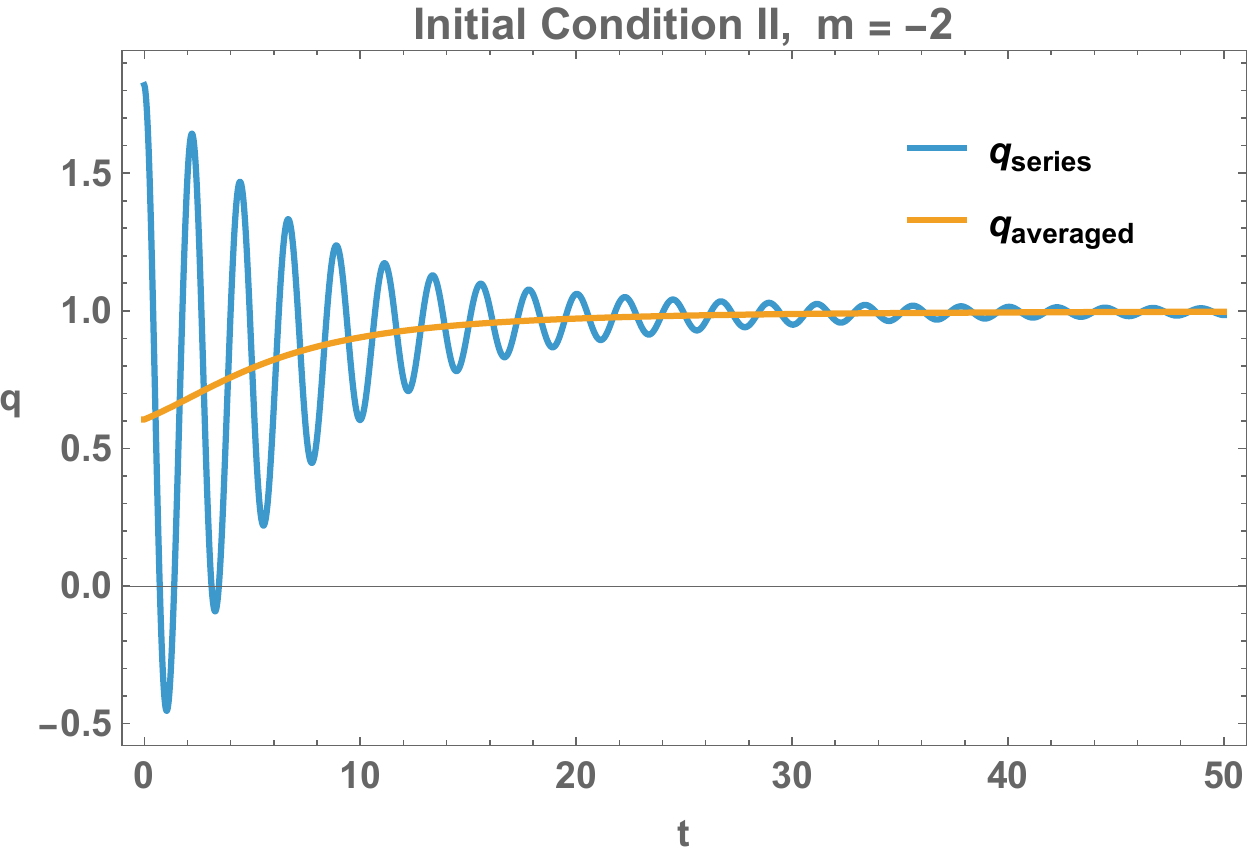}
    \includegraphics[scale=0.35]{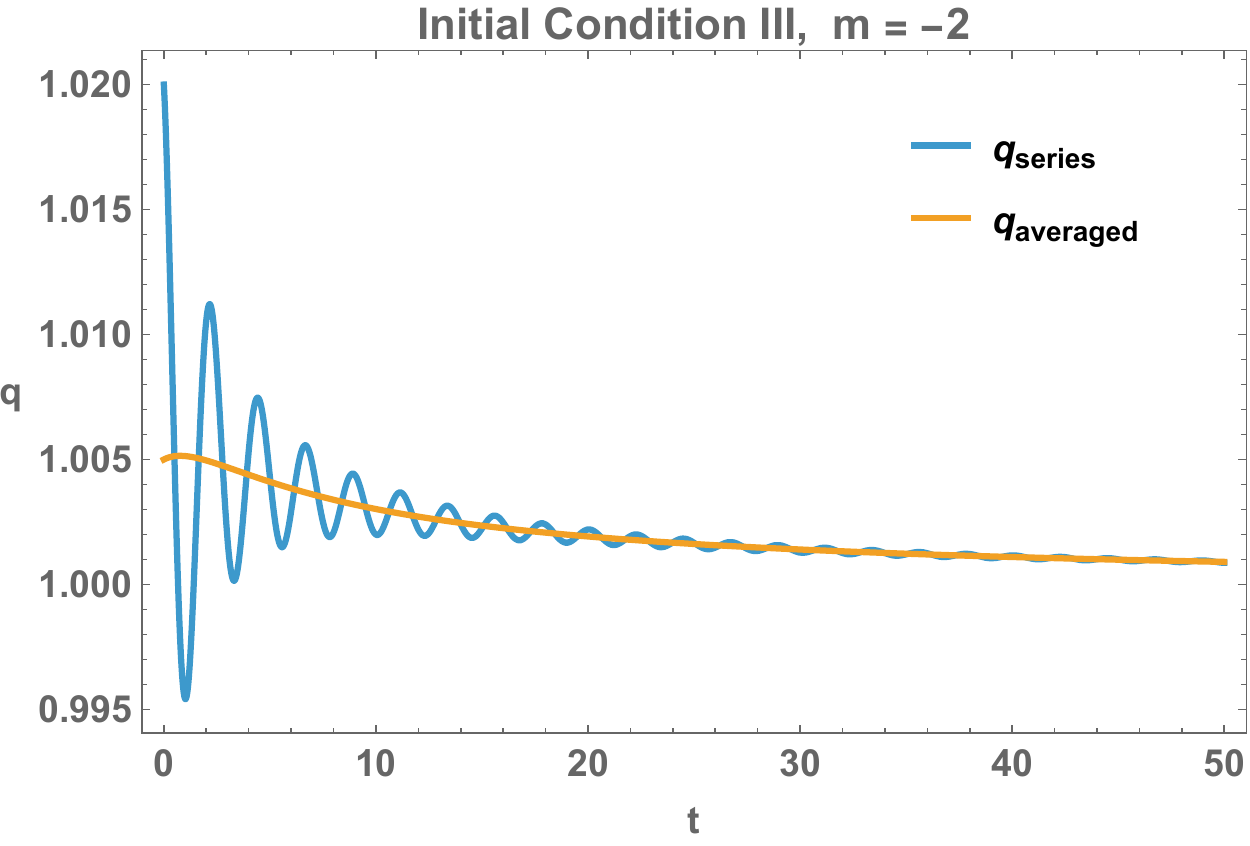}
    \includegraphics[scale=0.35]{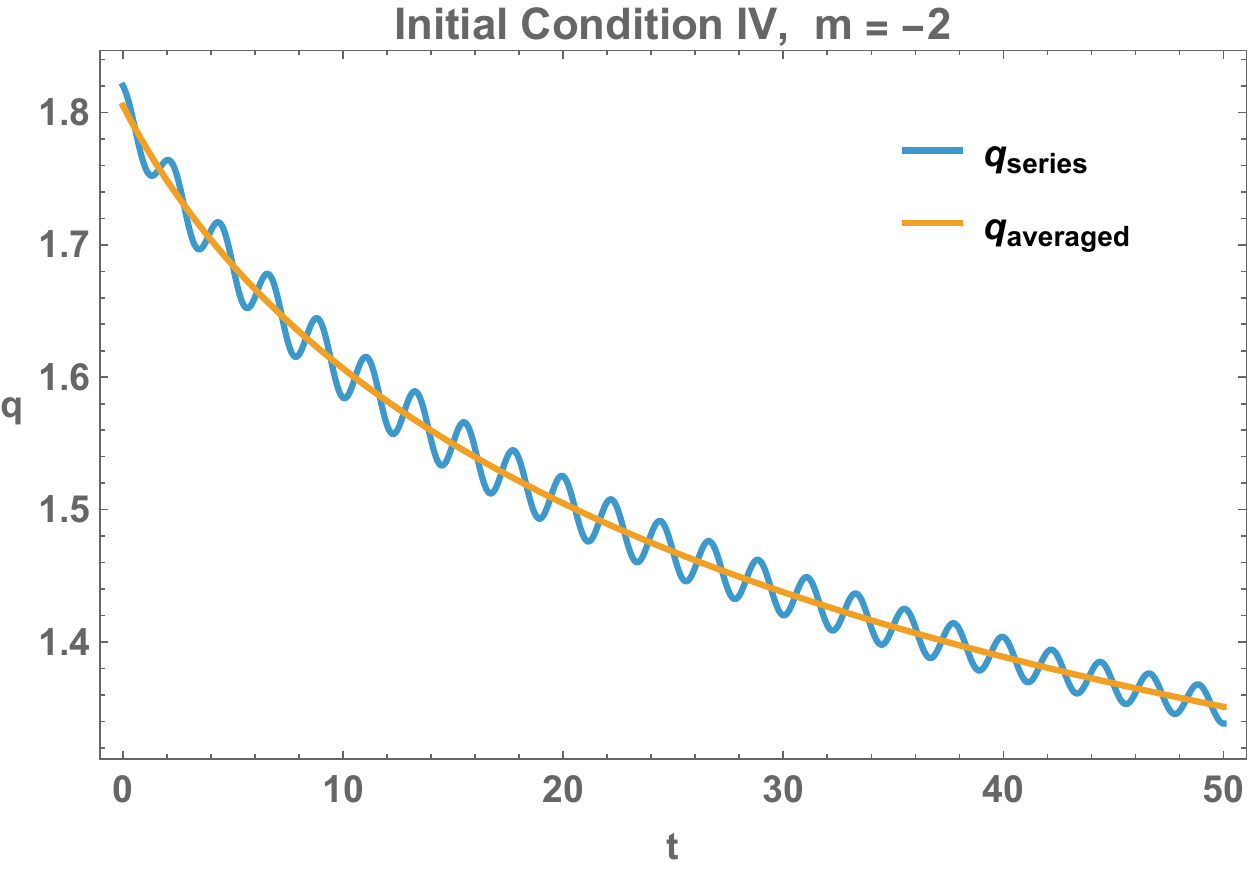}
    \includegraphics[scale=0.35]{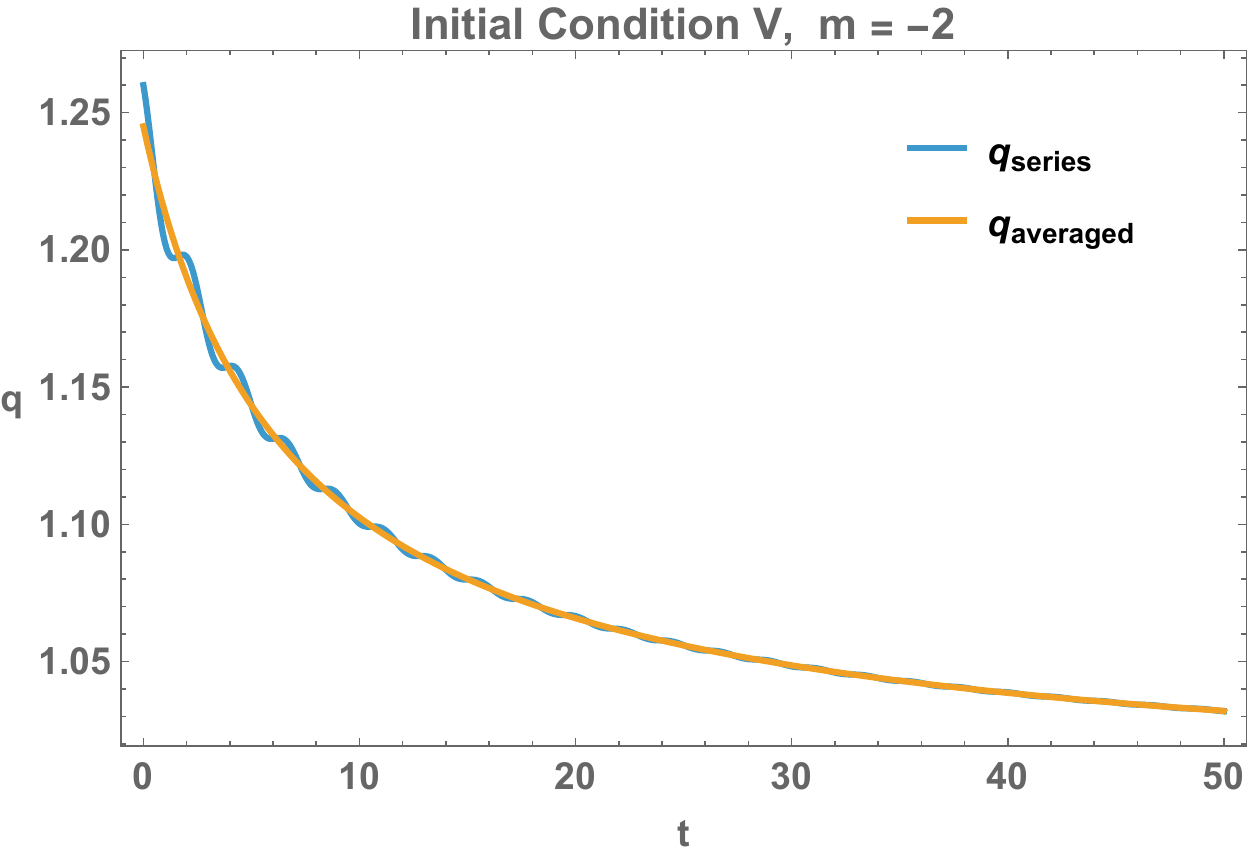}
    \caption{\textbf{Interaction 7 \eqref{int-7}}. Evolution of $q_{\text{series}}$ and $q_{\text{averaged}}$ for some of the initial conditions presented in Table \ref{Tab-final} for $m=-2$. We fixed the value $\gamma=\frac{4}{3}$, this describes a decelerating epoch.}
    \label{evo-7-q-b}
\end{figure}

In Figures \ref{evo-7-q-a} and \ref{evo-7-q-b} we considered Interaction 7 \eqref{int-7} and different values of the parameters involved. In particular, for Figure \ref{evo-7-q-a} we considered again the value $\gamma=\frac{1}{2}$ that produces a transition from deceleration to acceleration. In Figure \ref{evo-7-q-b}, we set $\gamma=\frac{4}{3},$ the evolution of the averaged deceleration parameter describes a decelerated regime as it does not cross the horizontal axis.

\begin{figure}[H]
    \centering
    \includegraphics[scale=0.35]{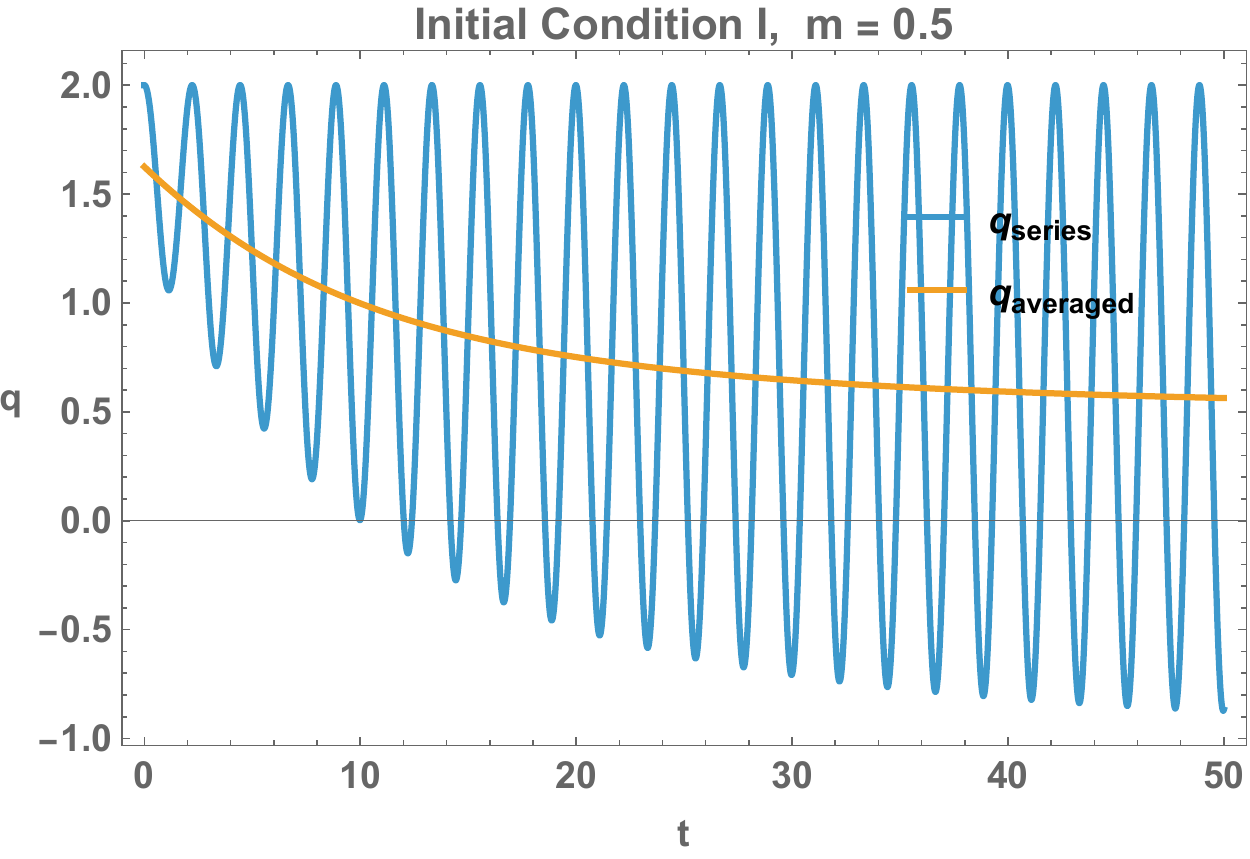}
        \includegraphics[scale=0.35]{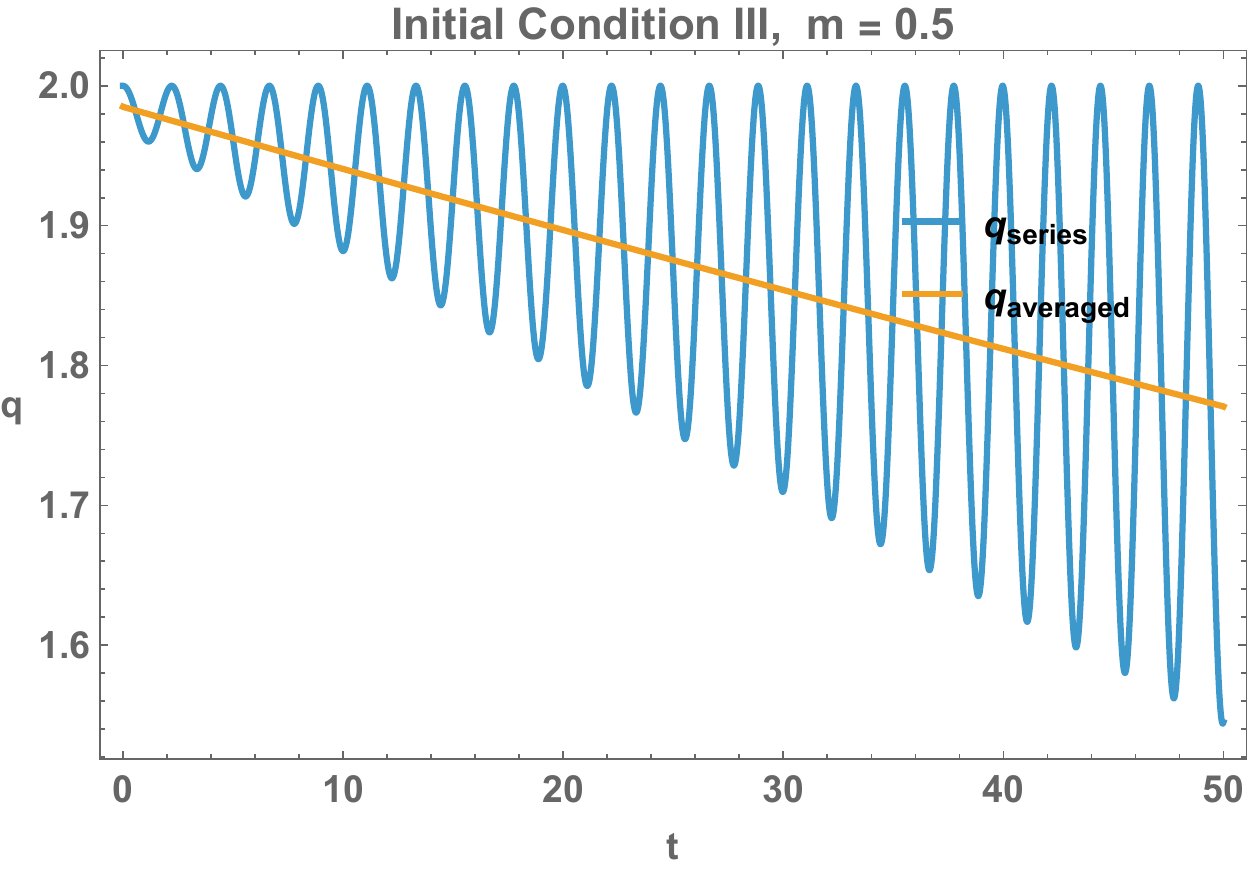}

    \caption{\textbf{Interaction 9 \eqref{int-9}}. Evolution of $q_{\text{series}}$ and $q_{\text{averaged}}$ for some of the initial conditions presented in Table \ref{Tab-final} for $m=0.5$. We fixed the value $\gamma=2$, this describes a decelerating epoch.}
    \label{evo-9-q-a}
\end{figure}

Finally, in Figure \ref{evo-9-q-a} we considered Interaction 9 \eqref{int-9} with the fixed value $\gamma=2.$ The value of the deceleration parameter stars at $q=2$ but $q_{\text{averaged}}$ converges slowly to $\frac{1}{2}.$

\FloatBarrier
\subsection{Numerical integration of original and averaged systems}
We continue our analysis by presenting three dimensional projections of the solutions of the original averaged systems, in order to evaluate our analytic approach.

\begin{figure}[H]
    \centering
    \includegraphics[scale=0.6]{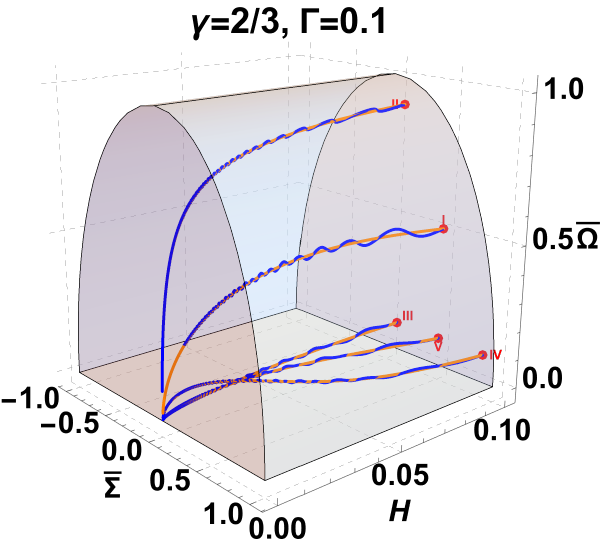}
    \includegraphics[scale=0.6]{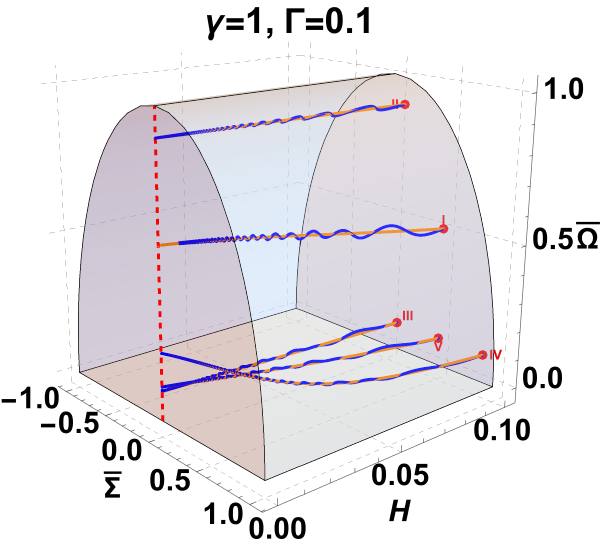}
\caption{\textbf{Interaction 1 \eqref{int-1}}. Three dimensional projections of the numerical solutions of the original system \eqref{comp-int-1-a}--\eqref{comp-int-1-d} (blue) and the averaged system \eqref{BI-prom-1}--\eqref{BI-prom-4} (orange) for $\gamma=2/3, 1$. The red dashed line is the attracting set for $\gamma=1$.}
\label{fig:petitA}
\end{figure}

\begin{figure}[H]
\centering
    \includegraphics[scale=0.7]{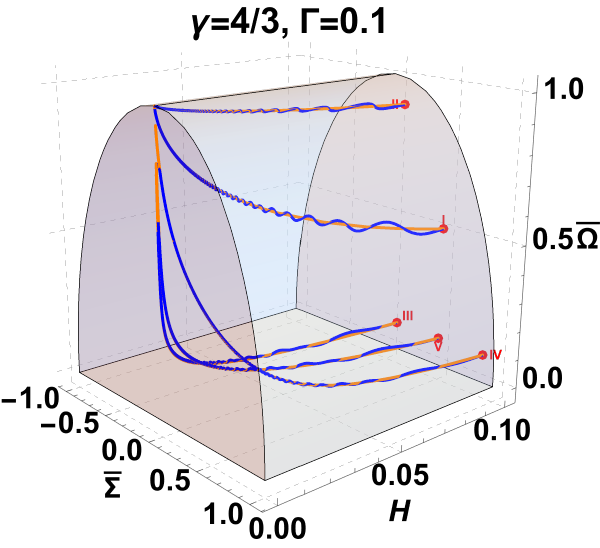}
    \includegraphics[scale=0.7]{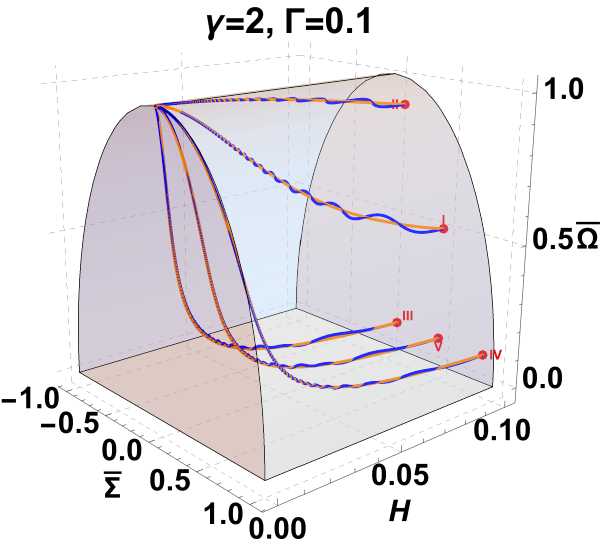}
    
    \caption{\textbf{Interaction 1 \eqref{int-1}}. Figure \ref{fig:petitA} continued for $\gamma=4/3,2$.}
    \label{fig:numéricas BI-Interaction-1}
\end{figure}

Figures \ref{fig:petitA} and \ref{fig:numéricas BI-Interaction-1} show that the numerical solutions of the original system \eqref{gen-redu-1}--\eqref{gen-redu-4}, shown in blue, and those of the averaged system \eqref{BI-prom-1}--\eqref{BI-prom-4}, shown in orange, exhibit equivalent asymptotic behavior in the presence of Interaction 1 \eqref{int-1}. As previously presented in \cite{Leon:2021rcx} for Bianchi I models without interaction, the averaged solutions suppress the oscillations present in the original system, thereby facilitating the analysis and study of the guiding system compared to the initial model.

\begin{figure}[H]
    \centering
    \includegraphics[scale=0.7]{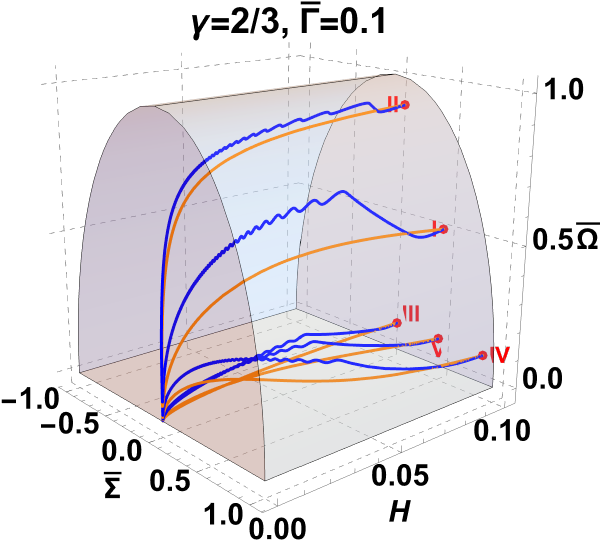}
    \includegraphics[scale=0.7]{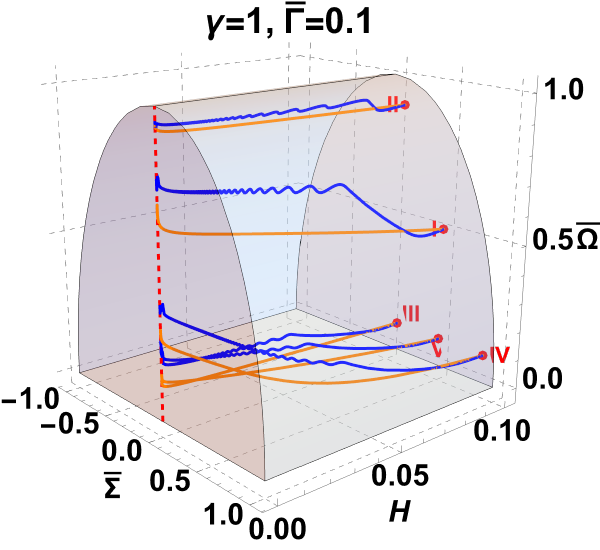}
    
\caption{\textbf{Interaction 2 \eqref{int-2}}. Three dimensional projections of the numerical solutions of the original system \eqref{comp-int-2-a}--\eqref{comp-int-2-d} (blue) and the averaged system \eqref{1-prom-int2}-\eqref{4-prom-int2} (orange) for $\gamma=2/3, 1$. The red dashed line is the attracting set for $\gamma=1$.}
\label{fig:petitB}
\end{figure}

\begin{figure}[H]
\centering
    \includegraphics[scale=0.7]{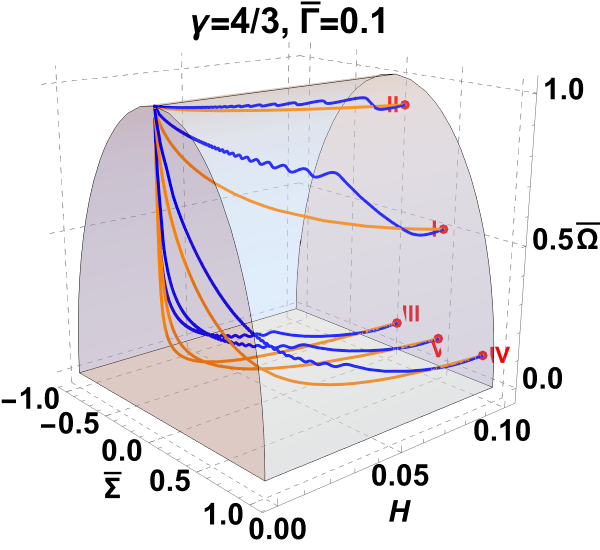}
    \includegraphics[scale=0.7]{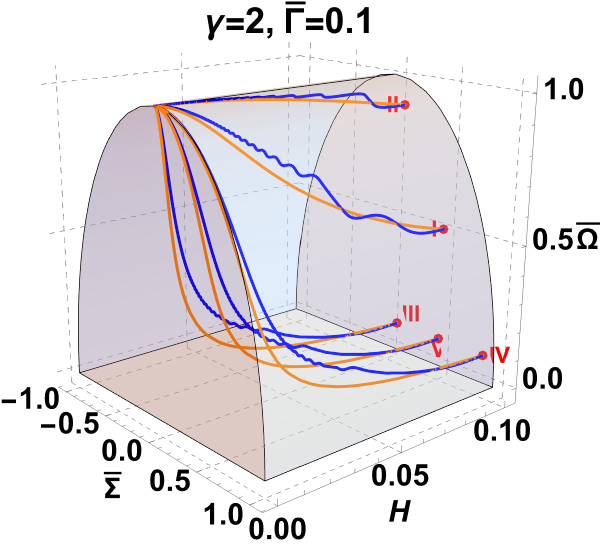}
    \caption{\textbf{Interaction 2 \eqref{int-1}}. Figure \ref{fig:petitB} continued for $\gamma=4/3,2$.}\label{fig:numéricas BI-Interaction-2}
\end{figure}

In Figures \ref{fig:petitB} and \ref{fig:numéricas BI-Interaction-2}, the numerical solutions of systems \eqref{comp-int-2-a}-\eqref{comp-int-2-d} and \eqref{1-prom-int2}-\eqref{4-prom-int2} for Interaction 2 \eqref{int-2}, exhibit initial discrepancies due to the tangent function in the equations of the original system, but both share the same late-time asymptotic behaviour, highlighting the usefulness of the averaged system.

\begin{figure}[H]
    \centering
    \includegraphics[scale=0.7]{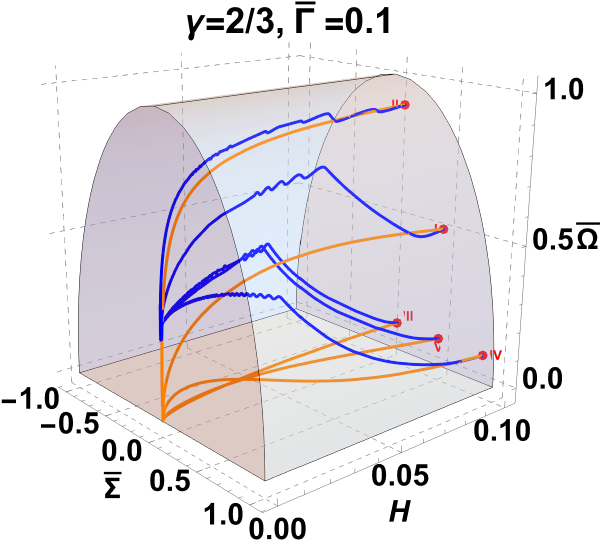}
    \includegraphics[scale=0.7]{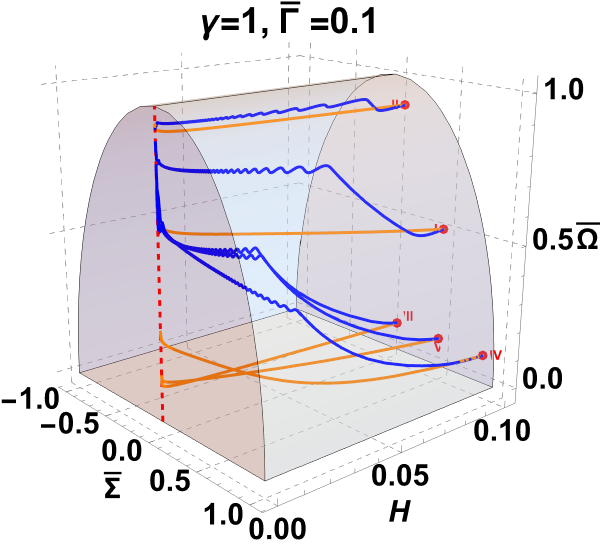}
        
\caption{\textbf{Interaction 3 \eqref{int-3}}. Three dimensional projections of the numerical solutions of the original system \eqref{comp-int-3-a}--\eqref{comp-int-3-d} (blue) and the averaged system \eqref{1-prom-int3}--\eqref{4-prom-int3} (orange). The red dashed line represents the sink set for the case $\gamma = 1$.}
\label{fig:petitC}
\end{figure}

\begin{figure}[H]
\centering
    \includegraphics[scale=0.7]{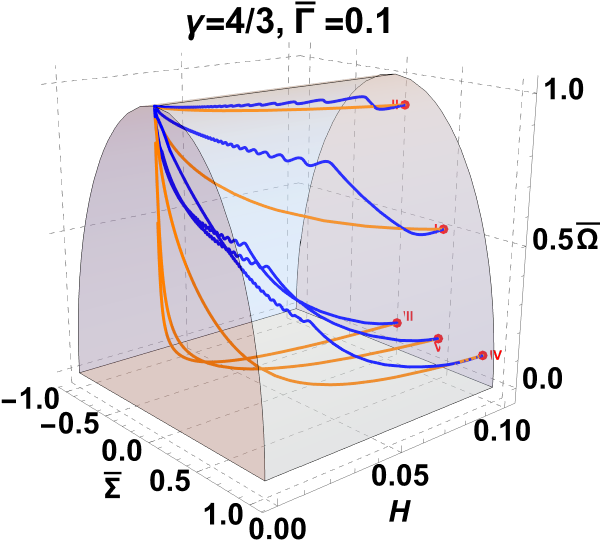}
    \includegraphics[scale=0.7]{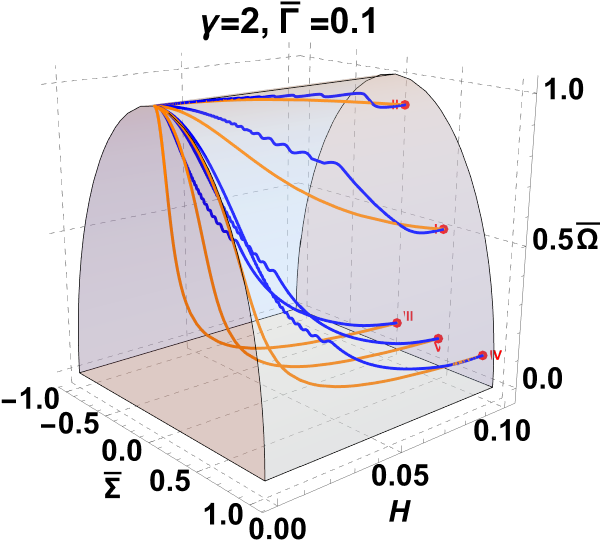}
    \caption{\textbf{Interaction 3 \eqref{int-3}}. Figure \ref{fig:petitC} continued for $\gamma=4/3,2$.}\label{fig:numéricas BI-Interaction-3}
\end{figure}

In Figures \ref{fig:petitC} and \ref{fig:numéricas BI-Interaction-3} we considered Interaction 3 \eqref{int-3}. An initial jump is observed due to the tangent function in the equations of the full system \eqref{comp-int-3-a}--\eqref{comp-int-3-d}, but both solutions tend toward the same attractor.

\begin{figure}[H]
    \centering
    \includegraphics[scale=0.7]{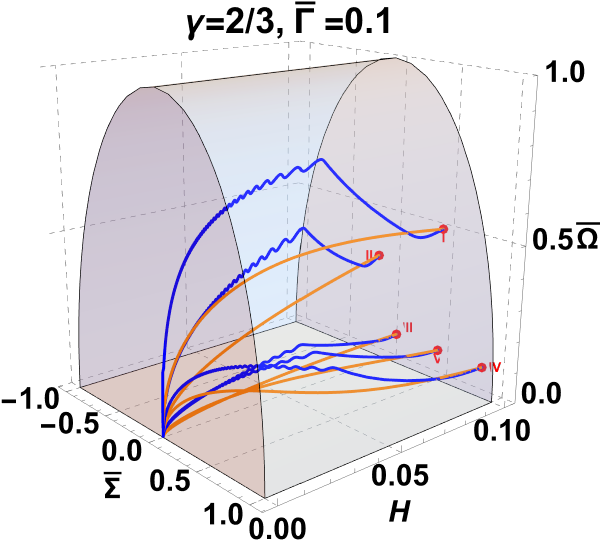}
    \includegraphics[scale=0.7]{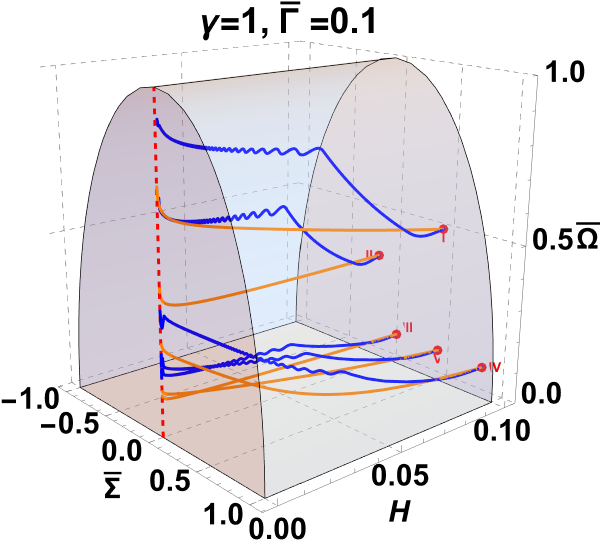}
        
\caption{\textbf{Interaction 4 \eqref{int-4}}. Three dimensional projections of the numerical solutions of the original system \eqref{comp-int-4-a}--\eqref{comp-int-4-d} (blue) and the averaged system \eqref{1-prom-int4}--\eqref{4-prom-int4} (orange). The red dashed line represents the sink set for the case $\gamma=1$.}
\label{fig:petitD}
\end{figure}

\begin{figure}[H]
\centering
    \includegraphics[scale=0.7]{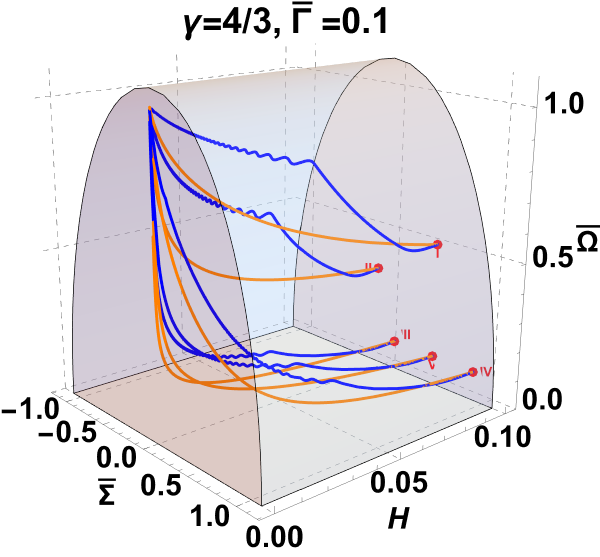}
    \includegraphics[scale=0.7]{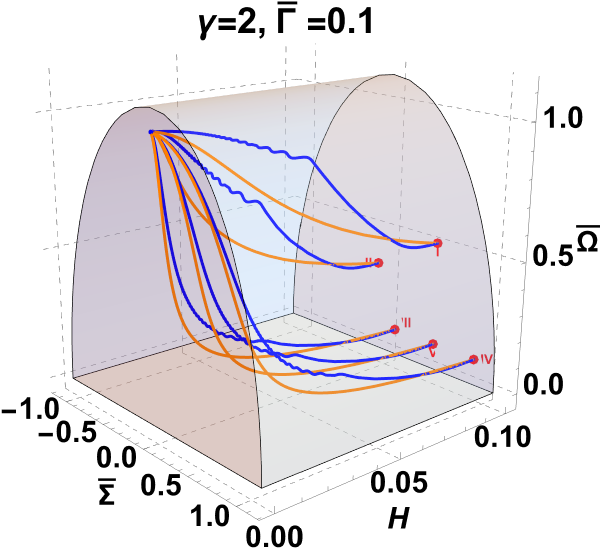}
    \caption{\textbf{Interaction 4 \eqref{int-4}}. Figure \ref{fig:petitD} continued for $\gamma=4/3,2$.}\label{fig:numéricas BI-Interaction-4}
\end{figure}

By considering Interaction 4 \eqref{int-4}, in  Figures \ref{fig:petitD} and \ref{fig:numéricas BI-Interaction-4}, an initial jump is observed due to the tangent function in the equations of the full system \eqref{comp-int-4-a}--\eqref{comp-int-4-d}, but both solutions converge to the same attractor.

\begin{figure}[H]
    \centering
    \includegraphics[scale=0.7]{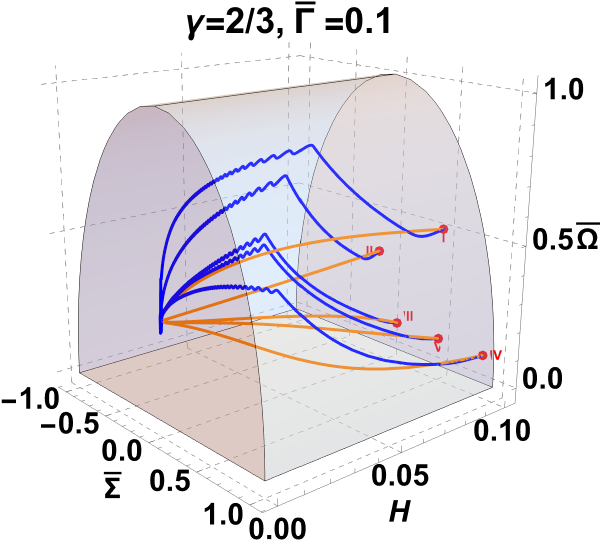}
    \includegraphics[scale=0.7]{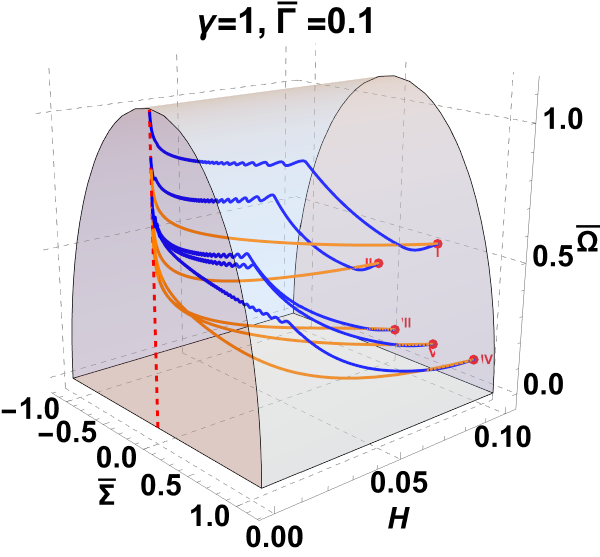}
        
\caption{\textbf{Interaction 5 \eqref{int-5}}. Three dimensional projections of the numerical solutions of the original system \eqref{comp-int-5-a}--\eqref{comp-int-5-d} (blue) and the averaged system \eqref{1-prom-int5}--\eqref{4-prom-int5} (orange). The red dashed line represents the sink set for the case $\gamma=1$.}
\label{fig:petitE}
\end{figure}

\begin{figure}[H]
\centering
    \includegraphics[scale=0.7]{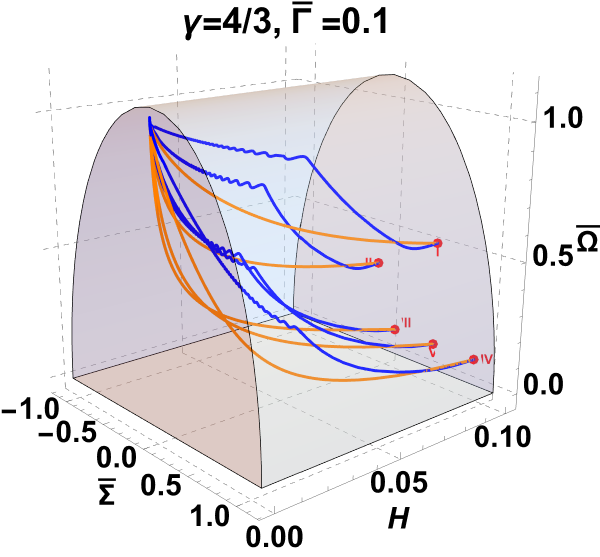}
    \includegraphics[scale=0.7]{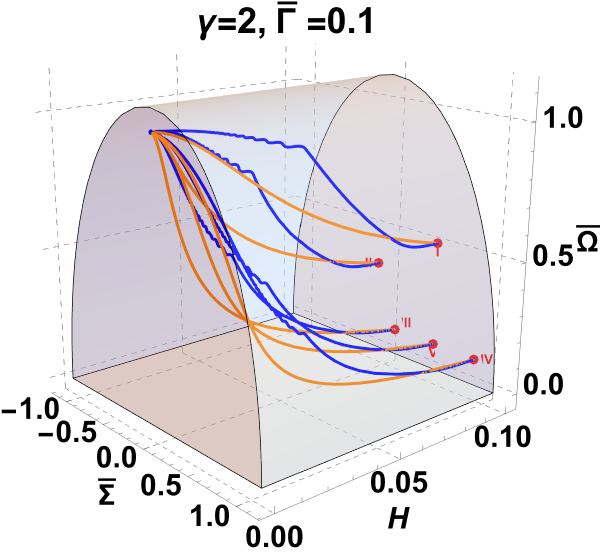}
    \caption{\textbf{Interaction 5 \eqref{int-5}}. Figure \ref{fig:petitE} continued for $\gamma=4/3,2$.}\label{fig:numéricas BI-Interaction-5}
\end{figure}

In Figures \ref{fig:petitE} and \ref{fig:numéricas BI-Interaction-5}, an initial jump (at early times) is again observed due to the presence of the tangent function in the equations of the full system \eqref{comp-int-5-a}--\eqref{comp-int-5-d}, arising from the choice of Interaction 5 \eqref{int-5}. However, both solutions converge to the same late-time attractor.

\begin{figure}[H]
    \centering
    \includegraphics[scale=0.7]{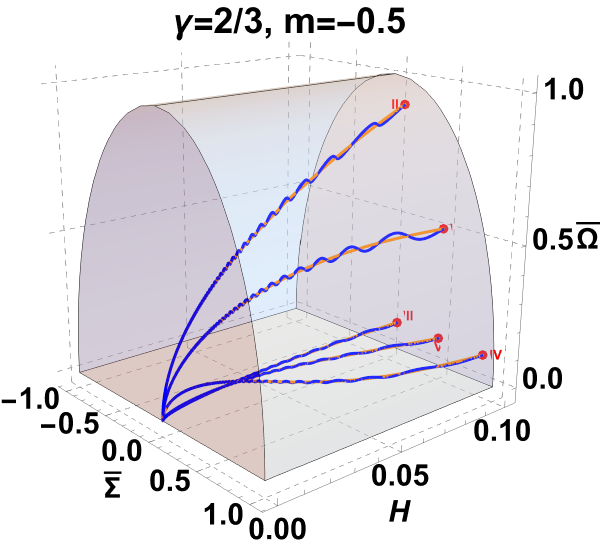}
    \includegraphics[scale=0.7]{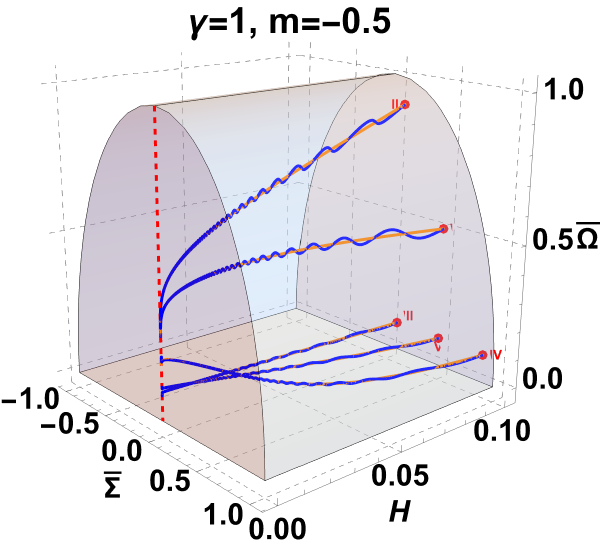}
        
\caption{\textbf{Interaction 6 \eqref{int-6}}. Projections of the numerical solutions of the original system \eqref{comp-int-6-a}–\eqref{comp-int-6-d} (blue) and the averaged system \eqref{1-prom-int6}–\eqref{4-prom-int6} (orange). The red dashed line represents the sink set for the case $\gamma = 1$.}
\label{fig:petitf}
\end{figure}

\begin{figure}[H]
    \centering
      \includegraphics[scale=0.7]{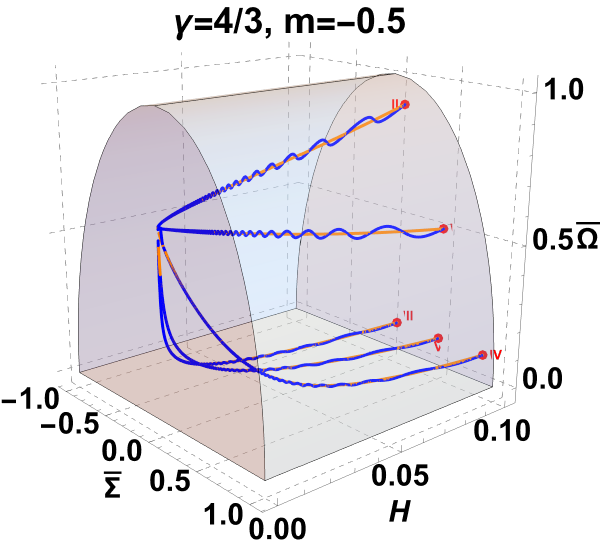}
    \includegraphics[scale=0.7]{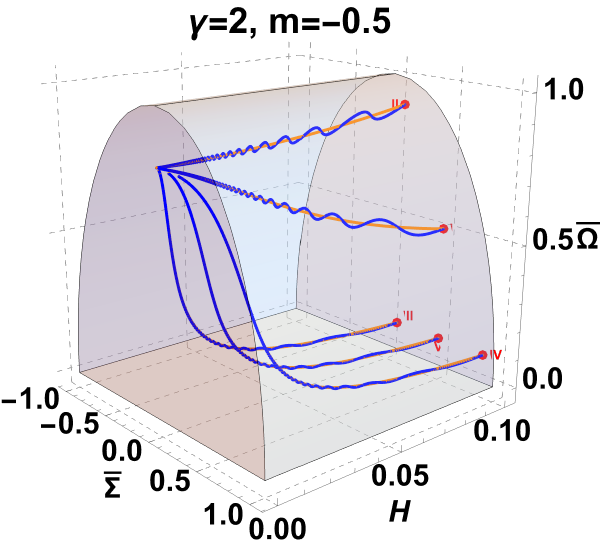}
        
\caption{\textbf{Interaction 6 \eqref{int-6}}. Figure \ref{fig:petitf} continued for $\gamma=4/3,2$.}
\label{fig:petitf-a}
\end{figure}

In Figures \ref{fig:petitf} and \ref{fig:petitf-a} the numerical solutions were calculated using Interaction 6 \eqref{int-6}. The behaviour is as expected, the oscillatory behaviour is suppressed during averaging process and the solutions have the same late-time behaviour.

\begin{figure}[H]
    \centering
    \includegraphics[scale=0.7]{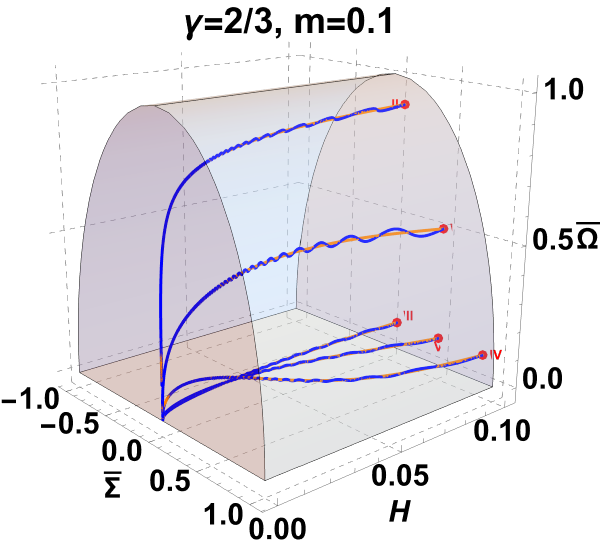}
    \includegraphics[scale=0.7]{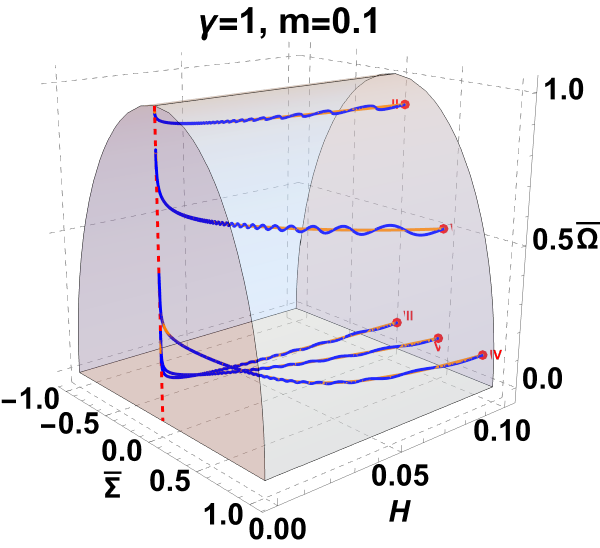}
        
\caption{\textbf{Interaction 7 \eqref{int-7}}. Projections of the numerical solutions of the original system \eqref{comp-int-7-a}–\eqref{comp-int-7-d} (blue) and the averaged system \eqref{1-prom-int7}–\eqref{4-prom-int7} (orange). The red dashed line represents the sink set for the case $\gamma = 1$.}
\label{fig:petitg}
\end{figure}

\begin{figure}[H]
    \centering
      \includegraphics[scale=0.7]{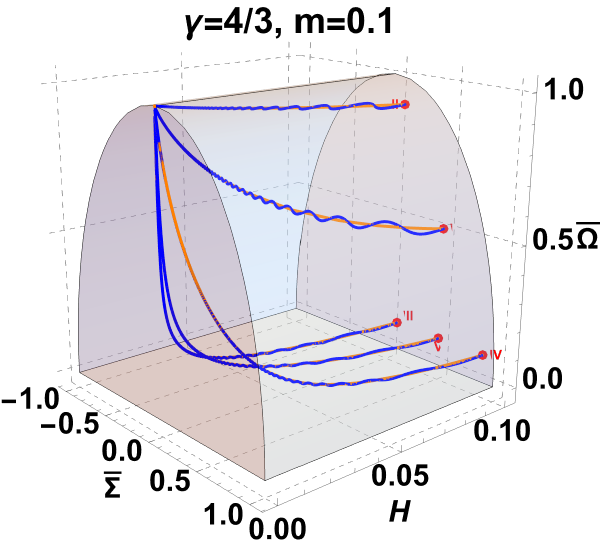}
    \includegraphics[scale=0.7]{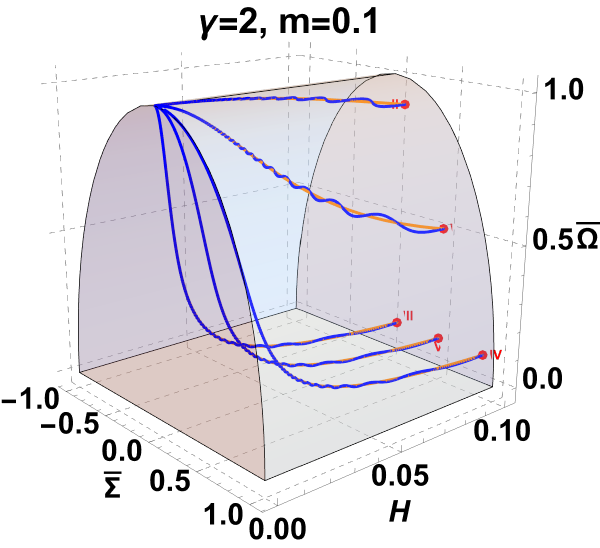}
        
\caption{\textbf{Interaction 7 \eqref{int-7}}. Figure \ref{fig:petitg} continued for $\gamma=4/3,2$.}
\label{fig:petitg-a}
\end{figure}

In Figures \ref{fig:petitg} and \ref{fig:petitg-a} the numerical solutions were calculated using Interaction 7 \eqref{int-7}. The behaviour is as expected, the oscillatory behaviour is suppressed in the averaging process and the solutions have the same late-time behaviour.

\begin{figure}[H]
    \centering
    \includegraphics[scale=0.7]{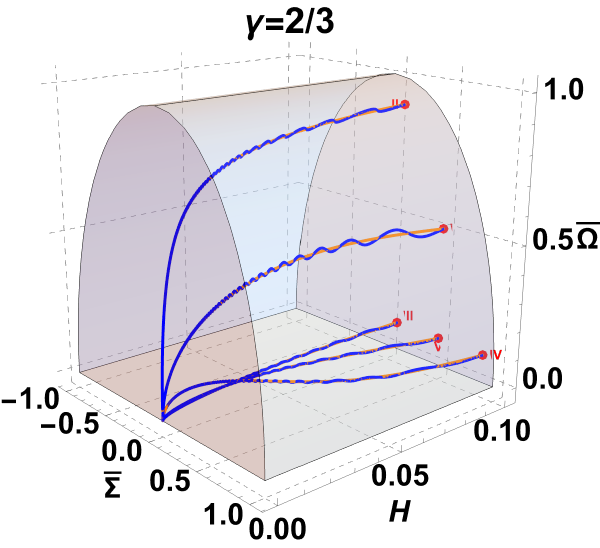}
    \includegraphics[scale=0.7]{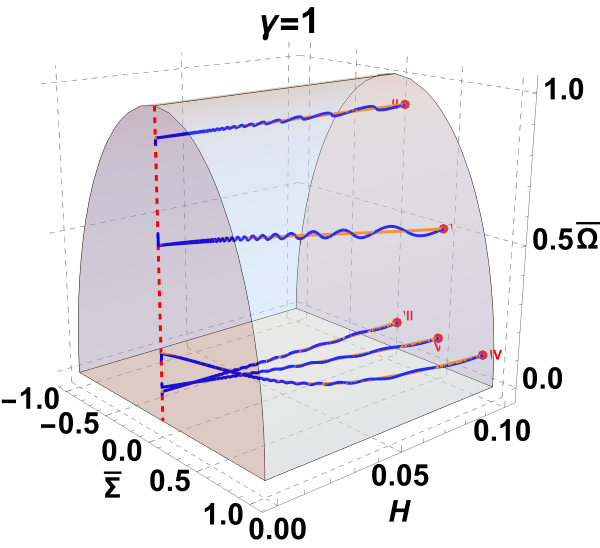}
        
\caption{\textbf{Interaction 8 \eqref{int-8}}. Projections of the numerical solutions of the original system \eqref{comp-int-8-a}–\eqref{comp-int-8-d} (blue) and the averaged system \eqref{1-prom-int8}–\eqref{4-prom-int8} (orange). The red dashed line represents the sink set for the case $\gamma = 1$.}
\label{fig:petith}
\end{figure}

\begin{figure}[H]
    \centering
      \includegraphics[scale=0.7]{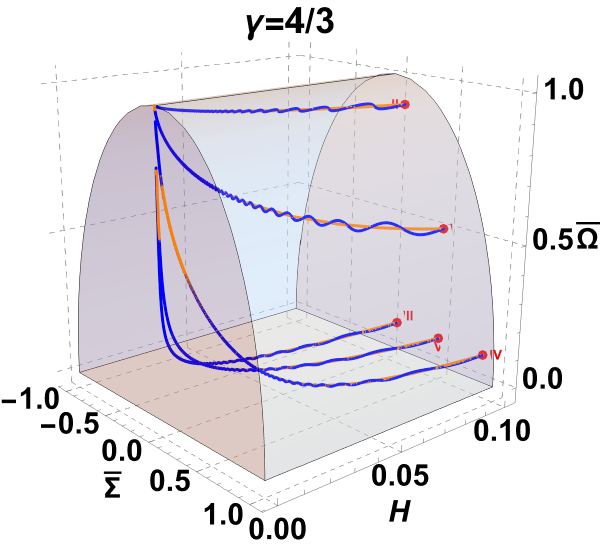}
    \includegraphics[scale=0.7]{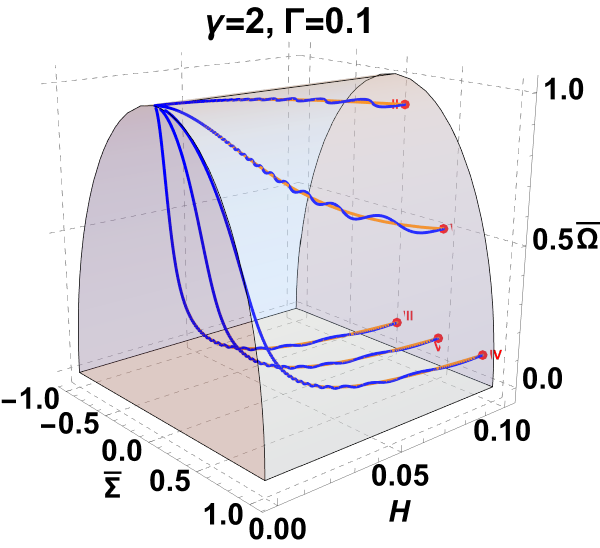}
        
\caption{\textbf{Interaction 8 \eqref{int-8}}. Figure \ref{fig:petith} continued for $\gamma=4/3,2$.}
\label{fig:petith-a}
\end{figure}

\begin{figure}[H]
    \centering
    \includegraphics[scale=0.7]{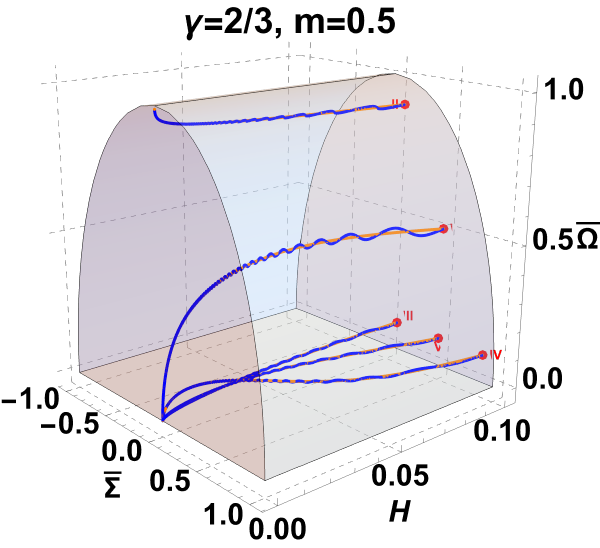}
    \includegraphics[scale=0.7]{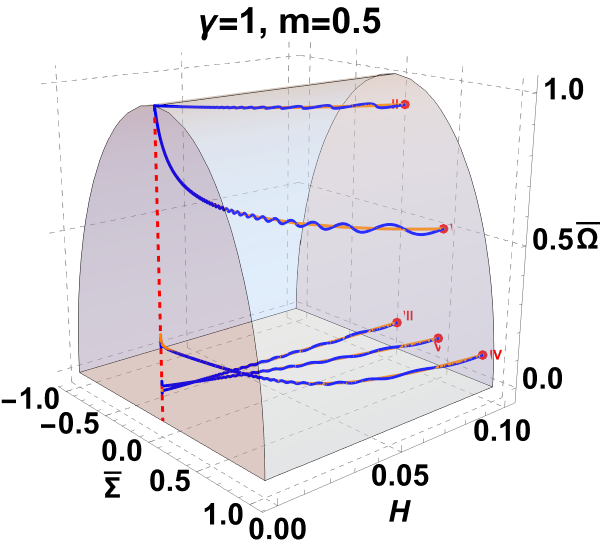}
        
\caption{\textbf{Interaction 9 \eqref{int-9}}. Projections of the numerical solutions of the original system \eqref{comp-int-9-a}–\eqref{comp-int-9-d} (blue) and the averaged system \eqref{1-prom-int9}–\eqref{4-prom-int9} (orange). The red dashed line represents the sink set for the case $\gamma = 1$.}
\label{fig:petiti}
\end{figure}

\begin{figure}[H]
    \centering
      \includegraphics[scale=0.7]{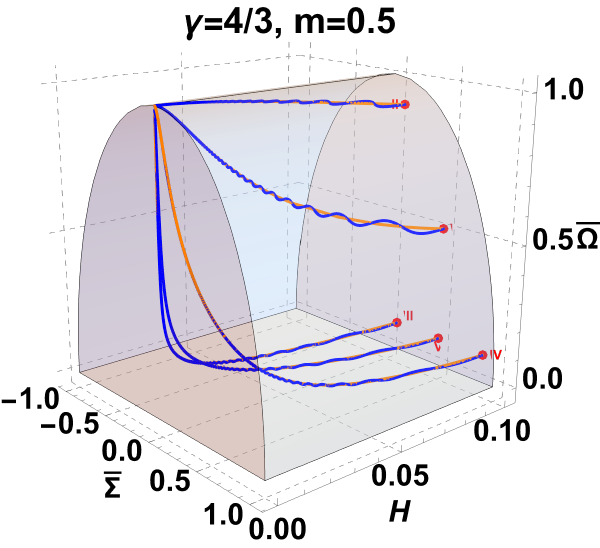}
    \includegraphics[scale=0.7]{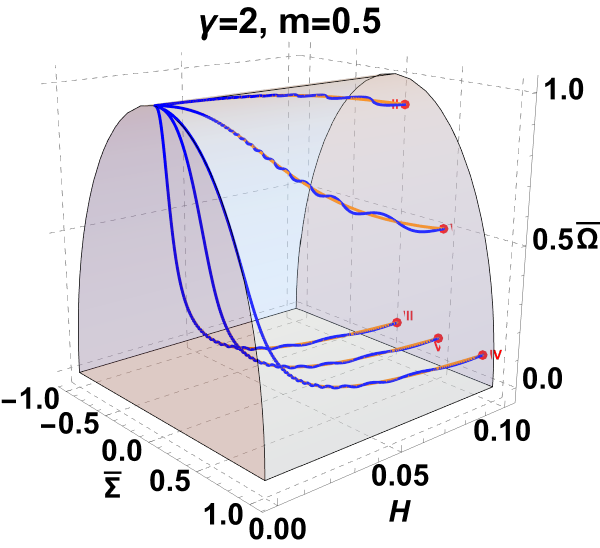}
        
\caption{\textbf{Interaction 9 \eqref{int-9}}. Figure \ref{fig:petiti} continued for $\gamma=4/3,2$.}
\label{fig:petiti-a}
\end{figure}

In Figures \ref{fig:petith} and \ref{fig:petith-a}, the numerical solutions were calculated using Interaction 8 \eqref{int-8}. As expected, the oscillatory behaviour in the original system is averaged out, and both systems exhibit the same qualitative behaviour at late times.
In Figures \ref{fig:petiti} and \ref{fig:petiti-a}, the numerical solutions correspond to Interaction 9 \eqref{int-9}. The results confirm that the averaging method effectively captures the long-term dynamics by removing fast oscillations while preserving the asymptotic behaviour.

\section{Conclusion}
\label{conclu-4}
In this research, the Locally Rotationally Symmetric Bianchi I spacetime with scalar field a nonzero matter component, with interaction was investigated. This analysis extends the previous studie \cite{Leon:2021rcx} by introducing the interactiong function $Q$ that characterizes the energy exchange between the different matter components of the system. The incorporation of $Q$ allows for the analysis of how these interactions impact the global dynamics and the stability of the equilibrium points, consequently, the evolution of the cosmological history and evolution. 

The function $Q$ used in this analysis is defined as:
\begin{equation}
    Q = \Gamma \left(\frac{H}{H_0}\right)^{1-\delta} \rho_m^\alpha \rho_\phi^{1-\alpha-\beta} (\rho_m + \rho_\phi)^\beta \dot{\phi}^\delta,
\end{equation}
\noindent
where $\Gamma$ is a coupling constant, $H$ is the Hubble function, $\rho_m$ and $\rho_\phi$ are the energy densities of matter and the scalar field, respectively, and $\dot{\phi}$ is the time derivative of the scalar field. The parameters $\alpha$, $\beta$, and $\delta$ determine how the different components contribute to the energy exchange.

Nine different models were considered:
\begin{itemize}
    \item \textbf{Interaction 1 \eqref{int-1}:} The dynamics coincide with those published in~\cite{Leon:2021rcx}, as the guiding system, critical points, and their stability remain unchanged, with no significant dynamical differences.
    \item \textbf{Interaction 2 \eqref{int-2}:} The rescaled parameter $\bar{\Gamma} = H_0 \Gamma$ affects the stability of the critical points. With $\bar{\Gamma} = \pm 0.1$, $P_1$ and $P_2$ exchange roles as attractors in the phase diagrams.
    \item \textbf{Interaction 3 \eqref{int-3}:} $\bar{\Gamma}$ influences the dynamics and the existence of a new equilibrium point, $P_5$, which is an attractor for $0 \leq \gamma < 1$ and $0 \leq \bar{\Gamma} < -3(\gamma - 1)$. However, the guiding system presents a singularity at $\Omega = 0$, which acts as an attractor when $\bar{\Gamma} = 0.1$, a behaviour that is not physically acceptable.
    \item \textbf{Interaction 4 \eqref{int-4}:} $\bar{\Gamma}$ determines the stability and the emergence of a new point $P_6$, which is an attractor for $1 < \gamma \leq 2$ and $-3(\gamma - 1) < \bar{\Gamma} \leq 0$.
    \item \textbf{Interaction 5 \eqref{int-5}:} A singularity at $\Omega = 0$ is again detected. Two new equilibrium points, $P_7$ and $P_8$, appear: $P_8$ is a sink for $0 \leq \gamma < 1$ and $0 \leq \bar{\Gamma} < \frac{3}{4}(1 - \gamma)$, while $P_7$ is a sink for $1 < \gamma \leq 2$ and $\frac{3}{4}(1 - \gamma) < \bar{\Gamma} \leq 0$. However, the issue of the line $\Omega = 0$ acting as an attractor persists when $\bar{\Gamma} \leq 0$.
    \item \textbf{Interaction 6 \eqref{int-6}:} A new equilibrium point appears in this case, point $P_9$ which is an attractor for $1<\gamma\leq 2,$ $m>0$ or $1<\gamma\leq 2,$ $m<0,$ where the parameter $m=\Gamma H_0$ influences the dynamics.
    \item \textbf{Interaction 7 \eqref{int-7}:} There are two new points $M_{1,2}.$ They only exist for $\gamma=2$ and belong to the line of equilibrium points defined by $\bar{\Omega}=0$ which is an attractor for $-\sqrt{\frac{m+1}{m}}<\bar{\Sigma}
   <\sqrt{\frac{m+1}{m}}.$
    \item \textbf{Interaction 8 \eqref{int-8}:} The equilibrium points and the dynamics are similar to that obtained for Interaction 1.
    \item \textbf{Interaction 9 \eqref{int-9}:} The new equilibrium point $N_1$ is an attractor for  $0\leq \gamma <1, \frac{-\gamma ^2+2
   \gamma -1}{\gamma -2}<m<1-\gamma $ or $
   1<\gamma \leq 2, m<1-\gamma .$
\end{itemize}

The tools of averaging theory and dynamical systems were successfully applied to these five new models, generalizing the analysis of the Bianchi I model \cite{Leon:2021rcx}. The results contributed to demonstrating the validity of the proposed models for analysing energy interactions in anisotropic cosmology. Specifically:
\begin{itemize}
    \item \textbf{Formulation of averaged equations:} These equations preserve the asymptotic behaviour of the original field equations, both in models without interaction between matter components and those with interaction.
    \item \textbf{Classification of equilibrium points:} A detailed classification of the equilibrium points in Bianchi I models with interaction between matter components was carried out, according to their stability.
\end{itemize}

This analysis, complemented by numerical solutions, shows how the parameters of $Q$ can modify the trajectories in the phase space and the asymptotic conditions of the system. The deceleration parameter was evaluated at each of the equilibrium points obtained. We can highlight the fact that there is an evolution in the behaviour of the deceleration parameter from a positive regime where $q_{\text{averaged}}$ describes deceleration and then it crosses the horizontal axis to describe an accelerated epoch as depicted in Figure \ref{evo-1-q-a}. Some numerical solutions were generated using a set of initial conditions summarised in Table \ref{Tab-final}. These solutions showed that the late-time behaviour of the solutions of both the original and averaged systems is the same. On the other hand, the role of $\delta$ was highlighted in modulating the oscillations of the scalar field and its influence on the model's attractor. 

In summary, this work introduces an extended theoretical framework to incorporate energy interactions in anisotropic cosmological models, providing tools to understand how these interactions affect the dynamics of the universe. Furthermore, it enables a more detailed analysis of existing systems and opens new possibilities for exploring nonlinear cosmological phenomena.

\begin{acknowledgments}
A. D. Millano was supported by Agencia Nacional de Investigación y Desarrollo -ANID- Subdirección de Capital Humano/Doctorado Nacional/año 2020- folio 21200837. The authors thank the support of VRIDT through Resoluci\'{o}n VRIDT No. 096/2022 and Resoluci\'{o}n VRIDT No. 098/2022. This work was financially supported by FONDECYT 1240514 ETAPA 2025.
\end{acknowledgments}
\bigskip

\bigskip

\bibliography{references}

\end{document}